\documentclass[11pt,a4paper,oneside]{article}
\pdfoutput=1

\usepackage[affil-it]{authblk}
\usepackage[a4paper,left=2.5cm,right=2.5cm,top=2.8cm,bottom=3.7cm]{geometry}
	
\usepackage[]{amsmath}
	\numberwithin{equation}{section}
\usepackage{amsthm}
\usepackage[]{amssymb}
\usepackage{amsfonts}
\usepackage{mathrsfs}

\DeclareMathOperator{\R}{\mathbb{R}}
\DeclareMathOperator{\C}{\mathbb{C}}
\DeclareMathOperator{\Z}{\mathbb{Z}}
\DeclareMathOperator{\cs}{\mathbb{S}}
\newcommand{\mz}{\mathcal{Z}}
\newcommand{\mf}{\mathcal{F}}
\newcommand{\mN}{\mathcal{N}}
\newcommand{\mC}{\mathcal{C}}
\newcommand{\mH}{\mathcal{H}}
\newcommand{\fC}{\mathfrak{C}}
\newcommand{\fwC}{\widetilde{\mathfrak{C}}}
\newcommand{\mM}{\mathscr{M}}
\newcommand{\mX}{\mathcal{X}}
\newcommand{\fX}{\mathfrak{X}}

\newcommand{\Gr}{\mathscr{G}\!\mathrm{r}}

\newcommand{\um}{\underline{m}}
\newcommand{\un}{\underline{\mathsf{w}}}
\newcommand{\uN}{\underline{N}}

\newcommand{\sn}{w}

\newcommand{\dd}{\mathrm{d}}
\newcommand{\ii}{\mathrm{i}}
\newcommand{\wt}{\mathrm{wt}}

\newcommand{\dbr}{\begin{color}{blue}\bullet\end{color}}
\newcommand{\nsbr}{\begin{color}{red}\bullet\end{color}}

\newcommand{\U}[1]{\mathrm{U}\!\left( #1 \right)}
\newcommand{\SU}[1]{\mathrm{SU}\!\left( #1 \right)}

\newcommand{\YM}{Yang--Mills}
\newcommand{\hk}{hyperK\"{a}hler}

\newcommand{\uz}{\underline{z}}
\newcommand{\ux}{\underline{x}}
\newcommand{\uy}{\underline{y}}

\newtheorem{mainthm}{Main result}

\theoremstyle{definition}
\newtheorem{algorit}{Algorithm}
\newtheorem{defin}{Definition}

\usepackage[utf8]{inputenc}
\usepackage[english]{babel}

\usepackage[pdftex]{graphicx}
\usepackage[dvipsnames]{xcolor}

\usepackage[linktocpage=true,colorlinks=true,citecolor=Blue,linkcolor=blue,urlcolor=Blue]{hyperref}

\usepackage[toc,page]{appendix}
\usepackage[nottoc]{tocbibind}
\usepackage{caption}
\usepackage{enumerate}

\usepackage{ytableau}

\usepackage{tikz}
	\usetikzlibrary{calc,shapes,backgrounds,er,positioning,arrows.meta,shapes.geometric}

\usepackage[numbers,compress,square,comma]{natbib}	

\begin{document}
\bibliographystyle{ourstyle}
\captionsetup[figure]{labelfont={bf,small},labelformat={default},labelsep=period,font=small}

{\pagenumbering{roman} 

		\renewcommand*{\thefootnote}{\fnsymbol{footnote}}
	\title{\textbf{\huge Crystal bases and three-dimensional $\mathcal{N}=4$ Coulomb branches}}

		\author[$\clubsuit$]{Leonardo Santilli\footnote{lsantilli@fc.ul.pt}}
	\affil[$\clubsuit$]{\small Grupo de F\'{\i}sica Matem\'{a}tica, Departamento de Matem\'{a}tica, Faculdade de Ci\^{e}ncias, Universidade de Lisboa, Campo Grande, Edif\'{\i}cio C6, 1749-016 Lisboa, Portugal.}
	
	\author[$\spadesuit$,$\clubsuit$]{Miguel Tierz\footnote{tierz@mat.ucm.es, tierz@fc.ul.pt}}
	\affil[$\spadesuit$]{\small Departamento de An\'alisis Matem\'atico y Matem\'atica Aplicada, Universidad Complutense de Madrid, Plaza de las Ciencias, 3, 28040 Madrid, Spain}

	\date{\vspace{0.5cm}}

	\maketitle
	\thispagestyle{empty}

	\begin{abstract}
	We establish and develop a correspondence between certain crystal bases (Kashiwara crystals) and the Coulomb branch of three-dimensional $ \mathcal{N} =4 $ gauge theories. The result holds for simply-laced, non-simply laced and affine quivers. Two equivalent derivations are given in the non-simply laced case, either by application of the axiomatic rules or by folding a simply-laced quiver. We also study the effect of turning on real masses and the ensuing simplification of the crystal. We present a multitude of explicit examples of the equivalence. Finally, we put forward a correspondence between infinite crystals and Hilbert spaces of theories with isolated vacua.
		
	\end{abstract}

	\clearpage
	\tableofcontents
	\thispagestyle{empty}
}

	\clearpage
	\pagenumbering{arabic}
	\setcounter{page}{1}
		\renewcommand*{\thefootnote}{\arabic{footnote}}
		\setcounter{footnote}{0}

\section{Introduction}

Quantum field theories with supersymmetry have been a staple of research in fundamental Physics for more than half a century. The interest in their study has surpassed the already formidable framework of attempting to understand the physical world, leading to very central discoveries in Mathematics. There is an extremely rich body of work on this remarkable interplay between Physics and Mathematics, but we shall focus on a specific corner of that interdisciplinary endeavor, the one dedicated to the study of moduli spaces of vacua of supersymmetric quiver gauge theories.\par
One far-reaching lesson from supersymmetric quantum field theories is that a wealth of field-theoretical questions can be rephrased solely in terms of algebraic varieties. A pivotal role in this program is played by quivers, combinatorial objects that encode information of both supersymmetric field theories \cite{Douglas:1996sw} and algebraic varieties \cite{Nakajima:1994nid}. Throughout this work, a quiver will always be assumed to be modelled after a symmetrizable Kac--Moody algebra \cite{kac1990infinite}.\par
\medskip
The central objects of study in the present work are three-dimensional quiver gauge theories with eight supercharges \cite{Seiberg:1996nz}, that is to say, with $\mN=4$ supersymmetry. Additionally, we restrict our attention to theories that are \emph{good} in the Gaiotto--Witten classification \cite{Gaiotto:2008ak}.\par
Typically, quiver theories with eight supercharges have moduli spaces of vacua containing two distinguished branches, called Coulomb and Higgs branch. As reviewed in Section \ref{sec:modulireview}, the Higgs branch is \hk~\cite{Hitchin:1986ea}, regardless of the dimension of the spacetime in which the theory lives. Conversely, the nature of the Coulomb branch changes as the spacetime dimension is varied.\par
\medskip
Three-dimensional $\mN=4$ Coulomb branches are \hk~varieties \cite{Seiberg:1996nz}, but their most general and comprehensive characterization remains elusive. The systematic study of these algebraic varieties building on their field-theoretical realization has been initiated in \cite{Bullimore:2015lsa}.\par
Alternatively, the algebro-geometric properties of the Coulomb branch can be accessed computing its Hilbert series from the gauge theory data. The monopole formula \cite{Cremonesi:2013lqa} yields the Coulomb branch Hilbert series of any three-dimensional $\mathcal{N}=4$ good theory in the form of a combinatorial expression. Efforts to interpret the monopole formula as the Poincar\'e polynomial of a cohomology led to the successful definition and study of Coulomb branches from a mathematical standpoint \cite{Nakajima:2015txa,Braverman:2016wma,Braverman:2016pwk} (related work includes but is not limited to \cite{Finkelberg:2017nbc,Braverman:2018gvt,Kamnitzer:2018,Muthiah:2019jif,Weekes:2019cwk,Nakajima:2019olw,Dancer:2020wll,Weekes:2020rgb,Zhou:2020bwa}). One of the main outcomes of this line of research is that three-dimensional $\mN=4$ Coulomb branches are slices in the affine Grassmannian, a result envisioned in \cite{Bullimore:2015lsa} and rigorously formalized in \cite{Braverman:2016pwk}. The quantization of the Coulomb branch coordinate ring then follows as a byproduct \cite{Bullimore:2015lsa,Braverman:2016wma}, with the quantized Coulomb branch amenable to be analyzed using localization techniques \cite{Dedushenko:2017avn,Dedushenko:2018icp} (see also \cite{Beem:2016cbd} for a complementary, bootstrap-oriented analysis).\par
We ought to mention that the first appearance of the affine Grassmannian in the study of three-dimensional Coulomb branches dates back to the work \cite{Kapustin:2006pk}, in which these moduli spaces are characterized as the moduli spaces of singular monopoles in $\R^3$. We elaborate further on the connection among the various approaches at the end of Section \ref{sec:modulireview}.\par
A convenient way to produce many (but certainly not all) three-dimensional $\mN=4$ quiver gauge theories relies on the Hanany--Witten brane setups in type IIB string theory \cite{Hanany:1996ie}. The connection among Hanany--Witten brane systems, the associated quiver gauge theories and the affine Grassmannian has been investigated in \cite{Bourget:2021siw}.\par
\medskip
Quivers appear in the study of quantum groups as well. Lusztig used them to construct bases, known as canonical bases, for the quantization of the associated enveloping algebras \cite{Lusztig90a,Lusztig90b,Lusztig91}. Parallel work of Kashiwara \cite{Kashiwara90,Kashiwara91,Kashiwara95} showed the existence of crystal bases on the same quantized enveloping algebras. The bases of Lusztig and Kashiwara are equivalent \cite{Grojnowski:93}.\par
\begin{figure}[htb]
\centering
	\begin{tikzpicture}
	\node[text=Black,draw,align=center,rectangle,rounded corners] (CB) at (0,0) {{\small 3d $\mN=4$}\\ {\small Coulomb branch}};
	\node (QV) at (-4,1.5) {\small Quiver variety};
	\node[align=center] (AG) at (4,1.5) {{\small Slice in the}\\ {\small affine Grassmannian}};
	\node[] (KC) at (0,-2) {\small Kashiwara crystal};
	\draw (KC)--(QV);
	\draw (KC)--(AG);
	\draw (QV)--(AG);
	\end{tikzpicture}
\caption{Three realizations of 3d $\mathcal{N}=4$ Coulomb branches.}
\label{fig:CBtriangle}
\end{figure}
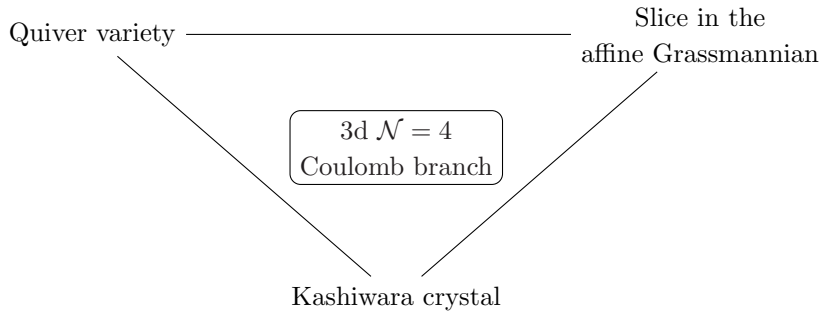

The existence of a Kashiwara crystal structure on slices in the affine Grassmannian was established by Braverman and Gaitsgory \cite{Braverman99}. We will rederive in this work this statement from the physics of three-dimensional $\mathcal{N}=4$ Coulomb branches. Conversely, such a mathematically rigorous result provides us with yet another tool to understand the Coulomb branches, and uncovers an additional characterization of these multi-faceted symplectic varieties, as schematically drawn in Figure \ref{fig:CBtriangle}.
\par
More specifically, in this work we establish a one-to-one correspondence between certain Kashiwara crystals and three-dimensional $\mN=4$ Coulomb branches. We give a direct and explicit derivation based on quantum field theory as well as on Hanany--Witten brane setups \cite{Hanany:1996ie}. The result is rigorously proved in two steps:
\begin{equation*}
	 \text{\small Coulomb branch} \xrightarrow{\quad \text{\cite{Braverman:2016pwk}} \quad }  \text{ \small Slice in the affine Grassmannian } \xrightarrow{\quad \text{\cite{Braverman99}} \quad } \text{\small Kashiwara crystal} .
\end{equation*}
We emphasize that the correspondence is not limited to type A quivers, but holds for finite, both simply and non-simply laced, and affine quivers. In particular, we have two derivations in the non-simply laced case: by application of the axiomatic rules or by folding a parent simply-laced quiver. The two procedures are shown to give the same output. In addition, we study the effect of turning on real masses, showing how the crystal simplifies for generic mass deformations, and how it changes discontinuously at positive-codimensional loci in parameter space.\par
\medskip
In our use of Kashiwara crystals, we have relied considerably on the recent presentation by Bump and Schilling \cite{BumpSchilling} where, based not only on the results of Kashiwara, but also on the later work by Stembridge \cite{stembridge2003local}, an axiomatic combinatorial approach to the construction of crystal bases associated to quantum-deformed enveloping algebras of finite-dimensional Lie algebras is developed. This approach can be of significant advantage when there is already a background or familiarity with the combinatorics of Young tableaux, such as in the construction of the irreducible representations of the symmetric and general linear groups, by means of such tableaux.\par
Finally, we stress that a considerable amount of the material in this paper might be already known to experts. Indeed, various of our results may be derived by chains of known relationships in algebraic geometry and geometric representation theory. One of the goals of the present work is to reorganize the salient material in the different areas involved in such a way as to hopefully illuminate further on the logic of the subject and to make it accessible to a different audience, while providing an explicit and algorithmic presentation, which makes the Coulomb branch physics transparent and is easily implemented in a computer algebra system such as \textsc{sage} \cite{sage}.\par
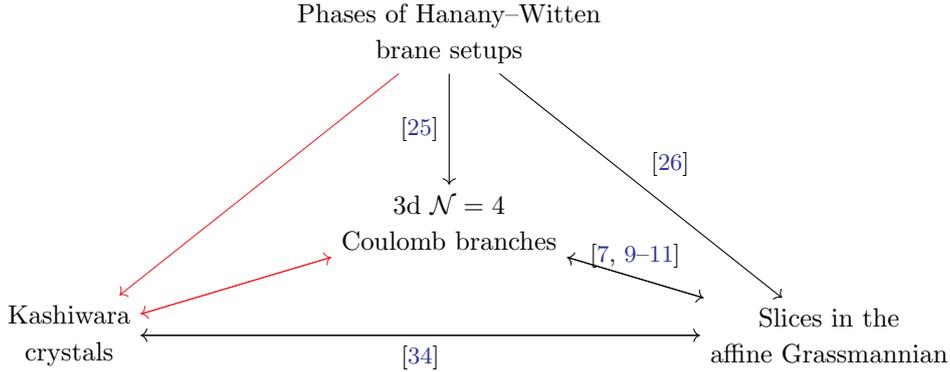
\begin{figure}[htb]
\centering
\begin{tikzpicture}[bend angle=10]
	\node[align=center] (CB) at (0,0) {{\small 3d $\mN=4$}\\ {\small Coulomb branches}};
	\node[align=center] (HW) at (0,2.5) {{\small Phases of Hanany--Witten}\\ {\small brane setups}};
	\node[align=center] (AG) at (5,-1.5) {{\small Slices in the}\\ {\small affine Grassmannian}};
	\node[align=center] (KC) at (-5,-1.5) {{\small Kashiwara}\\ {\small crystals}};
	
	\path[Red,->] (KC) edge (CB);
	\path[Red,->] (CB) edge (KC);
	\path[->] (AG) edge node[pos=0.5,anchor=south] {\footnotesize \cite{Bullimore:2015lsa,Nakajima:2015txa,Braverman:2016wma,Braverman:2016pwk}} (CB);
	\path[->] (CB) edge (AG);
	\path[->] (HW) edge node[anchor=east] {\footnotesize \cite{Hanany:1996ie}} (CB);
	\path[->] (HW) edge node[anchor=south west] {\footnotesize \cite{Bourget:2021siw}} (AG);
	\path[Red,->] (HW) edge (KC);
	\path[->] (AG) edge node[pos=0.5,anchor=north] {\footnotesize \cite{Braverman99}} (KC);
	\path[->] (KC) edge (AG);
	\end{tikzpicture}
\caption{Summary of the topics covered in the body of the paper, and their web of relations.}
\label{fig:CBsummary}
\end{figure}\par
\medskip
A summary of the topics encompassed by our analysis, with the web of their interrelations, is depicted in Figure \ref{fig:CBsummary}. The main novelties introduced by our work are highlighted in red.\par
The paper is organized as follows. Section \ref{sec:3dN4CBreview} gives a brief introduction to three-dimensional $\mathcal{N}=4$ theories and their Coulomb branches. After that, we present our main result in Section \ref{sec:C=C}. In Section \ref{sec:CrystalAxioms} we axiomatically introduce a family of Kashiwara crystals and state their correspondence with Coulomb branches. Then, in Sections \ref{sec:quiversub}-\ref{sec:crystalBrane} this correspondence is derived directly from the Coulomb branch physics and from brane setups, respectively. Section \ref{sec:CBKCAexample} contains several examples that illustrate our claims in type A. The notion of resolved crystal is introduced in Section \ref{sec:massdef}, and therein exploited to study the effect of mass deformations. We then show explicitly how the previous general results apply to type D quivers in Section \ref{sec:Dquivercrystal}, to non-simply laced quivers in Section \ref{sec:BCquivercrystal}, and to affine type A quivers in Section \ref{sec:affinecrystalquiver}.\par
In Section \ref{sec:Demazure} we discuss a different class of crystals, which we relate to the Hilbert spaces of theories with isolated vacua, through their Coulomb branch Verma module structure \cite{Bullimore:2016hdc}. We conclude in Section \ref{sec:outlook} with an outlook on potential avenues for future research, mostly aimed at expanding upon Section \ref{sec:Demazure}. The text is complemented with two appendices. Appendix \ref{app:ExamplesC} substantiates the results of the main text with an extensive list of detailed examples. Appendix \ref{app:indlim} briefly discusses an inductive limit on the crystals.\par

\section{3d $\mathcal{N}=4$ Coulomb branches and the affine Grassmannian}
\label{sec:3dN4CBreview}
The purpose of this introductory section is to present our notation and conventions. We begin in Section \ref{sec:notation} with basic mathematical definitions that will be needed in the rest of the work. Section \ref{sec:modulireview} contains a lightning overview of the moduli spaces of vacua of 3d $\mathcal{N}=4$ theories \cite{Seiberg:1996nz,Gaiotto:2008ak}, while Section \ref{sec:param} reviews the parameter spaces of interest.

\subsection{Setup and notation}
\label{sec:notation}
\subsubsection{Groups and lattices}
Let $G$ be a complex, reductive, algebraic group of rank $r$, and let $\mathfrak{g}$ be the corresponding Lie algebra. $\mathfrak{g}$ will be assumed to be a symmetrizable Kac--Moody algebra \cite{kac1990infinite}. In most cases it will be taken to be a finite classical or affine type A Lie algebra. The Langlands dual group to $G$ is ${}^{\scriptstyle L} G$ and the corresponding Lie algebra is denoted by ${}^{\scriptstyle L} \mathfrak{g}$. For example, if $G=\mathrm{SL}(r+1)$, then ${}^{\scriptstyle L} G = \mathrm{PSL} (r+1)$ and ${}^{\scriptstyle L} \mathfrak{g} \cong \mathfrak{g} = \mathfrak{su} (r+1)$.\par
Let $\mathbb{T}_G$ be the maximal torus of $G$. The weight lattice of $G$ is $\Lambda_{\mathrm{w}} = \mathrm{Hom} (\mathbb{T}_G, \mathbb{C}^{\ast})$ and coweight lattice of $G$ is denoted by $\Lambda_{\mathrm{w}} ^{\vee}$ \cite{Bumpbook}. The weight and coweight lattices of ${}^{\scriptstyle L} G$ are $\Lambda_{\mathrm{w}} ^{\vee}$ and $\Lambda_{\mathrm{w}} $, respectively \cite{Bumpbook}. Besides, $\Phi $ will denote the set of roots of $G$, with $\triangle $ being the subset of simple positive roots. Langlands duality exchanges roots and coroots \cite{Bumpbook}.\par

\subsubsection{The affine Grassmannian}
Let $\mathscr{O} = \mathbb{C} [\![ t ]\!]$ denote the ring of Taylor series in $t$ and $\mathscr{K} = \mathbb{C} (\!( t )\!)$ the field of Laurent series in $t$. That is, $\mathscr{K}$ contains functions of the form $\sum_{k \in \mathbb{Z}} c_k t^k$ with arbitrarily but finitely many $c_k \ne 0$ with $k \le 0$. For a matrix group $G$, the notation $G (\mathscr{K})$ (resp. $G (\mathscr{O})$) indicates the group of matrices with entries from $\mathscr{K}$ (resp. from $\mathscr{O}$) taking values in $G$.\par
\begin{defin}The \emph{affine Grassmannian} of $G$ is 
\begin{equation*}
	\Gr _G = G (\mathscr{K}) / G (\mathscr{O}) .
\end{equation*}
\end{defin}
It consists of $\pi_1 ( G )$ many connected components.

\subsubsection{Quivers}
Let $\mathsf{Q}$ be a framed quiver and denote by $\mathsf{Q}^{\circ} \subset \mathsf{Q}$ the naturally associated unframed quiver. We adopt the standard notation for theories with eight supercharges, in which unoriented edges mean pairs of complex conjugate edges.\par
The quiver $\mathsf{Q}^{\circ}$ will be taken to be shaped like the Dynkin diagram of a symmetrizable Kac--Moody algebra $\mathfrak{g}$. We denote by $r= \mathrm{rk} (\mathfrak{g})$ the number of nodes of $\mathsf{Q}^{\circ}$, and we will use an index $j= 1, \dots, r$ running over them.\par
With this in mind, $\mathsf{Q}$ is characterized by a generalized Dynkin diagram together with two arrays of integers, 
\begin{equation}
\label{eq:notationNn}
	\uN = (N_1, \dots , N_r ) , \qquad \un = (\mathsf{w}_1, \dots, \mathsf{w}_r )
\end{equation}
specifying the ranks of gauge and flavour nodes.\footnote{In the mathematical literature, $(N_1, \dots, N_r)$ is more often denoted $(\mathsf{v}_1, \dots, \mathsf{v}_r)$.}
\begin{defin}
    A quiver $ \mathsf{Q} $ is called \emph{balanced} if \cite{Gaiotto:2008ak} 
\begin{equation}
\label{eq:balancecond}
	2 N_j - \sum_{j^{\prime}  :  (j \to j^{\prime}) \in \mathrm{Edges} (\mathsf{Q}^{\circ})  } N_{j^{\prime}} = \mathsf{w}_j , \qquad \forall j \in \mathrm{Nodes} (\mathsf{Q}^{\circ}) 
\end{equation}
with the sum running over the nodes $j^{\prime} \in \mathrm{Nodes} (\mathsf{Q}^{\circ}) $ that are connected to the node $j$. For $A (\mathfrak{g})$ the generalized Cartan matrix of $\mathfrak{g}$, relation \eqref{eq:balancecond} is written $ A (\mathfrak{g}) \uN = \un $.
\end{defin}
For more on quivers and quiver varieties, we refer to \cite{Nakajima:1994nid,Ginzburg:2009}.\par

\subsection{Moduli spaces of vacua of 3d $\mathcal{N}=4$ theories}
\label{sec:modulireview}
A 3d $\mathcal{N}=4$ supersymmetric gauge theory is specified by two pieces of data: a gauge group $\mathbf{G}$ and a (generically, reducible) quaternionic representation $\mathfrak{R}$ of it. These choices are conveniently encoded in a framed quiver $\mathsf{Q}$. The unframed quiver $\mathsf{Q}^{\circ} \subset \mathsf{Q}$ captures the gauge group, while the framing captures the flavour symmetry. We will always assume that the gauge group $\mathbf{G}$ is a product of unitary groups,
\begin{equation}
\label{eq:GaugeGroupU}
	\mathbf{G} = \prod_{j \in \mathrm{Nodes} \left( \mathsf{Q}^{\circ} \right) }  \U{N_j} .
\end{equation}\par
The field content of the gauge theory consists of a vector multiplet, in the adjoint representation of the Lie algebra of $\mathbf{G}$, and a hypermultiplet determined by the representation $\mathfrak{R}$. The R-symmetry is $\SU{2}_C \times \SU{2}_H$. Deformations of the theory are parametrized by $\SU{2}_C$ triplets of masses $\vec{m}$ and by $\SU{2}_H$ triplets of Fayet--Iliopoulos (FI) parameters $\vec{\zeta}$.\par
These theories admit intricate moduli spaces of vacua, parametrizing flat directions for scalar zero-modes in the vector and hypermultiplet. The moduli spaces have two distinguished branches.
\begin{itemize}
	\item The Coulomb branch, $\mC$. It is a \hk~manifold with $\mathrm{SU}(2)_C$ action that rotates the triplet of complex structures. For ADE quivers, $\mC$ is a slice in the corresponding affine Grassmannian \cite{Bullimore:2015lsa,Braverman:2016pwk,Bourget:2021siw}. It is a holomorphic-symplectic singularity \cite{Braverman:2016wma,Weekes:2020rgb} parametrized by dressed magnetic monopoles \cite{Tong:1998fa}.
	\item The Higgs branch, $\mH$, along which the gauge group is broken to a finite (possibly trivial) subgroup. It is a singular \hk~quotient \cite{Hitchin:1986ea} with $\mathrm{SU}(2)_H$ rotating the three complex structures. Besides, it is protected against quantum corrections. It follows from early mathematical works \cite{KP,KS} (and also from \cite{Namikawa}) that, for a wealth of ADE quivers, $\mH \left[ \mathsf{Q} \right]$ is the closure of a nilpotent orbit of the Lie algebra $\mathfrak{g}$ into which $\mathsf{Q}^{\circ}$ is shaped \cite{Gaiotto:2008ak}. The singularity structure of closures of nilpotent orbits is then reinterpreted as the phase diagram of the gauge theory, a claim tested and successfully reproduced in a vast class of examples \cite{Hanany:2016gbz,Cabrera:2016vvv,Cabrera:2017ucb,Hanany:2018uzt,Hanany:2019tji}.
\end{itemize}
Additionally, the moduli spaces of vacua include the so-called mixed branches, that intersect $\mC$ and $\mH$ non-trivially at singular loci of the two branches. Physically, a mixed branch corresponds to a partial Higgsing.\par
As emphasized in \cite{Bourget:2019aer,Grimminger:2020dmg}, the singularity structure of Higgs and Coulomb branches is described by the symplectic foliation of a symplectic singularity \cite{Beauville,Kaledin}. Geometrically, real masses partially resolve the Coulomb branch singularity, because hypermultiplet modes become massless at separated points on $\mC$, rendering the singularity less severe. For generic masses, the Higgs branch is lifted.\par
Three-dimensional $\mN=4$ theories enjoy mirror symmetry \cite{Intriligator:1996ex}, an IR duality that relates certain pairs of quivers $\mathsf{Q}, \mathsf{Q}^{\vee}$ according to 
\begin{equation*}
	\mathcal{H} \left[ \mathsf{Q} \right]   \cong  \mathcal{C}  \left[ \mathsf{Q}^{\vee} \right] , \qquad  \mathcal{C} \left[ \mathsf{Q} \right]  \cong  \mathcal{H}  \left[ \mathsf{Q}^{\vee} \right] , \qquad  \left( \vec{m}, \vec{ \zeta} \right) = \left(  \vec{ \zeta}^{\vee} , \vec{m}^{\vee}\right) .
\end{equation*}

\subsubsection{Monopoles, effective field theories and crystals}
Having reviewed the structure of $\mC$, let us briefly comment on three approaches to study it.\par
A direct approach consists in analyzing the effective field theory (EFT) on the 3d $\mN=4$ Coulomb branch \cite{Seiberg:1996nz}. The singularities of $\mC$ are then understood in this picture in terms of modes that are massless at positive codimensional loci on $\mC$. The EFT description breaks down at these loci and, moving onto them, we get a new EFT with less moduli. In what follows we will implicitly think of the Kashiwara crystal construction as a realization of this EFT perspective.\par
The description of $\mC$ as a slice in $\Gr_{{}^L G}$, in turn, stems from studying the Coulomb branch chiral ring, generated by dressed $\mathbf{G}$-monopoles \cite{Bullimore:2015lsa}, where $\mathbf{G}$ is the gauge group \eqref{eq:GaugeGroupU}.\par
A connection between $\mC$ and $\Gr_{{}^L G}$ was established earlier in \cite{Kapustin:2006pk}. The approach of \cite{Kapustin:2006pk} realizes the Coulomb branch as the moduli space of singular $G$-monopoles in $\R^3$ with monopole charge $\un$. The singularities of $\mC$ are due to collision of monopoles, and the induced monopole bubbling effect reproduces the stratification of slices in $\Gr_{{}^L G}$ \cite{Kapustin:2006pk}. This latter perspective laid the groundwork to explore the interrelations between moduli spaces of $G$-monopoles, quantum groups $U_q ({}^{\scriptstyle L } \mathfrak{g})$ and the KZ equation (see \cite{Aganagic:2020olg,Aganagic:2021ubp} and references therein). The appearance of Kashiwara crystals may equivalently be foreseen from this alternative approach.

\subsection{Parameter space}
\label{sec:param}
Let $G_F$ be the flavour symmetry group. Our choice \eqref{eq:GaugeGroupU} of gauge group implies that\footnote{The diagonal factor $\U{1}_{\text{diag}}$ coincides with a gauged $\U{1} \subset \mathbf{G}$ and hence is not a genuine global symmetry.} 
\begin{equation*}
G_F = \left[  \U{\mathsf{w}_1} \times  \U{\mathsf{w}_2} \times \cdots \times  \U{\mathsf{w}_r} \right] / \U{1}_{\text{diag}} = \mathrm{PS} \left[  \U{\mathsf{w}_1} \times  \U{\mathsf{w}_2} \times \cdots \times  \U{\mathsf{w}_r} \right] ,
\end{equation*}
and we let
\begin{equation*}
	n=  \sum_{j=1} ^{r} \mathsf{w}_j
\end{equation*}
in the above notation. In what follows, we will only turn on real masses.\par
The parameter space $\mM^{(n)}$ of real masses is stratified:
\begin{equation*}
	\mM ^{(n)} = \left\{ \um = (m_1, \dots, m_n) \in \mathbb{R}^{n} \ : \  m_1 \ge m_2 \ge \cdots \ge m_n \text{ and } \sum_{j=1}^{n} m_j = 0 \right\} \equiv \bigsqcup _{\lambda \in \mathbb{Y}_n } \mM_{\lambda} ^{(n)} ,
\end{equation*}
where $\mathbb{Y}_n$ denotes the set of all partitions of $n$. $\mM_{\lambda} ^{(n)}$ is the parameter space with exactly $\lambda_1$ hypers of equal mass $m_{1}= \dots = m_{\lambda_1}$, exactly $\lambda_2$ hypers of equal mass $m_{\lambda_1 + 1}= \dots = m_{\lambda_1 + \lambda_2}$ (but $m_{\lambda_1 } \ne m_{\lambda_1 + 1}$), and so on.\par
The singularity of $\mC$ is partly resolved in this setup, with the massless case corresponding to $\lambda = (n)$, whereas generic masses correspond to $\lambda = (1^n)$. We denote the corresponding partial resolution as 
\begin{equation}
\label{eq:defresol}
	\mX_{\lambda} [\mathsf{Q}] \to \mC [\mathsf{Q}] , \qquad \text{ if } \um \in \mM_{\lambda} ^{(n)}  .
\end{equation}
To lighten the notation, we will often omit the dependence on $\mathsf{Q}$. Furthermore, we will denote by $\mX = \mX (\um)$ the piecewise-constant function that gives the variety $\mX_{\lambda}$ when $\um \in \mM_{\lambda} ^{(n)}$.\par
A given $\lambda$ breaks the flavour group $G_F$ to a block-diagonal subgroup, with the flavour bundle splitting accordingly. In addition, whenever $\lambda$ has $s \in \mathbb{N}$ rows of equal length there is the action of a discrete automorphism group $S_{s}$ permuting the $s$ groups of hypermultiplets.\par
Notice that $\mM^{(n)} \cong H^2 (\mathcal{C} [\mathsf{Q}], \mathbb{R}) $ is the restriction of the Cartan subalgebra $\mathfrak{t}_F$ of the flavour symmetry algebra to the principal Weyl chamber, and $\lambda \ne (1^n)$ specifies a wall of such chamber.

\section{Coulomb branches are Kashiwara crystals}
\label{sec:C=C}

In this section we introduce Kashiwara crystals \cite{Kashiwara90,Kashiwara91,Kashiwara95} and present their relationship with $\mC$. Crystal bases appear in the crystal limit $q \to 0$ of the representation theory of quantum groups $U_q ({}^{\scriptstyle L } \mathfrak{g})$ \cite{Kashiwara95,Hong}. We will adopt their combinatorial definition \cite{BumpSchilling}.
\begin{mainthm}
\label{thm:1}
	The symplectic foliation of $\mC$ is described by a Kashiwara crystal $\fC$.
\end{mainthm}\par
Before delving into the details of this correspondence, we ought to emphasize that Result \ref{thm:1} is not genuinely new. Let us sketch results in the literature that overlap with ours.
\begin{itemize}
	\item The connection between slices in the affine Grassmannian and Kashiwara crystals was already analyzed in \cite{Braverman99}. Then, our first result follows from \cite{Braverman99} together with the statement that $\mC[\mathsf{Q}]$ corresponds to a slice in the affine Grassmannian \cite{Braverman:2016pwk}. Nevertheless, our aim is to make the connection with field theory as transparent as possible. The approach of \cite{Braverman99} and the one in the present work are compared in Section \ref{sec:CrystalAxioms}.
	\item A class of crystals to be discussed in Section \ref{sec:Demazure} (and differing from the ones discussed in the rest of the text) admits a geometric construction hinged on quiver varieties \cite{Kashiwara:97}.
	\item There exists a triangle of isomorphisms \cite{MV03}
		\begin{equation*}
			\begin{tikzpicture}
				\node (N) at (0,1) {Nakajima quiver variety};
				\node[anchor=east] (G) at (-1,0) {Slice in the affine Grassmannian};
				\node[anchor=west] (O) at (1,0) {Closure of nilpotent orbit};
				\path[->] (N) edge node[anchor=east,pos=0.35] {$ \scriptstyle \cong$ \hspace{4pt} } (G);
				\path[->] (G) edge node[above] {$\scriptstyle \cong$} (O);
				\path[->] (O) edge node[anchor=west,pos=0.65] { \hspace{4pt}  $\scriptstyle \cong $} (N);
			\end{tikzpicture}
		\end{equation*}
		As argued by Dranowski \cite{Dranowski}, one can label strata of nilpotent orbit closures by Young tableaux, and then track the images of the tableaux across the isomorphisms. This mechanism underpins a crystal structure for bases in each vertex of the triangle. The presentation in Section \ref{sec:Demazure} is akin to the work \cite{Dranowski}, and slightly overlaps with it, as elucidated in due course.
\end{itemize}
Underlying these many facets of the bond between quivers and crystals is the observation that quiver varieties encode the structure of quantum groups and their canonical bases \cite{Lusztig91}, together with the equivalence between Lusztig's canonical bases and Kashiwara's crystal bases \cite{Grojnowski:93,Dranowski}.\par
Our goal is to give an explicit, combinatorial construction that makes manifest the relation among crystals, Coulomb branches, brane configurations and the affine Grassmannian, adding one more layer to the analysis of \cite{Bourget:2021siw}. In this section we mostly focus on type A quivers, although we keep the definitions general for later convenience.

\subsection{Crystals from axioms and Coulomb branches}
\label{sec:CrystalAxioms}
We now define the Kashiwara crystals, closely following the monograph \cite{BumpSchilling}, up to few mild variations that will make the correspondence with Coulomb branches more manifest. Then, we construct crystals that capture the geometry of Coulomb branches.\par

\subsubsection{Definition of Kashiwara crystals}
Fix $n \in \mathbb{N}$ and let $\sigma = (\sigma_1, \sigma_2, \dots, \sigma_{\ell} )$ be a partition of $n$, identified with the Young diagram whose $j^{\text{th}}$ row consists of $\sigma_j$ boxes. Besides, fix an alphabet $[r] \equiv \left\{ 0,1,2, \dots, r , r+1 \right\}$ (note the inclusion of $0$ and $r+1$). A semi-standard Young tableau $T$ of shape $\sigma $ is a filling of the Young diagram $\sigma $ with letters from the alphabet $[r]$ such that (i) each row is non-decreasing from left to right and (ii) each column is strictly increasing from top to bottom. Moreover, there exists a weight function $\wt$ on tableaux, which is defined as 
\begin{equation*}
	\wt (T) = \left( \mathsf{w}_1, \dots, \mathsf{w}_r \right) , \qquad \mathsf{w}_i = \text{ \# of times the letter $i$ appears in $T$} ,
\end{equation*}
and takes values in a certain weight lattice $\Lambda$ to be identified below.
\begin{defin}
	Fix a root system $ \Phi $ and a weight lattice $\Lambda_{\mathrm{w}}$. Denote by $j$ the index running over the index set of $ \Phi $. A \emph{Kashiwara crystal} is a non-empty collection $\fC _{\sigma}$ of semi-standard Young tableaux of shape $\sigma$ together with maps
	\begin{equation*}
	\begin{aligned}
		e_j , f_j  & : \fC _{\sigma} \to \fC _{\sigma} \sqcup \left\{ 0 \right\} \\
		\varepsilon_j  , \varphi_j & : \fC _{\sigma} \to \mathbb{Z} \sqcup \left\{ - \infty \right\} \\
		\wt & : \fC _{\sigma} \to \Lambda_{\mathrm{w}}
	\end{aligned}
	\end{equation*}
	satisfying the following conditions:
	\begin{enumerate}[(i)]
		\item $e_j (T^{\prime}) =T \Leftrightarrow f_j (T) = T^{\prime} $ for any two $T, T^{\prime} \in \fC_{\sigma}$. If this holds, then 
		\begin{equation*}
			\wt (T) = \wt (T^{\prime}) + \alpha_j 
		\end{equation*}
		and $\varepsilon_j (T) = \varepsilon_j (T^{\prime}) -1$, $\varphi_j (T) = \varphi_j (T^{\prime}) -1$.
		\item $\varphi_j (T) = \varepsilon_j (T) + \left( \wt (T), \alpha_j \right)$ for all $T \in \fC_{\sigma}$.
	\end{enumerate}
	The operators $e_j , f_j $ are called \emph{Kashiwara operators}.
\end{defin}
In what follows we will only be interested in the so-called \emph{normal} crystals \cite{BumpSchilling}, that satisfy 
\begin{equation*}
	\varepsilon _j (T) = \max \left\{ k \ : \ e_j ^{k} (T) \ne 0 \right\} , \qquad \varphi _j (T) = \max \left\{ k \ : \ f_j ^{k} (T) \ne 0 \right\} .
\end{equation*}
Thus, we henceforth neglect the maps $\varepsilon_j, \varphi_j$ as they are uniquely fixed by the rest of the data.\par
\begin{defin}
	Let $\mu \in \Lambda_{\mathrm{w}}$ be a highest weight. A Kashiwara crystal $\fC$ is called a \emph{highest weight} crystal of highest weight $\mu$ if there exists a tableaux $T_{\mu} \in \fC$ such that (i) $\wt (T_{\mu}) = \mu$, (ii) $e_j (T_{\mu}) = 0$ $\forall j$ and (iii) every $ T^{\prime} \in \fC$ is reached acting on $T_{\mu}$ with some sequence of Kashiwara operators $f_j$. Besides, we will call $T_{\mu}$ the highest weight tableaux of $\fC$.
\end{defin}
If $\fC$ is a normal and highest weight crystal of weight $\mu$, then necessarily $\mu$ is a dominant weight \cite{Braverman99,BumpSchilling}.

\subsubsection{Crystals from quivers}
Now that we have set the stage, we associate a Kashiwara crystal $\fC \left[ \mathsf{Q} \right]$ to any framed quiver $\mathsf{Q}$.
\begin{algorit}\label{algorithm1}
	Let $\mathsf{Q}$ be a framed quiver shaped like the Dynkin diagram of $\mathfrak{g}$, with gauge and flavour nodes specified by $\uN =(N_1, \dots, N_r)$ and $\un = (\mathsf{w}_1, \dots, \mathsf{w}_r)$, respectively. Associate a Kashiwara crystal $\fC [\mathsf{Q}]$ to it as follows.
	\begin{enumerate}[1.]
		\item $\fC [\mathsf{Q}]$ consists of tableaux shaped as the single-row partition $(n)$ of $n= \sum_{j=1}^{r} \mathsf{w}_j$.
		\item The alphabet is $[r]$.
		\item The Kashiwara operators are built from the root system $\Phi^{\vee}$ and the weight lattice $\Lambda_{\mathrm{w}} ^{\vee}$ of ${}^{\scriptstyle L} G$.
		\item $\fC [\mathsf{Q}]$ is the highest weight normal crystal of highest weight $\un$.
		\item Amputate the crystal after $N_j$ transitions involving the letter $j$ have been performed, $\forall j=1, \dots, r$. If $\mathsf{Q}$ is balanced, this step has no effect.
	\end{enumerate}
\end{algorit}
Notice the appearance of Langlands duality. On the geometric side, this aspect was discussed in \cite{Braverman99,BK,Krylov}. The underlying physical reason is that Coulomb branches are moduli spaces of magnetic objects \cite{Tong:1998fa,Cremonesi:2013lqa}.\par
As a further remark, we observe that, acting on the highest weight tableaux with the simple roots $\alpha_1 ^{\vee}, \alpha_r ^{\vee}$ of ${}^{\scriptstyle L} G $ will give rise to the letters $0,r+1$. These are not accounted for by the weight function $\wt$, since they do not contribute to form a ${}^{\scriptstyle L} G $-weight.\par
\medskip
At this point, we are ready to reformulate in a rigorous manner our main result. To this aim, we introduce the notion of phase diagram of a symplectic singularity.
\begin{defin}
\label{def:PhaseDiag}
	Let $X$ be any Poisson variety. The \emph{phase diagram} $\mathscr{P} (X)$ of $X$ is the oriented graph whose vertices are the leaves of the symplectic foliation of $X$ and whose arrows are the transverse slices among the leaves, oriented toward decreasing dimension.
\end{defin}
When $X$ is $\mC [\mathsf{Q}]$, this definition encapsulates the standard physical definition of phase diagram of a 3d $\mN=4$ theory on its Coulomb branch. Besides, our definition of phase diagram is intimately related to the notion of Hasse diagram, which is a diagram representing the partial order of a set. When applied to Poisson varieties, and more specifically to branches of the moduli spaces of vacua of a 3d $\mN=4$ theory (as e.g. in \cite{Bourget:2019aer,Grimminger:2020dmg}), the Hasse diagram agrees with the phase diagram as defined above.\footnote{The Hasse diagram of a Poisson variety $X$ carries in fact more information than $\mathscr{P} (X)$, as it encodes the geometry of the transverse slices. In Definition \ref{def:PhaseDiag} we only require the phase diagram to discern the distinct transverse slices. In practice, the geometry of the transverse slices can be fixed comparing with a handful of known examples. Let us also mention that the edges in the Hasse diagram are usually unoriented.}
\begin{mainthm}[Precise statement of Result \ref{thm:1}]
\label{thm:precise}
Let $\mathsf{Q}$ be a quiver shaped as the generalized Dynkin diagram of a symmetrizable Kac--Moody algebra $\mathfrak{g}$. Denote by $G$ the reductive algebraic group associated to $\mathfrak{g}$. Let $\mC [\mathsf{Q}]$ be the Coulomb branch of the associated 3d $\mN=4$ gauge theory and $\fC [\mathsf{Q}]$ the Kashiwara crystal yielded by Algorithm \ref{algorithm1}. Then $ \fC [\mathsf{Q}] \cong \mathscr{P} \left( \mC [\mathsf{Q}] \right)$ is an isomorphism of oriented graphs.
\end{mainthm}
As mentioned above, a mathematical proof of this statement can be given, combining \cite{Braverman99} (and subsequent works) with the recent results on the mathematical formulation of 3d $\mN=4$ Coulomb branches \cite{Braverman:2016wma,Braverman:2016pwk,Nakajima:2019olw}. In summary, we are simply giving a different presentation of the crystals in \cite{Braverman99}, which describe $\Gr _{{}^L G}$ slices. The latter, in turn, capture $\mC [\mathsf{Q}]$. Notice that this reasoning would give a stronger result than a congruence of graphs, as it deals directly with algebraic varieties.\par
We do not give further details, since this would not add to the existing literature. Instead, our aim is to present our algorithmic construction in a variety of examples, making the physics more transparent.\par
The starting point of Algorithm \ref{algorithm1} is the highest weight tableau
\begin{equation*}
	\ytableausetup{smalltableaux} 
	\begin{tikzpicture}
	\node (tr) at (4,0) {$\begin{ytableau} {\scriptstyle 1 } & {\scriptstyle 1} & \none[ \scriptstyle \cdots ] & {\scriptstyle 1} & {\scriptstyle 2} & \none[ \scriptstyle \cdots ] & {\scriptstyle 2} & \none[ \scriptstyle \cdots ] & {\scriptstyle r} & \none[ \scriptstyle \cdots ] & {\scriptstyle r} \end{ytableau} $};
	\node (t3) at (4,-0.3) {$\underbrace{\begin{ytableau} \none[ \ ] & \none[ \ ] & \none[ \ ] & \none[ \ ] \end{ytableau}}_{\mathsf{w}_1}  \underbrace{ \begin{ytableau} \none[ \ ] & \none[ \ ] & \none[ \ ] \end{ytableau} }_{\mathsf{w}_2} \begin{ytableau} \none[ \ ]  \end{ytableau} \underbrace{ \begin{ytableau} \none[ \ ] & \none[ \ ] & \none[ \ ] \end{ytableau} }_{\mathsf{w}_r}$};
	\end{tikzpicture}
\end{equation*}
of weight $\wt = (\mathsf{w}_1, \dots, \mathsf{w}_r)$. Then we start changing $\wt$ either with a simple root $\alpha_j ^{\vee} \in \triangle ^{\vee}$ of ${}^{\scriptstyle L} G$ (i.e. a coroot of $G$) or with a combination $\sum_j \mathsf{a}_j \alpha_j ^{\vee} \in \Phi^{\vee}$, $ (\mathsf{a}_1, \dots, \mathsf{a}_r) \in \mathbb{N}^r$. However, each such combination can be reached in various steps, each step being a change of $\wt$ by a simple root or by a combination $\sum_{j= k} ^{k+p} \alpha_j ^{\vee}$ for some $k, k+p \in \left\{ 1, \dots, r \right\} $. Therefore we see that, by construction, Kashiwara operators are in correspondence with transverse slices to the symplectic singularity, either being a du Val singularity $A_{p}$ or the closure of a minimal nilpotent orbit $a_p,b_p,c_p,d_p, e_{6,7,8}$.\footnote{We only consider unitary gauge groups \eqref{eq:GaugeGroupU}, thus we never get du Val singularities other than type A.}\par

\subsubsection{A convenient bijection}
For a more convenient visualization, as well as to insist on the analogy with quivers, we will adopt a different presentation of tableaux, in which we replace a string of $\mathsf{w}_j$ boxes filled with the letter $j$ by a single box labelled with $w_j \equiv \mathsf{w}_j +1$. In other words, we draw the weight $\wt (T)$, with entries shifted by 1, instead of $T$ itself. With this in mind, we get the bijection 
\begin{equation}
	\ytableausetup{smalltableaux} 
	\begin{tikzpicture}
	\node (tl) at (-2,0) {$ \begin{ytableau} {\scriptscriptstyle w_1 } & \none[\scriptstyle \otimes] & {\scriptscriptstyle w_2} & \none[\scriptstyle \otimes] &\none[ \scriptstyle \cdots ] & \none[\scriptstyle \otimes] & {\scriptscriptstyle w_r} \end{ytableau}$};
	\node (map) at (0,0) {$\overset{ 1:1 }{\longleftrightarrow}$};
	\node (tr) at (2.5,0) {$\begin{ytableau} {\scriptstyle 1 } & {\scriptstyle 1} & \none[ \scriptstyle \cdots ] & {\scriptstyle 1} & {\scriptstyle 2} & \none[ \scriptstyle \cdots ] & {\scriptstyle 2} & \none[ \scriptstyle \cdots ] & {\scriptstyle r} & \none[ \scriptstyle \cdots ] & {\scriptstyle r} \end{ytableau} $};
	\node (t3) at (2.5,-0.3) {$\underbrace{\begin{ytableau} \none[ ] & \none[ ] & \none[ ] & \none[ ] \end{ytableau}}_{\mathsf{w}_1}  \underbrace{ \begin{ytableau} \none[ ] & \none[ ] & \none[  ] \end{ytableau} }_{\mathsf{w}_2} \begin{ytableau} \none[  ]  \end{ytableau} \underbrace{ \begin{ytableau} \none[  ] & \none[  ] & \none[  ] \end{ytableau} }_{\mathsf{w}_r}$};
	\end{tikzpicture}
	\label{eq:bijcrys}
\end{equation}
together with the mapping of the Kashiwara operators $e_j,f_j$ on the right into operators that have the same action, but change the filling of the tableau rather than the weight. Notice that, under the map \eqref{eq:bijcrys}, the roles of $r$ and $n$ are exchanged.\par
For instance, take the $A_1$ quiver with $\mathsf{w}_1=n$ flavours and $N= \lfloor \frac{n}{2} \rfloor$, so that it is either balanced, for $n$ even, or minimally unbalanced, for $n$ odd. The corresponding crystals are:
\begin{equation}
	\ytableausetup{nosmalltableaux}
		\begin{tikzpicture}
		\node (pe) at (0,1) {$n$ even:};
		\node (po) at (4,1) {$n$ odd:};

		\node (t1c) at (0,0) {$\begin{ytableau} {\scriptstyle n +1} \end{ytableau}$};
		\node (t2c) at (0,-1.5) {$\begin{ytableau}{\scriptstyle n-1} \end{ytableau}$};
		\node (t3c) at (0,-3) {$\vdots$};
		\node (t4c) at (0,-4.5) {$\begin{ytableau}{\scriptstyle 3 } \end{ytableau}$};
		\node (t5c) at (0,-6) {$\begin{ytableau}{\scriptstyle 1 } \end{ytableau}$};

		\path[->] (t1c) edge node[left] {${\scriptstyle n-1}$} (t2c);
		\path[->] (t2c) edge node[left] {${\scriptstyle n-3}$} (t3c);
		\path[->] (t3c) edge node[left] {${\scriptstyle 3}$} (t4c);
		\path[->] (t4c) edge node[left] {${\scriptstyle 1}$} (t5c);

		\node (t1r) at (4,0) {$\begin{ytableau} {\scriptstyle n+1} \end{ytableau}$};
		\node (t2r) at (4,-1.5) {$\begin{ytableau} {\scriptstyle n-1} \end{ytableau}$};
		\node (t3r) at (4,-3) {$\vdots$};
		\node (t5r) at (4,-5.25) {$\begin{ytableau} {\scriptstyle 2 }\end{ytableau}$};

		\path[->] (t1r) edge node[right] {${\scriptstyle n-1}$} (t2r);
		\path[->] (t2r) edge node[right] {${\scriptstyle n-3}$} (t3r);
		\path[->] (t3r) edge node[right] {${\scriptstyle 2}$} (t5r);
		
		\end{tikzpicture}
\label{eq:PSU2ex}
\end{equation}
which reproduce the $\pi_1 (\mathrm{PSL(2)}) =2$ components of $\Gr _{\mathrm{PSL(2)}}$. In this way, the coweight of $G$ associated to each component is directly read off from the content of the tableau.\par
As another warm-up example, consider the $A_2$ quiver $\U{2} \times \U{2} $ with two hypermultiplets at each node. The corresponding crystal $\fC \left[ \overset{ 2}{\Box} -  \overset{ 2}{\circ}-  \overset{ 2}{\circ} - \overset{ 2}{\Box} \right]$ is 
\begin{equation*}
\ytableausetup{smalltableaux}
	\begin{tikzpicture}
		\node (t1) at (0,0) {$\begin{ytableau} 3 & \none[ \scriptstyle \otimes ] & 3 \end{ytableau}$};
		\node (t2a) at (-1,-1) {$\begin{ytableau} 4 & \none[ \scriptstyle \otimes ] & 1 \end{ytableau}$};
		\node (t2b) at (1,-1) {$\begin{ytableau} 1 & \none[ \scriptstyle \otimes ] & 4 \end{ytableau}$};
		\node (t3) at (0,-2) {$\begin{ytableau} 2 & \none[ \scriptstyle \otimes ] & 2 \end{ytableau}$};
		\node (t4) at (0,-3) {$\begin{ytableau} 1 & \none[ \scriptstyle \otimes ] & 1 \end{ytableau}$};
		
		\path[->] (t1) edge node[left,anchor=south east] {${\scriptstyle A_1}$} (t2a);
		\path[->] (t1) edge node[right,anchor=south west] {${\scriptstyle A_1}$} (t2b);
		\path[->] (t2a) edge node[left] {${\scriptstyle A_2}$} (t3);
		\path[->] (t2b) edge node[right] {${\scriptstyle A_2}$} (t3);
		\path[->] (t3) edge node[right] {${\scriptstyle a_2}$} (t4);

		\node (v1) at (-5.5,0) {$\begin{ytableau} 1 &1 & 2 & 2 \end{ytableau}$};
		\node (v2a) at (-7,-1) {$\begin{ytableau} 1 & 1  & 1 & 3 \end{ytableau}$};
		\node (v2b) at (-4,-1) {$\begin{ytableau} 0 & 2 & 2 & 2  \end{ytableau}$};
		\node (v3) at (-5.5,-2) {$\begin{ytableau} 0 &1 & 2 & 3 \end{ytableau}$};
		\node (v4) at (-5.5,-3) {$\begin{ytableau} 0 & 0 & 3 & 3 \end{ytableau}$};
		
		\path[->] (v1) edge (v2a);
		\path[->] (v1) edge (v2b);
		\path[->] (v2a) edge (v3);
		\path[->] (v2b) edge (v3);
		\path[->] (v3) edge (v4);
		
		
		\node (a1) at (-2.5,0) {$\longleftrightarrow$};
		\node (a2) at (-2.5,-1) {$\longleftrightarrow$};
		\node (a3) at (-2.5,-2) {$\longleftrightarrow$};
		\node (a3) at (-2.5,-3) {$\longleftrightarrow$};
		\node[anchor=south] (l1) at (-2.5,0) {\footnotesize \eqref{eq:bijcrys}};
				
	\end{tikzpicture}
\end{equation*}
which agrees with the slice in $\Gr _{\mathrm{PSL} (3)}$ for the given quiver.

\subsection{Crystals from quiver subtraction}
\label{sec:quiversub}

A useful way to visualize leaves and slices in the affine Grassmannian is quiver subtraction \cite{Cabrera:2018ann}. The idea is roughly as follows. Starting with the quiver $\mathsf{Q}$, minimal transitions correspond to identify a ``minimal'' sub-quiver of $\mathsf{Q}$ and to subtract it, so producing a new quiver whose Coulomb branch is a lower-dimensional slice in $\Gr _{{}^{L} G}$.\par
The rigorous statement underneath the quiver subtraction is that slices in the affine Grassmannian are stratified and each stratum is a quiver variety \cite{MV03}.\par
In the Kashiwara crystal setting, to each quiver $\mathsf{Q}$ we associate the highest weight tableau $T$ in the crystal $\fC [\mathsf{Q}]$ and, if a quiver $\mathsf{Q}^{\prime}$ is one of the possible outputs of subtracting a minimal quiver to $\mathsf{Q}$, the corresponding highest weight tableau $T^{\prime}$ descends from $T$ acting with a suitable Kashiwara operator. The consistency of this procedure follows immediately from Algorithm \ref{algorithm1}. In turn, this allows a direct and visual check of all our results against those obtained via quiver subtraction.
\begin{mainthm}
	Consider a quiver $\mathsf{Q} $ and all its descendant via quiver subtraction. There is a one-to-one correspondence between tableaux $T \in \fC \left[ \mathsf{Q} \right]$ and descendant quivers of $\mathsf{Q}$. Besides, transitions $T \to T^{\prime}$ in $\fC \left[ \mathsf{Q} \right]$ are in one-to-one correspondence with quivers that describe transverse slices to the singular loci in $\mC \left[ \mathsf{Q} \right]$.
\end{mainthm}
Examples are provided below.

\subsection{Crystals from branes}
\label{sec:crystalBrane}

A convenient way to describe 3d $\mN=4$ theories on their Coulomb branches is via an arrangement of D3 branes suspended between NS5 branes in type IIB string theory \cite{Hanany:1996ie}. Consider the following setup \cite{Hanany:1996ie}:
\begin{equation*}
	\begin{tabular}{l|c|c|c|c|c|c|c|c|c|c|}
		 $ \ $ & $x^{0}$ & $x^{1}$& $x^{2}$& $x^{3}$& $x^{4}$& $x^{5}$& $x^{6}$& $x^{7}$& $x^{8}$& $x^{9}$ \\
		 \hline
		 D3 & $\bullet$ & $\bullet$ & $\bullet$ & $ \ $ & $ \ $ &$ \ $ &$ \bullet $ &$ \ $ &$ \ $ &$ \ $ \\
		 \hline
		 \begin{color}{red} NS5 \end{color} & $\bullet$ & $\bullet$ & $\bullet$  & $ \bullet $ & $\bullet$ & $\bullet$ & $ \ $ &$ \ $ &$ \ $ &$ \ $ \\
		 \hline
		 \begin{color}{blue} D5 \end{color} & $\bullet$ & $\bullet$ & $\bullet$  & $ \ $ & $ \ $ &$ \ $ & $ \ $ & $\bullet$ & $\bullet$ & $\bullet$  \\
		 \hline
	\end{tabular}
\end{equation*}
In the graphical representation of the Hanany--Witten configurations we adopt the color code of \cite{Grimminger:2020dmg,Bourget:2021siw}: NS5 branes are depicted in red, D5 branes in blue and D3 branes in black. We limit ourselves to discuss type A quivers in type IIB string theory, but other classical types could be included.\par
The construction of $\fC$ as given in Algorithm \ref{algorithm1} can be recast in an algorithm for the brane arrangements. In turn, from any given phase of the brane configuration we read off a tableau, and we connect any two tableaux that are related through a phase transition of the brane system (called a Kraft--Procesi transition in \cite{Cabrera:2016vvv}).
\begin{algorit}
	Consider an arrangement of $r+1$ NS5 branes, with $\mathsf{w}_j$ D5 branes in the $j^{\text{th}}$ interval between NS5 branes and $N_j$ D3 branes suspended between the $j^{\text{th}}$ and $(j+1)^{\text{th}}$ NS5 brane. To such configuration we associate the highest weight tableau $T$ of shape $(n)$ and $\wt (T)= (\mathsf{w}_1, \dots, \mathsf{w}_r)$, and apply \eqref{eq:bijcrys}: 
	\begin{equation*}
\ytableausetup{nosmalltableaux}
\begin{tikzpicture}
	\path[red,thick] (-5,1) edge (-5,-0.5);
	\path[red,thick] (-3,1) edge (-3,-0.5);
	\path[red,thick] (-1,1) edge (-1,-0.5);
	\path[red,thick] (5,1) edge (5,-0.5);
	\path[red,thick] (3,1) edge (3,-0.5);
	\path[red,thick] (1,1) edge (1,-0.5);
	
	\node (d1) at (-4,0.5) {$\overbrace{\dbr \dbr \cdots \dbr \dbr}^{\mathsf{w}_1}$};
	\node (d2) at (-2,0.5) {$\overbrace{\dbr \dbr \cdots \dbr \dbr}^{\mathsf{w}_2}$};
	\node (d3) at (2,0.5) {$\overbrace{\dbr \dbr \cdots \dbr \dbr}^{\mathsf{w}_{r-1}}$};
	\node (d4) at (4,0.5) {$\overbrace{\dbr \dbr \cdots \dbr \dbr}^{\mathsf{w}_{r}}$};
	\node (dots) at (0,0.25) {$\cdots $};
	
	\node (t1) at (-4,-1.5) {$\begin{ytableau} \scriptstyle \sn_1  \end{ytableau}$};
	\node (t2) at (-2,-1.5) {$\begin{ytableau} \scriptstyle \sn_2 \end{ytableau}$};
	\node (2dot) at (0,-1.5) {$\cdots $};
	\node (t3) at (2,-1.5) {$\begin{ytableau} \scriptstyle \sn_{ r-1}  \end{ytableau}$};
	\node (t4) at (4,-1.5) {$\begin{ytableau} \scriptstyle \sn_{r} \end{ytableau}$};
	
	\node (o1) at (-3,-1.5) {$ \otimes $};
	\node (o2) at (-1,-1.5) {$ \otimes $};
	\node (o3) at (1,-1.5) {$ \otimes $};
	\node (o4) at (3,-1.5) {$ \otimes $};
\end{tikzpicture}
\end{equation*}
	where $w_j \equiv \mathsf{w}_j +1$. Then, construct a crystal $\fC$ by connecting any two tableaux obtained from brane configurations that are related by one of the following moves: 
	\begin{equation*}
\ytableausetup{nosmalltableaux}
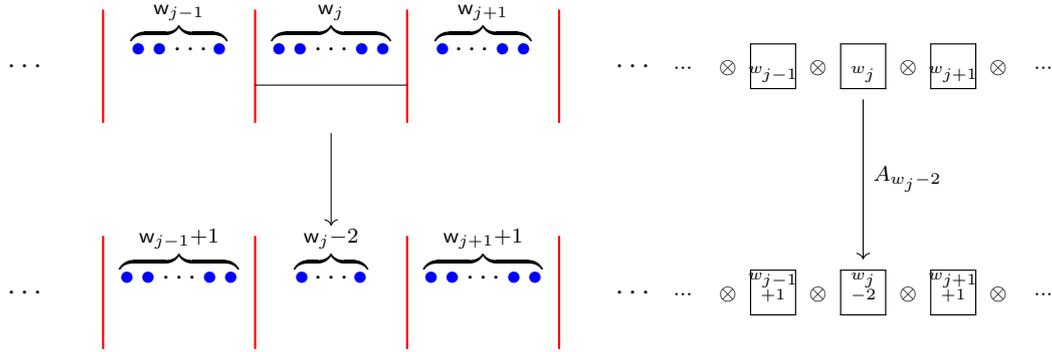
\begin{tikzpicture}
	\path[red,thick] (-6,1) edge (-6,-0.5);
	\path[red,thick] (-4,1) edge (-4,-0.5);
	\path[red,thick] (-2,1) edge (-2,-0.5);
	\path[red,thick] (0,1) edge (0,-0.5);
	\path (-4,0) edge (-2,0);
	
	\node (d1) at (-5,0.75) {$\overbrace{\dbr \dbr \cdots \dbr}^{\mathsf{w}_{j-1}}$};
	\node (d2) at (-3,0.75) {$\overbrace{\dbr \dbr \cdots \dbr \dbr}^{\mathsf{w}_{j}}$};
	\node (d3) at (-1,0.75) {$\overbrace{\dbr \cdots \dbr \dbr}^{\mathsf{w}_{j+1}}$};
	\node (dot1) at (1,0.25) {$\cdots $};
	\node (dot2) at (-7,0.25) {$\cdots $};
	
	\path[red,thick] (-6,-2) edge (-6,-3.5);
	\path[red,thick] (-4,-2) edge (-4,-3.5);
	\path[red,thick] (-2,-2) edge (-2,-3.5);
	\path[red,thick] (0,-2) edge (0,-3.5);
	
	\node (D1) at (-5,-2.25) {$\overbrace{\dbr \dbr \cdots \dbr \dbr}^{\mathsf{w}_{j-1} +1}$};
	\node (D2) at (-3,-2.25) {$\overbrace{\dbr  \cdots \dbr}^{\mathsf{w}_{j}-2}$};
	\node (D3) at (-1,-2.25) {$\overbrace{\dbr \dbr \cdots \dbr \dbr}^{\mathsf{w}_{j+1}+1}$};
	\node (Dot1) at (1,-2.75) {$\cdots $};
	\node (Dot2) at (-7,-2.75) {$\cdots $};
	
	\node (aux1) at (-3,-0.5) {};
	\node (aux2) at (-3,-2) {};
	\path[->] (aux1) edge (aux2);

	\node (t1) at (4,0.25) {$\begin{ytableau} \none[\scriptstyle \cdots ] &  \none[\scriptstyle \otimes ] & { \scriptscriptstyle \sn_{j-1} } &  \none[\scriptstyle \otimes ] & {\scriptscriptstyle \sn_j } &  \none[\scriptstyle \otimes ] & { \scriptscriptstyle \sn_{j+1} } &  \none[\scriptstyle \otimes ] &  \none[\scriptstyle \cdots ] \end{ytableau}$};
	\node (t2) at (4,-2.75) {$\begin{ytableau} \none[\scriptstyle \cdots ] &  \none[\scriptstyle \otimes ] & { \scriptscriptstyle  \overset{ \sn_{j-1}}{+1} } &  \none[\scriptstyle \otimes ] &  {\scriptscriptstyle  \overset{\sn_j}{-2} }  &  \none[\scriptstyle \otimes ] & {\scriptscriptstyle  \overset{\sn_{j+1}}{+1}  } &  \none[\scriptstyle \otimes ] &  \none[\scriptstyle \cdots ] \end{ytableau}$};
	
	\path[->] (t1) edge node[right] {$\scriptstyle A_{w_{j} -2} $} (t2);
\end{tikzpicture}
\end{equation*}
for the $A_p$, and for the $a_p$ 
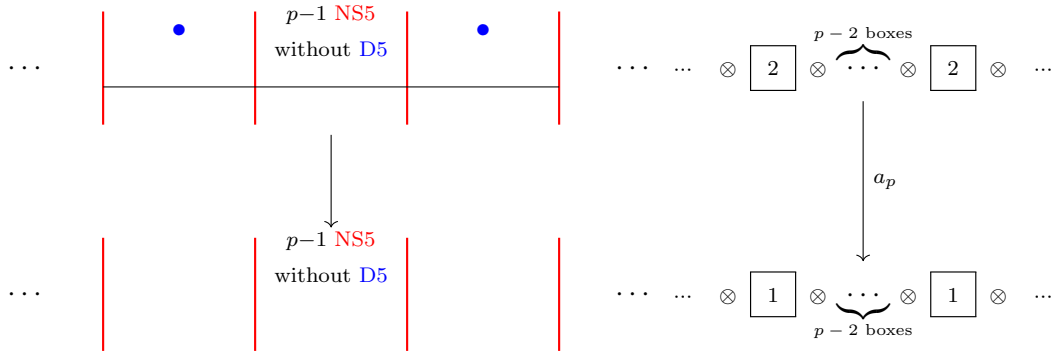
\begin{equation*}
\ytableausetup{nosmalltableaux}
\begin{tikzpicture}
	\path[red,thick] (-6,1) edge (-6,-0.5);
	\path[red,thick] (-4,1) edge (-4,-0.5);
	\path[red,thick] (-2,1) edge (-2,-0.5);
	\path[red,thick] (0,1) edge (0,-0.5);
	\path (-6,0) edge (0,0);
	
	\node (d1) at (-5,0.75) {$ \dbr $};
	\node (d2) at (-3,0.75) {$\scriptstyle \begin{matrix} { \scriptstyle p-1 \text{ \begin{color}{red}NS5\end{color}} } \\ { \scriptstyle \text{without \begin{color}{blue}D5\end{color}} } \end{matrix} $};
	\node (d3) at (-1,0.75) {$\dbr $};
	\node (dot1) at (1,0.25) {$\cdots $};
	\node (dot2) at (-7,0.25) {$\cdots $};
	
	\path[red,thick] (-6,-2) edge (-6,-3.5);
	\path[red,thick] (-4,-2) edge (-4,-3.5);
	\path[red,thick] (-2,-2) edge (-2,-3.5);
	\path[red,thick] (0,-2) edge (0,-3.5);
	
	\node (D2) at (-3,-2.25) {$\scriptstyle \begin{matrix} { \scriptstyle p-1 \text{ \begin{color}{red}NS5\end{color}} } \\ { \scriptstyle \text{without \begin{color}{blue}D5\end{color}} } \end{matrix} $};
	\node (Dot1) at (1,-2.75) {$\cdots $};
	\node (Dot2) at (-7,-2.75) {$\cdots $};
	
	\node (aux1) at (-3,-0.5) {};
	\node (aux2) at (-3,-2) {};
	\path[->] (aux1) edge (aux2);

	\node (t1) at (4,0.25) {$\begin{ytableau} \none[\scriptstyle \cdots ] &  \none[\scriptstyle \otimes ] & { \scriptstyle 2 } &  \none[\scriptstyle \otimes ] & \none[ { \scriptscriptstyle \overbrace{ \cdots }^{\scriptscriptstyle \text{$p-2$ boxes} } } ] &  \none[\scriptstyle \otimes ] & { \scriptstyle 2 } &  \none[\scriptstyle \otimes ] &  \none[\scriptstyle \cdots ] \end{ytableau}$};
	\node (t2) at (4,-2.75) {$\begin{ytableau} \none[\scriptstyle \cdots ] &  \none[\scriptstyle \otimes ] & {\scriptstyle  1 } &  \none[\scriptstyle \otimes ] & \none[ { \scriptscriptstyle \underbrace{ \cdots }_{\scriptscriptstyle \text{$p-2$ boxes} } } ] &  \none[\scriptstyle \otimes ] & {\scriptstyle  1 } &  \none[\scriptstyle \otimes ] &  \none[\scriptstyle \cdots ] \end{ytableau}$};
	
	\path[->] (t1) edge node[right] {${ \scriptstyle a_{p} }$} (t2);
\end{tikzpicture}
\end{equation*}
\end{algorit}\par
Notice that we always perform a pair of Hanany--Witten transitions
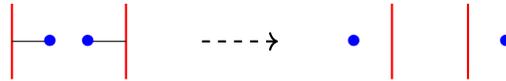
\begin{equation*}
    \begin{tikzpicture}
	\path[red,thick] (-2,1) edge (-2,0);
	\draw (-2,0.5)--(-1.5,0.5);
	\node at (-1.5,0.5) {$\dbr$};
	\path[red,thick] (-0.5,1) edge (-0.5,0);
	\draw (-1,0.5)--(-0.5,0.5);
	\node at (-1,0.5) {$\dbr$};
	
	\path[thick,->,dashed] (0.5,0.5) edge (1.5,0.5);

	\path[red,thick] (3,1) edge (3,0);
	\path[red,thick] (4,1) edge (4,0);
	\node at (4.5,0.5) {$\dbr$};
	\node at (2.5,0.5) {$\dbr$};
	\end{tikzpicture}
\end{equation*}
after a phase (or Kraft--Procesi) transition, and refer to this combined move as a transition.\par
Besides, we point out that the letters in the alphabet $[r] $ are associated with intervals between two consecutive NS5 branes, including from the leftmost NS5 to $- \infty$, labelled by the letter $0$, and from the rightmost NS5 to $+ \infty$, labelled by the letter $r+1$. This simple observation is further discussed and utilized in Appendix \ref{app:indlim}.\par

\subsection{Type A examples}
\label{sec:CBKCAexample}
In this subsection we work out a few explicit examples of $A_r$ quivers with unitary gauge nodes. We assume that the gauge ranks are large enough to admit all transitions. Otherwise, one simply amputates from $\fC$ all legs after $N_{j}$ operations have been performed on the $j^{\text{th}}$ box.\par
In the examples, we juxtapose $\fC$ with its Hasse diagram, that we have computed from quiver subtraction. Dots in the Hasse diagram represent symplectic leaves, with the quaternionic dimension explicitly written.

\subsubsection{$A_2, \un=(5,2)$}
$\fC \left[  \overset{ 5}{\Box} -  \overset{ 4}{\circ}-  \overset{ 3}{\circ} - \overset{ 2}{\Box} \right] $ is 
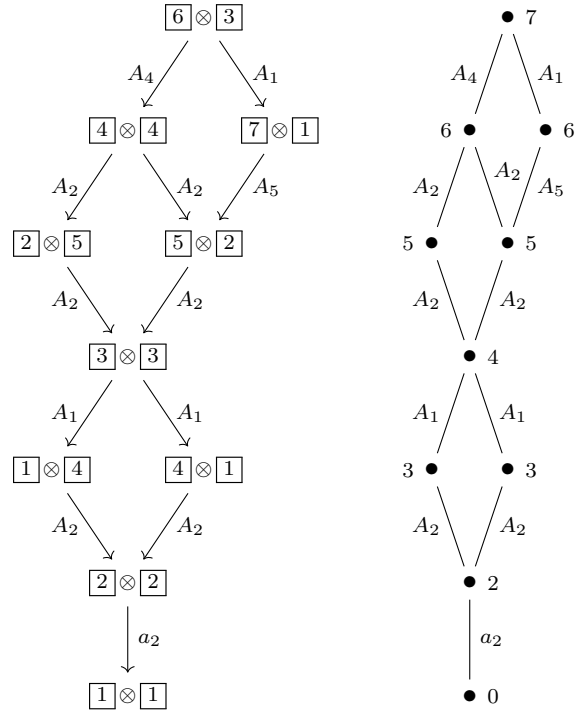
\begin{equation}
	\ytableausetup{smalltableaux}
	\begin{tikzpicture}
		\node (t1) at (0,0) {$\begin{ytableau} 6 & \none[ \scriptstyle \otimes ] & 3 \end{ytableau}$};
		\node (t2a) at (-1,-1.5) {$\begin{ytableau} 4 & \none[ \scriptstyle \otimes ] & 4 \end{ytableau}$};
		\node (t2b) at (1,-1.5) {$\begin{ytableau} 7 & \none[ \scriptstyle \otimes ] & 1 \end{ytableau}$};
		\node (t3a) at (-2,-3) {$\begin{ytableau} 2 & \none[ \scriptstyle \otimes ] & 5 \end{ytableau}$};
		\node (t3b) at (0,-3) {$\begin{ytableau} 5 & \none[ \scriptstyle \otimes ] & 2 \end{ytableau}$};
		\node (t4) at (-1,-4.5) {$\begin{ytableau} 3 & \none[ \scriptstyle \otimes ] & 3 \end{ytableau}$};
		\node (t5a) at (-2,-6) {$\begin{ytableau} 1 & \none[ \scriptstyle \otimes ] & 4 \end{ytableau}$};
		\node (t5b) at (0,-6) {$\begin{ytableau} 4 & \none[ \scriptstyle \otimes ] & 1 \end{ytableau}$};
		\node (t6) at (-1,-7.5) {$\begin{ytableau} 2 & \none[ \scriptstyle \otimes ] & 2 \end{ytableau}$};
		\node (t7) at (-1,-9) {$\begin{ytableau} 1 & \none[ \scriptstyle \otimes ] & 1 \end{ytableau}$};
		
		\path[->] (t1) edge node[left] {${\scriptstyle A_4}$} (t2a);
		\path[->] (t1) edge node[right] {${\scriptstyle A_1}$} (t2b);
		\path[->] (t2b) edge node[right] {${\scriptstyle A_5}$} (t3b);
		\path[->] (t2a) edge node[left] {${\scriptstyle A_2}$} (t3a);
		\path[->] (t2a) edge node[right] {${\scriptstyle A_2}$} (t3b);
		\path[->] (t3a) edge node[left] {${\scriptstyle A_2}$} (t4);
		\path[->] (t3b) edge node[right] {${\scriptstyle A_2}$} (t4);
		\path[->] (t4) edge node[left] {${\scriptstyle A_1}$} (t5a);
		\path[->] (t4) edge node[right] {${\scriptstyle A_1}$} (t5b);
		\path[->] (t5a) edge node[left] {${\scriptstyle A_2}$} (t6);
		\path[->] (t5b) edge node[right] {${\scriptstyle A_2}$} (t6);
		\path[->] (t6) edge node[right] {${\scriptstyle a_2}$} (t7);

		\node (h1) at (4,0) {$\bullet $};
		\node (h2a) at (3.5,-1.5) {$\bullet $};
		\node (h2b) at (4.5,-1.5) {$\bullet $};
		\node (h3a) at (3,-3) {$\bullet $};
		\node (h3b) at (4,-3) {$\bullet $};
		\node (h4) at (3.5,-4.5) {$\bullet $};
		\node (h5a) at (3,-6) {$\bullet $};
		\node (h5b) at (4,-6) {$\bullet $};
		\node (h6) at (3.5,-7.5) {$\bullet $};
		\node (h7) at (3.5,-9) {$\bullet $};
		
		\node[anchor=west] (l1) at (4.1,0) {${\scriptstyle 7}$};
		\node[anchor=east] (l2a) at (3.4,-1.5) {${\scriptstyle 6}$};
		\node[anchor=west] (l2b) at (4.6,-1.5) {${\scriptstyle 6}$};
		\node[anchor=east] (l3a) at (2.9,-3) {${\scriptstyle 5}$};
		\node[anchor=west] (l3b) at (4.1,-3) {${\scriptstyle 5}$};
		\node[anchor=west] (l4) at (3.6,-4.5) {${\scriptstyle 4}$};
		\node[anchor=east] (l5a) at (2.9,-6) {${\scriptstyle 3}$};
		\node[anchor=west] (l5b) at (4.1,-6) {${\scriptstyle 3}$};
		\node[anchor=west] (l6) at (3.6,-7.5) {${\scriptstyle 2}$};
		\node[anchor=west] (l7) at (3.6,-9) {${\scriptstyle 0}$};
		
		\path[] (h1) edge node[left] {${\scriptstyle A_4}$} (h2a);
		\path[] (h1) edge node[right] {${\scriptstyle A_1}$} (h2b);
		\path[] (h2b) edge node[right] {${\scriptstyle A_5}$} (h3b);
		\path[] (h2a) edge node[left] {${\scriptstyle A_2}$} (h3a);
		\path[] (h2a) edge node[right,pos=0.3] {${\scriptstyle A_2}$} (h3b);
		\path[] (h3a) edge node[left] {${\scriptstyle A_2}$} (h4);
		\path[] (h3b) edge node[right] {${\scriptstyle A_2}$} (h4);
		\path[] (h4) edge node[left] {${\scriptstyle A_1}$} (h5a);
		\path[] (h4) edge node[right] {${\scriptstyle A_1}$} (h5b);
		\path[] (h5a) edge node[left] {${\scriptstyle A_2}$} (h6);
		\path[] (h5b) edge node[right] {${\scriptstyle A_2}$} (h6);
		\path[] (h6) edge node[right] {${\scriptstyle a_2}$} (h7);

	\end{tikzpicture}
\label{eq:A2n52}
\end{equation}
For clarity, in this first example we show the full $\mathrm{PSL}(3)$ crystal isomorphic to \eqref{eq:A2n52}:
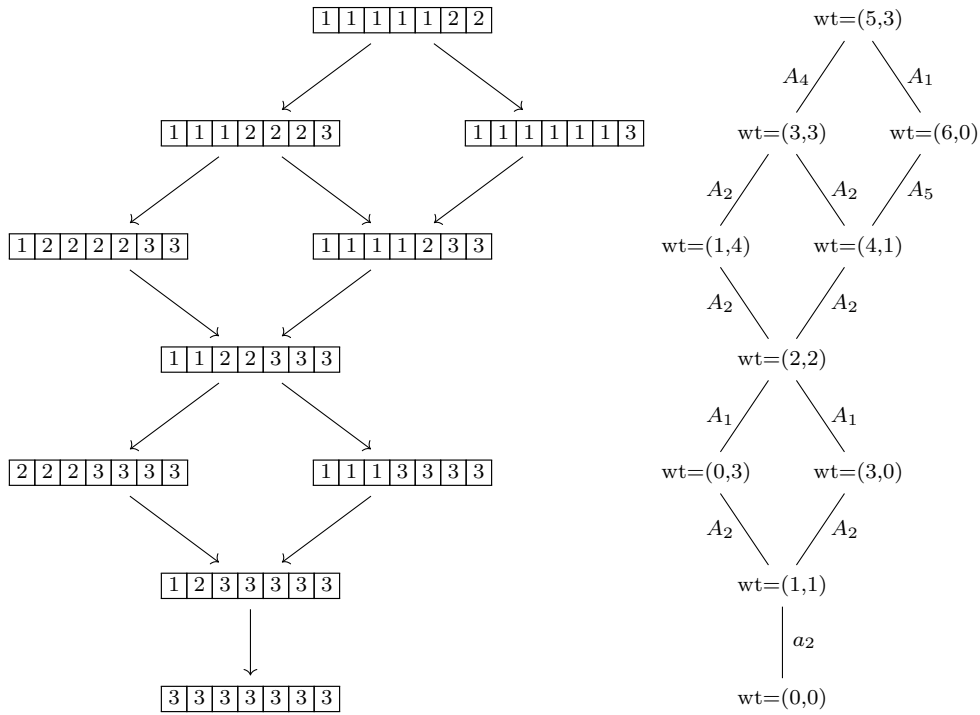
\begin{equation*}
\ytableausetup{smalltableaux} 
	\begin{tikzpicture}
		
		\node (t0) at (2,1.5) {\begin{ytableau}  1 & 1 & 1 & 1 & 1 & 2 & 2 \end{ytableau}};  
		\node (t1b) at (4,0) {\begin{ytableau}  1 & 1 & 1 & 1 &1 &1 & 3\end{ytableau}}; 
		\node (t1) at (0,0) {\begin{ytableau}  1 & 1 & 1 &  2 & 2 & 2 & 3 \end{ytableau}};
		\node (t2a) at (-2,-1.5) {\begin{ytableau}  1 & 2 & 2 & 2 & 2 & 3 & 3 \end{ytableau}};
		\node (t2b) at (2,-1.5) {\begin{ytableau}  1 & 1 & 1 & 1 & 2 & 3& 3 \end{ytableau}};
		\node (t3) at (0,-3) {\begin{ytableau}  1 & 1 & 2 & 2 & 3 & 3 & 3 \end{ytableau}};
		\node (t4a) at (-2,-4.5) {\begin{ytableau}  2 & 2 & 2 & 3 & 3 & 3 & 3 \end{ytableau}};
		\node (t4b) at (2,-4.5) {\begin{ytableau}  1 & 1 & 1 & 3 & 3 & 3 & 3 \end{ytableau}};
		\node (t5) at (0,-6) {\begin{ytableau}  1 & 2 & 3 & 3 & 3 & 3 & 3 \end{ytableau}};
		\node (t6) at (0,-7.5) {\begin{ytableau}  3 & 3 & 3 & 3 & 3 & 3 & 3 \end{ytableau}};  
		
		\path[->] (t0) edge  (t1);
		\path[->] (t0) edge  (t1b);
		\path[->] (t1) edge  (t2b);
		\path[->] (t1) edge  (t2a);
		\path[->] (t1b) edge  (t2b);
		\path[->] (t2a) edge  (t3);
		\path[->] (t2b) edge  (t3);
		\path[->] (t3) edge  (t4a);
		\path[->] (t3) edge  (t4b);
		\path[->] (t4a) edge  (t5);
		\path[->] (t4b) edge  (t5);
		\path[->] (t5) edge  (t6);

		\node (h0) at (8,1.5) {${ \scriptstyle \wt = (5,3) } $};
		\node (h1) at (7,0) {${ \scriptstyle \wt = (3,3) } $};
		\node (h1b) at (9,0) {${ \scriptstyle \wt = (6,0) } $};
		\node (h2a) at (6,-1.5) {${ \scriptstyle \wt = (1,4) } $};
		\node (h2b) at (8,-1.5) {${ \scriptstyle \wt = (4,1) } $};
		\node (h3) at (7,-3) {${ \scriptstyle \wt = (2,2) } $};
		\node (h4a) at (6,-4.5) {${ \scriptstyle \wt = (0,3) } $};
		\node (h4b) at (8,-4.5) {${ \scriptstyle \wt = (3,0) } $};
		\node (h5) at (7,-6) {${ \scriptstyle \wt = (1,1) } $};
		\node (h6) at (7,-7.5) {${ \scriptstyle \wt = (0,0) } $};

		\path[] (h0) edge node[left] {${\scriptstyle A_4}$} (h1);
		\path[] (h0) edge node[right] {${\scriptstyle A_1}$} (h1b);
		\path[] (h1b) edge node[right] {${\scriptstyle A_5}$} (h2b);
		\path[] (h1) edge node[left] {${\scriptstyle A_2}$} (h2a);
		\path[] (h1) edge node[right] {${\scriptstyle A_2}$} (h2b);
		\path[] (h2a) edge node[left] {${\scriptstyle A_2}$} (h3);
		\path[] (h2b) edge node[right] {${\scriptstyle A_2}$} (h3);
		\path[] (h3) edge node[left] {${\scriptstyle A_1}$} (h4a);
		\path[] (h3) edge node[right] {${\scriptstyle A_1}$} (h4b);
		\path[] (h4a) edge node[left] {${\scriptstyle A_2}$} (h5);
		\path[] (h4b) edge node[right] {${\scriptstyle A_2}$} (h5);
		\path[] (h5) edge node[right] {${\scriptstyle a_2}$} (h6);

	\end{tikzpicture}
\end{equation*}
The outcome matches $\mC \left[  \overset{ 5}{\Box} -  \overset{ 4}{\circ}-  \overset{ 3}{\circ} - \overset{ 2}{\Box} \right] $ as computed both via subsequent phases of brane configurations: 
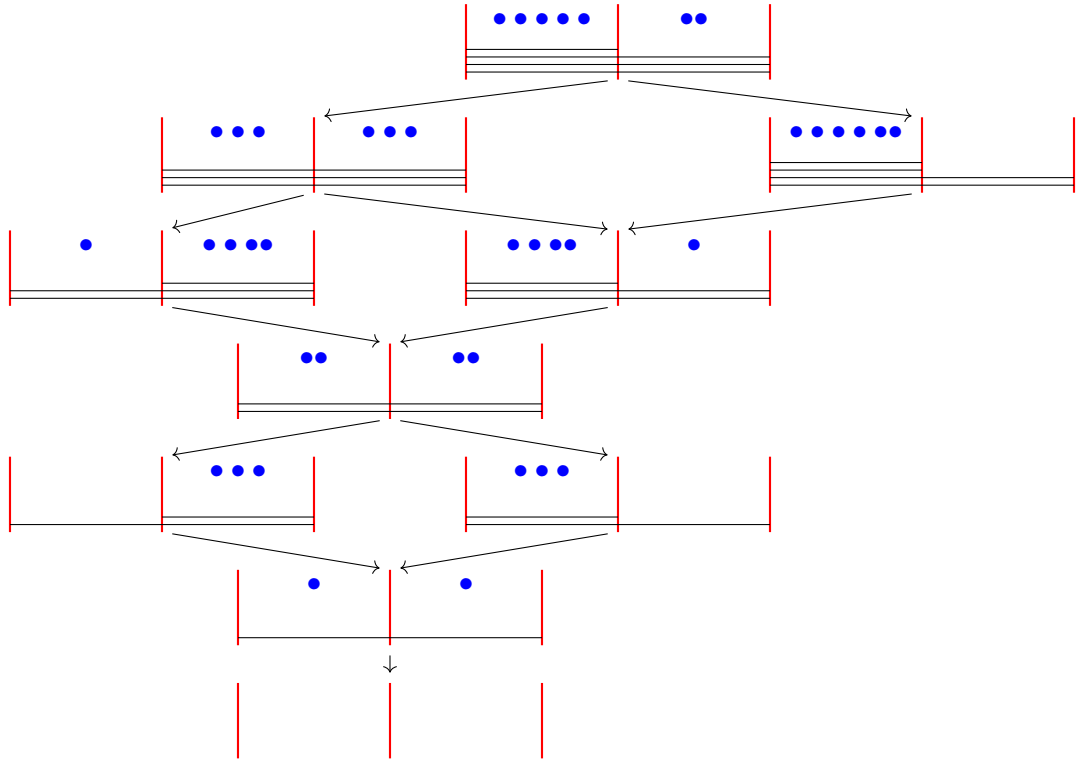
\begin{equation*}
\begin{tikzpicture}
	
		\path[red,thick] (-5,1) edge (-5,0);
		\path[red,thick] (-3,1) edge (-3,0);
		\path[red,thick] (-1,1) edge (-1,0);
		\path (-5,0.1) edge (-1,0.1);
		\path (-5,0.2) edge (-1,0.2);
		\path (-5,0.3) edge (-1,0.3);
		\path (-5,0.4) edge (-3,0.4);
		
		\node (d51l) at (-4,0.8) {${ \dbr \dbr  \dbr \dbr  \dbr }$};
		\node (d51r) at (-2,0.8) {${ \dbr \dbr   }$};
		
		\node (aux1b) at (-3,0) {};

		\path[red,thick] (-5,-0.5) edge (-5,-1.5);
		\path[red,thick] (-7,-0.5) edge (-7,-1.5);
		\path[red,thick] (-9,-0.5) edge (-9,-1.5);
		\path (-5,-1.4) edge (-9,-1.4);
		\path (-5,-1.3) edge (-9,-1.3);
		\path (-5,-1.2) edge (-9,-1.2);
		
		\node (d52la) at (-8,-0.7) {${ \dbr \dbr  \dbr  }$};
		\node (d52ra) at (-6,-0.7) {${ \dbr \dbr  \dbr }$};
		
		\node (aux2at) at (-7,-0.5) {};
		\node (aux2ab) at (-7,-1.5) {};

		\path[red,thick] (3,-0.5) edge (3,-1.5);
		\path[red,thick] (1,-0.5) edge (1,-1.5);
		\path[red,thick] (-1,-0.5) edge (-1,-1.5);
		\path (3,-1.4) edge (-1,-1.4);
		\path (3,-1.3) edge (-1,-1.3);
		\path (1,-1.2) edge (-1,-1.2);
		\path (1,-1.1) edge (-1,-1.1);
		
		\node (d52lb) at (0,-0.7) {${ \dbr \dbr  \dbr \dbr  \dbr \dbr }$};
		
		\node (aux2bt) at (1,-0.5) {};
		\node (aux2bb) at (1,-1.5) {};

		\path[red,thick] (-7,-2) edge (-7,-3);
		\path[red,thick] (-9,-2) edge (-9,-3);
		\path[red,thick] (-11,-2) edge (-11,-3);
		\path (-7,-2.9) edge (-11,-2.9);
		\path (-7,-2.8) edge (-11,-2.8);
		\path (-7,-2.7) edge (-9,-2.7);
		
		\node (d53la) at (-10,-2.2) {${ \dbr   }$};
		\node (d53ra) at (-8,-2.2) {${ \dbr \dbr  \dbr \dbr }$};
		
		\node (aux3at) at (-9,-2) {};
		\node (aux3ab) at (-9,-3) {};

		\path[red,thick] (-5,-2) edge (-5,-3);
		\path[red,thick] (-3,-2) edge (-3,-3);
		\path[red,thick] (-1,-2) edge (-1,-3);
		\path (-5,-2.9) edge (-1,-2.9);
		\path (-5,-2.8) edge (-1,-2.8);
		\path (-5,-2.7) edge (-3,-2.7);
		
		\node (d53lb) at (-4,-2.2) {${ \dbr \dbr  \dbr \dbr   }$};
		\node (d53rb) at (-2,-2.2) {${ \dbr   }$};
		
		\node (aux3bt) at (-3,-2) {};
		\node (aux3bb) at (-3,-3) {};

		\path[red,thick] (-8,-3.5) edge (-8,-4.5);
		\path[red,thick] (-6,-3.5) edge (-6,-4.5);
		\path[red,thick] (-4,-3.5) edge (-4,-4.5);
		\path (-8,-4.4) edge (-4,-4.4);
		\path (-8,-4.3) edge (-4,-4.3);
		
		\node (d41l) at (-7,-3.7) {${ \dbr \dbr }$};
		\node (d41r) at (-5,-3.7) {${ \dbr \dbr }$};
		
		\node (aux4t) at (-6,-3.5) {};
		\node (aux4b) at (-6,-4.5) {};

		\path[red,thick] (-7,-5) edge (-7,-6);
		\path[red,thick] (-9,-5) edge (-9,-6);
		\path[red,thick] (-11,-5) edge (-11,-6);
		\path (-7,-5.9) edge (-11,-5.9);
		\path (-7,-5.8) edge (-9,-5.8);
		
		\node (d55ra) at (-8,-5.2) {${ \dbr \dbr  \dbr }$};
		
		\node (aux5at) at (-9,-5) {};
		\node (aux5ab) at (-9,-6) {};

		\path[red,thick] (-5,-5) edge (-5,-6);
		\path[red,thick] (-3,-5) edge (-3,-6);
		\path[red,thick] (-1,-5) edge (-1,-6);
		\path (-5,-5.9) edge (-1,-5.9);
		\path (-5,-5.8) edge (-3,-5.8);
		
		\node (d55lb) at (-4,-5.2) {${ \dbr \dbr  \dbr   }$};
		
		\node (aux5bt) at (-3,-5) {};
		\node (aux5bb) at (-3,-6) {};

		\path[red,thick] (-8,-6.5) edge (-8,-7.5);
		\path[red,thick] (-6,-6.5) edge (-6,-7.5);
		\path[red,thick] (-4,-6.5) edge (-4,-7.5);
		\path (-8,-7.4) edge (-4,-7.4);
		
		\node (d61l) at (-7,-6.7) {${ \dbr }$};
		\node (d61r) at (-5,-6.7) {${ \dbr }$};
		
		\node (aux6t) at (-6,-6.5) {};
		\node (aux6b) at (-6,-7.5) {};
		
		\path[red,thick] (-8,-8) edge (-8,-9);
		\path[red,thick] (-6,-8) edge (-6,-9);
		\path[red,thick] (-4,-8) edge (-4,-9);
		
		\node (auxf) at (-6,-8) {};

		\path[->] (aux1b) edge (aux2at);
		\path[->] (aux1b) edge (aux2bt);
		\path[->] (aux2ab) edge (aux3at);
		\path[->] (aux2ab) edge (aux3bt);
		\path[->] (aux2bb) edge (aux3bt);		
		\path[->] (aux3ab) edge (aux4t);
		\path[->] (aux3bb) edge (aux4t);
		\path[->] (aux4b) edge (aux5at);
		\path[->] (aux4b) edge (aux5bt);
		\path[->] (aux5ab) edge (aux6t);
		\path[->] (aux5bb) edge (aux6t);
		\path[->] (aux6b) edge (auxf);

\end{tikzpicture}
\end{equation*}
and quiver subtraction: 
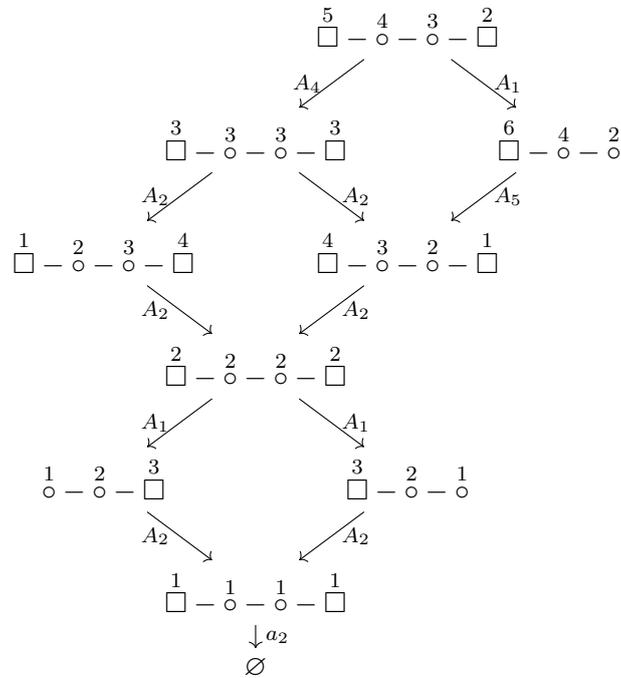
\begin{equation*}
\begin{tikzpicture}
	\node (q1) at (0,0) {$\overset{ 5}{\Box} -  \overset{ 4}{\circ}-  \overset{ 3}{\circ} - \overset{ 2}{\Box}$};
	\node (q2l) at (-2,-1.5) {$\overset{ 3}{\Box} -  \overset{ 3}{\circ}-  \overset{ 3}{\circ} - \overset{ 3}{\Box}$};
	\node (q2r) at (2,-1.5) {$\overset{ 6}{\Box} -  \overset{ 4}{\circ}-  \overset{ 2}{\circ} $};
	\node (q3l) at (-4,-3) {$\overset{ 1}{\Box} -  \overset{ 2}{\circ}-  \overset{ 3}{\circ} - \overset{ 4}{\Box}$};
	\node (q3r) at (0,-3) {$\overset{ 4}{\Box} -  \overset{ 3}{\circ}-  \overset{ 2}{\circ} - \overset{ 1}{\Box} $};
	\node (q4) at (-2,-4.5) {$\overset{ 2}{\Box} -  \overset{ 2}{\circ}-  \overset{ 2}{\circ} - \overset{ 2}{\Box}$};
	\node (q5l) at (-4,-6) {$\overset{ 1}{\circ}-  \overset{ 2}{\circ} - \overset{ 3}{\Box}$};
	\node (q5r) at (0,-6) {$\overset{ 3}{\Box} -  \overset{ 2}{\circ}-  \overset{ 1}{\circ} $};
	\node (q6) at (-2,-7.5) {$\overset{ 1}{\Box} -  \overset{ 1}{\circ}-  \overset{ 1}{\circ} - \overset{ 1}{\Box}$};
	\node (q7) at (-2,-8.5) {$ \varnothing $};

		\path[->] (q1) edge node[left] {${\scriptstyle A_4}$} (q2l);
		\path[->] (q1) edge node[right] {${\scriptstyle A_1}$} (q2r);
		\path[->] (q2r) edge node[right] {${\scriptstyle A_5}$} (q3r);
		\path[->] (q2l) edge node[left] {${\scriptstyle A_2}$} (q3l);
		\path[->] (q2l) edge node[right] {${\scriptstyle A_2}$} (q3r);
		\path[->] (q3l) edge node[left] {${\scriptstyle A_2}$} (q4);
		\path[->] (q3r) edge node[right] {${\scriptstyle A_2}$} (q4);
		\path[->] (q4) edge node[left] {${\scriptstyle A_1}$} (q5l);
		\path[->] (q4) edge node[right] {${\scriptstyle A_1}$} (q5r);
		\path[->] (q5l) edge node[left] {${\scriptstyle A_2}$} (q6);
		\path[->] (q5r) edge node[right] {${\scriptstyle A_2}$} (q6);
		\path[->] (q6) edge node[right] {${\scriptstyle a_2}$} (q7);

\end{tikzpicture}
\end{equation*}

\subsubsection{$A_5, T[\mathrm{SU}(6)]$}
The next example is a $T[\SU{n}]$ theory \cite{Gaiotto:2008ak} with $n=6$, described by the $A_5$ quiver 
\begin{equation*}
    \overset{1}{\circ} - \overset{2}{\circ} - \overset{3}{\circ} - \overset{4}{\circ} - \overset{5}{\circ} - \overset{6}{\Box} .
\end{equation*}
Notice that we can realize $\mC \left[ T[\SU{n}] \right]$ as a slice in the Coulomb branch $\mC \left[ \mathsf{Q} \right]$ of a higher rank quiver $\mathsf{Q}$. The  Coulomb branch of the mirror quiver $ \overset{6}{\Box} - \overset{5}{\circ}  - \overset{4}{\circ}  - \overset{3}{\circ} - \overset{2}{\circ} - \overset{1}{\circ}$ will be realized as a different slice of $\mC \left[ \mathsf{Q} \right]$. Of course, the two slices are isomorphic symplectic varieties.\par
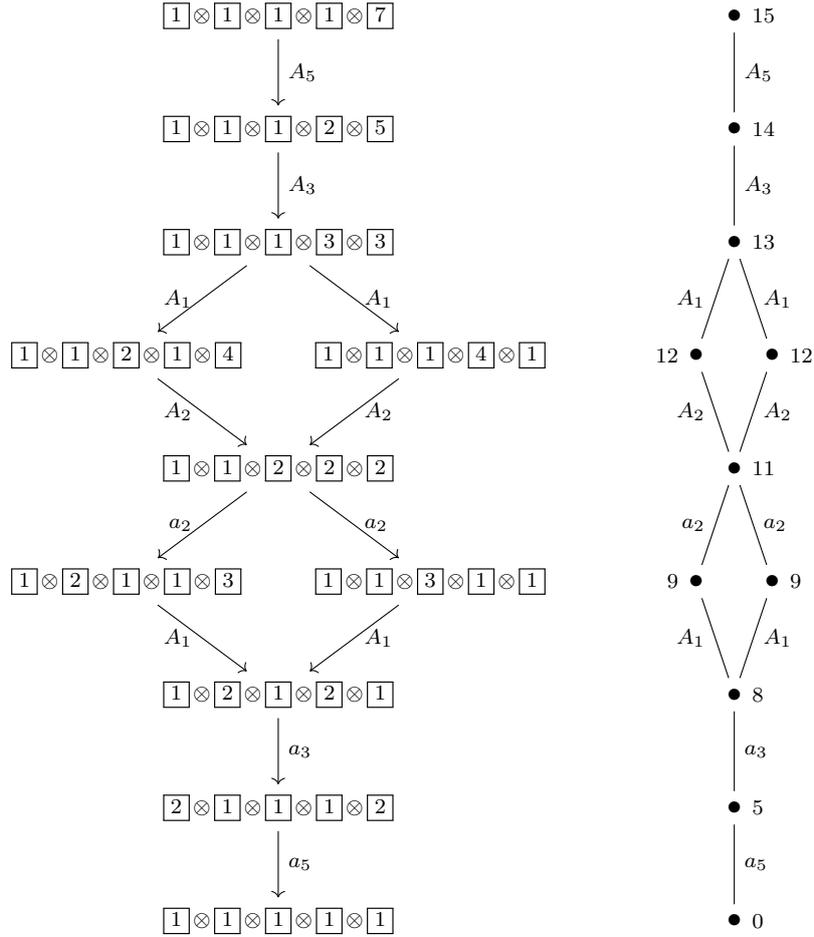
\begin{equation*}
\ytableausetup{smalltableaux}
	\begin{tikzpicture}
		\node (t1) at (-4,0) {$\begin{ytableau} 1 & \none[ \scriptstyle \otimes ] & 1 & \none[ \scriptstyle \otimes ] & 1 & \none[ \scriptstyle \otimes ] & 1  & \none[ \scriptstyle \otimes ] & 7 \end{ytableau}$};
		\node (t2) at (-4,-1.5) {$\begin{ytableau} 1 & \none[ \scriptstyle \otimes ] & 1 & \none[ \scriptstyle \otimes ] & 1 & \none[ \scriptstyle \otimes ] & 2  & \none[ \scriptstyle \otimes ] & 5 \end{ytableau}$};
		\node (t3) at (-4,-3) {$\begin{ytableau} 1 & \none[ \scriptstyle \otimes ] & 1 & \none[ \scriptstyle \otimes ] & 1 & \none[ \scriptstyle \otimes ] & 3  & \none[ \scriptstyle \otimes ] & 3 \end{ytableau}$};
		\node (t4a) at (-6,-4.5) {$\begin{ytableau} 1 & \none[ \scriptstyle \otimes ] & 1 & \none[ \scriptstyle \otimes ] & 2 & \none[ \scriptstyle \otimes ] & 1  & \none[ \scriptstyle \otimes ] & 4 \end{ytableau}$};
		\node (t4b) at (-2,-4.5) {$\begin{ytableau} 1 & \none[ \scriptstyle \otimes ] & 1 & \none[ \scriptstyle \otimes ] & 1 & \none[ \scriptstyle \otimes ] & 4  & \none[ \scriptstyle \otimes ] & 1 \end{ytableau}$};
		\node (t5) at (-4,-6) {$\begin{ytableau} 1 & \none[ \scriptstyle \otimes ] & 1 & \none[ \scriptstyle \otimes ] & 2 & \none[ \scriptstyle \otimes ] & 2  & \none[ \scriptstyle \otimes ] & 2 \end{ytableau}$};
		\node (t6a) at (-6,-7.5) {$\begin{ytableau} 1 & \none[ \scriptstyle \otimes ] & 2 & \none[ \scriptstyle \otimes ] & 1 & \none[ \scriptstyle \otimes ] & 1  & \none[ \scriptstyle \otimes ] & 3 \end{ytableau}$};
		\node (t6b) at (-2,-7.5) {$\begin{ytableau} 1 & \none[ \scriptstyle \otimes ] & 1 & \none[ \scriptstyle \otimes ] & 3 & \none[ \scriptstyle \otimes ] & 1  & \none[ \scriptstyle \otimes ] & 1 \end{ytableau}$};
		\node (t7) at (-4,-9) {$\begin{ytableau} 1 & \none[ \scriptstyle \otimes ] & 2 & \none[ \scriptstyle \otimes ] & 1 & \none[ \scriptstyle \otimes ] & 2  & \none[ \scriptstyle \otimes ] & 1 \end{ytableau}$};
		\node (t8) at (-4,-10.5) {$\begin{ytableau} 2 & \none[ \scriptstyle \otimes ] & 1 & \none[ \scriptstyle \otimes ] & 1 & \none[ \scriptstyle \otimes ] & 1  & \none[ \scriptstyle \otimes ] & 2 \end{ytableau}$};
		\node (t9) at (-4,-12) {$\begin{ytableau} 1 & \none[ \scriptstyle \otimes ] & 1 & \none[ \scriptstyle \otimes ] & 1 & \none[ \scriptstyle \otimes ] & 1  & \none[ \scriptstyle \otimes ] & 1 \end{ytableau}$};

		\path[->] (t1) edge node[right] {$\scriptstyle A_5$} (t2);
		\path[->] (t2) edge node[right] {$\scriptstyle A_3$} (t3);
		\path[->] (t3) edge node[left] {$\scriptstyle A_1$} (t4a);
		\path[->] (t3) edge node[right] {$\scriptstyle A_1$} (t4b);
		\path[->] (t4a) edge node[left] {$\scriptstyle A_2$} (t5);
		\path[->] (t4b) edge node[right] {$\scriptstyle A_2$} (t5);
		\path[->] (t5) edge node[left] {$\scriptstyle a_2$} (t6a);
		\path[->] (t5) edge node[right] {$\scriptstyle a_2$} (t6b);
		\path[->] (t6a) edge node[left] {$\scriptstyle A_1$} (t7);
		\path[->] (t6b) edge node[right] {$\scriptstyle A_1$} (t7);
		\path[->] (t7) edge node[right] {$\scriptstyle a_3$} (t8);
		\path[->] (t8) edge node[right] {$\scriptstyle a_5$} (t9);

		\node (h1) at (2,0) {$\bullet $};
		\node (h2) at (2,-1.5) {$\bullet $};
		\node (h3) at (2,-3) {$\bullet $};
		\node (h4a) at (1.5,-4.5) {$\bullet $};
		\node (h4b) at (2.5,-4.5) {$\bullet $};
		\node (h5) at (2,-6) {$\bullet $};
		\node (h6a) at (1.5,-7.5) {$\bullet $};
		\node (h6b) at (2.5,-7.5) {$\bullet $};
		\node (h7) at (2,-9) {$\bullet $};
		\node (h8) at (2,-10.5) {$\bullet $};
		\node (h9) at (2,-12) {$\bullet $};
		
		\node[anchor=west] (l1) at (2.1,0) {${\scriptstyle 15 }$};
		\node[anchor=west] (l2) at (2.1,-1.5) {${\scriptstyle 14 }$};
		\node[anchor=west] (l3) at (2.1,-3) {${\scriptstyle 13 }$};
		\node[anchor=east] (l4a) at (1.4,-4.5) {${\scriptstyle 12 }$};
		\node[anchor=west] (l4b) at (2.6,-4.5) {${\scriptstyle 12 }$};
		\node[anchor=west] (l5) at (2.1,-6) {${\scriptstyle 11 }$};
		\node[anchor=east] (l6a) at (1.4,-7.5) {${\scriptstyle 9 }$};
		\node[anchor=west] (l6b) at (2.6,-7.5) {${\scriptstyle 9 }$};
		\node[anchor=west] (l7) at (2.1,-9) {${\scriptstyle 8 }$};
		\node[anchor=west] (l8) at (2.1,-10.5) {${\scriptstyle 5 }$};
		\node[anchor=west] (l9) at (2.1,-12) {${\scriptstyle 0 }$};

		\path[] (h1) edge node[right] {$\scriptstyle A_5$} (h2);
		\path[] (h2) edge node[right] {$\scriptstyle A_3$} (h3);
		\path[] (h3) edge node[left] {$\scriptstyle A_1$} (h4a);
		\path[] (h3) edge node[right] {$\scriptstyle A_1$} (h4b);
		\path[] (h4a) edge node[left] {$\scriptstyle A_2$} (h5);
		\path[] (h4b) edge node[right] {$\scriptstyle A_2$} (h5);
		\path[] (h5) edge node[left] {$\scriptstyle a_2$} (h6a);
		\path[] (h5) edge node[right] {$\scriptstyle a_2$} (h6b);
		\path[] (h6a) edge node[left] {$\scriptstyle A_1$} (h7);
		\path[] (h6b) edge node[right] {$\scriptstyle A_1$} (h7);
		\path[] (h7) edge node[right] {$\scriptstyle a_3$} (h8);
		\path[] (h8) edge node[right] {$\scriptstyle a_5$} (h9);

	\end{tikzpicture}
\end{equation*}
The brane configurations read off from $\fC [T[\SU{6}]]$ agree with the phases of the Hanany--Witten setup for $\mC [ T[\SU{6}] ]$: 
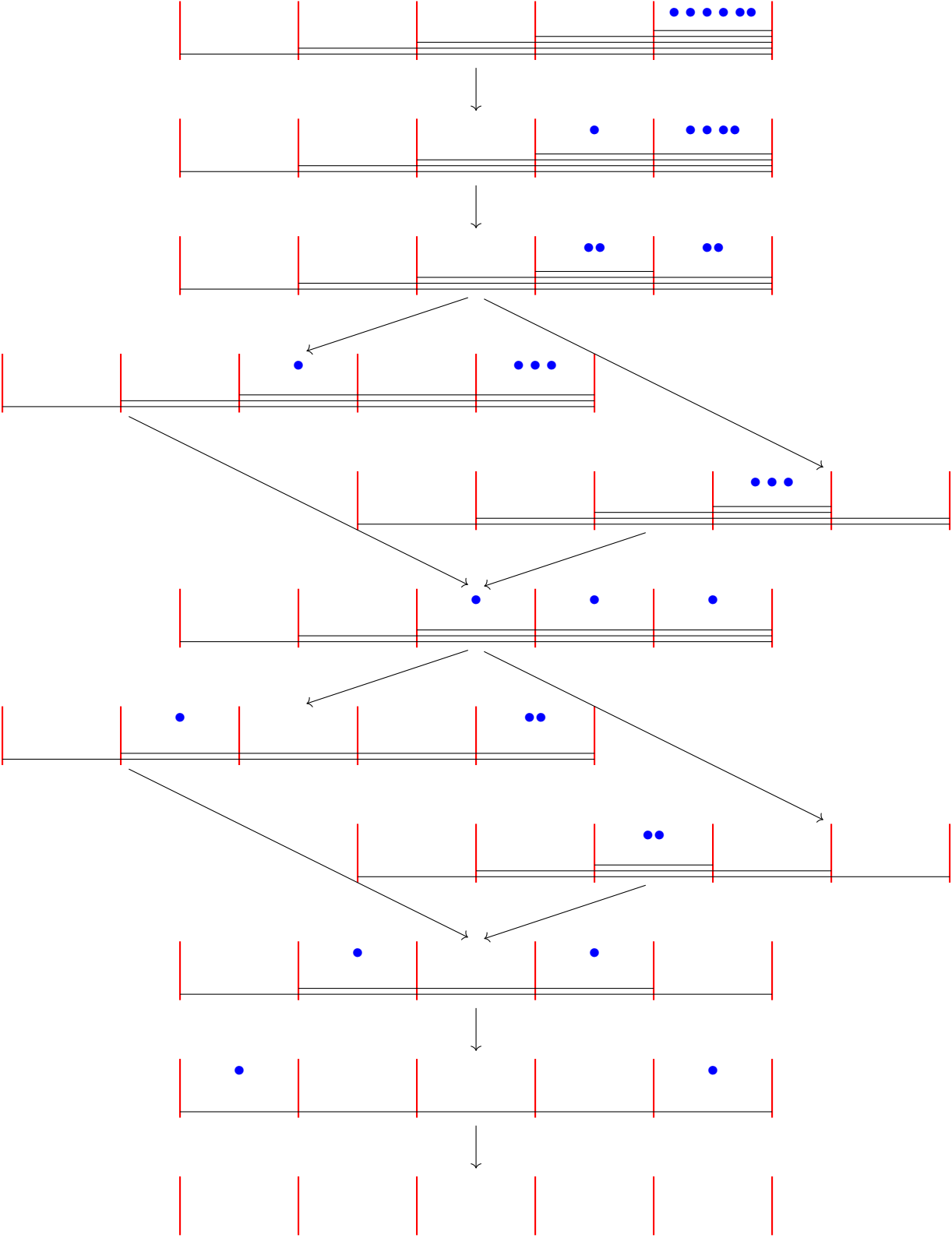
\begin{equation*}
	\begin{tikzpicture}
		\path[red,thick] (-5,1) edge (-5,0);
		\path[red,thick] (-3,1) edge (-3,0);
		\path[red,thick] (-1,1) edge (-1,0);
		\path[red,thick] (1,1) edge (1,0);
		\path[red,thick] (3,1) edge (3,0);
		\path[red,thick] (5,1) edge (5,0);
		\path (-5,0.1) edge (-3,0.1);
		\path (-3,0.1) edge (-1,0.1);
		\path (-3,0.2) edge (-1,0.2);
		\path (-1,0.1) edge (1,0.1);
		\path (-1,0.2) edge (1,0.2);
		\path (-1,0.3) edge (1,0.3);
		\path (1,0.1) edge (3,0.1);
		\path (1,0.2) edge (3,0.2);
		\path (1,0.3) edge (3,0.3);
		\path (1,0.4) edge (3,0.4);
		\path (3,0.1) edge (5,0.1);
		\path (3,0.2) edge (5,0.2);
		\path (3,0.3) edge (5,0.3);
		\path (3,0.4) edge (5,0.4);
		\path (3,0.5) edge (5,0.5);
		
		\node (b1) at (4,0.8) {${ \dbr \dbr  \dbr \dbr  \dbr \dbr }$};
		\node (a1b) at (0,0) {};
		
		\path[red,thick] (-5,-1) edge (-5,-2);
		\path[red,thick] (-3,-1) edge (-3,-2);
		\path[red,thick] (-1,-1) edge (-1,-2);
		\path[red,thick] (1,-1) edge (1,-2);
		\path[red,thick] (3,-1) edge (3,-2);
		\path[red,thick] (5,-1) edge (5,-2);
		\path (-5,-1.9) edge (-3,-1.9);
		\path (-3,-1.9) edge (-1,-1.9);
		\path (-3,-1.8) edge (-1,-1.8);
		\path (-1,-1.9) edge (1,-1.9);
		\path (-1,-1.8) edge (1,-1.8);
		\path (-1,-1.7) edge (1,-1.7);
		\path (1,-1.9) edge (3,-1.9);
		\path (1,-1.8) edge (3,-1.8);
		\path (1,-1.7) edge (3,-1.7);
		\path (1,-1.6) edge (3,-1.6);
		\path (3,-1.9) edge (5,-1.9);
		\path (3,-1.8) edge (5,-1.8);
		\path (3,-1.7) edge (5,-1.7);
		\path (3,-1.6) edge (5,-1.6);
		
		\node (b2a) at (4,-1.2) {${ \dbr \dbr   \dbr \dbr }$};
		\node (b2b) at (2,-1.2) {${ \dbr }$};
		\node (a2t) at (0,-1) {};
		\node (a2b) at (0,-2) {};
		\path[->] (a1b) edge (a2t);

		\path[red,thick] (-5,-3) edge (-5,-4);
		\path[red,thick] (-3,-3) edge (-3,-4);
		\path[red,thick] (-1,-3) edge (-1,-4);
		\path[red,thick] (1,-3) edge (1,-4);
		\path[red,thick] (3,-3) edge (3,-4);
		\path[red,thick] (5,-3) edge (5,-4);
		\path (-5,-3.9) edge (-3,-3.9);
		\path (-3,-3.9) edge (-1,-3.9);
		\path (-3,-3.8) edge (-1,-3.8);
		\path (-1,-3.9) edge (1,-3.9);
		\path (-1,-3.8) edge (1,-3.8);
		\path (-1,-3.7) edge (1,-3.7);
		\path (1,-3.9) edge (3,-3.9);
		\path (1,-3.8) edge (3,-3.8);
		\path (1,-3.7) edge (3,-3.7);
		\path (1,-3.6) edge (3,-3.6);
		\path (3,-3.9) edge (5,-3.9);
		\path (3,-3.8) edge (5,-3.8);
		\path (3,-3.7) edge (5,-3.7);
		
		\node (b3a) at (4,-3.2) {${ \dbr \dbr  }$};
		\node (b3b) at (2,-3.2) {${ \dbr \dbr  }$};
		\node (a3t) at (0,-3) {};
		\node (a3b) at (0,-4) {};
		\path[->] (a2b) edge (a3t);

		\path[red,thick] (-8,-5) edge (-8,-6);
		\path[red,thick] (-6,-5) edge (-6,-6);
		\path[red,thick] (-4,-5) edge (-4,-6);
		\path[red,thick] (-2,-5) edge (-2,-6);
		\path[red,thick] (0,-5) edge (0,-6);
		\path[red,thick] (2,-5) edge (2,-6);
		\path (-8,-5.9) edge (-6,-5.9);
		\path (-6,-5.9) edge (-4,-5.9);
		\path (-6,-5.8) edge (-4,-5.8);
		\path (-4,-5.9) edge (-2,-5.9);
		\path (-4,-5.8) edge (-2,-5.8);
		\path (-4,-5.7) edge (-2,-5.7);
		\path (-2,-5.9) edge (0,-5.9);
		\path (-2,-5.8) edge (0,-5.8);
		\path (-2,-5.7) edge (0,-5.7);
		\path (0,-5.9) edge (2,-5.9);
		\path (0,-5.8) edge (2,-5.8);
		\path (0,-5.7) edge (2,-5.7);
		
		\node (b4a) at (1,-5.2) {${ \dbr \dbr  \dbr }$};
		\node (b4b) at (-3,-5.2) {${ \dbr  }$};
		\node (a4t) at (-3,-5) {};
		\node (a4b) at (-6,-6) {};
		\path[->] (a3b) edge (a4t);

		\path[red,thick] (-2,-7) edge (-2,-8);
		\path[red,thick] (0,-7) edge (0,-8);
		\path[red,thick] (2,-7) edge (2,-8);
		\path[red,thick] (4,-7) edge (4,-8);
		\path[red,thick] (6,-7) edge (6,-8);
		\path[red,thick] (8,-7) edge (8,-8);
		\path (-2,-7.9) edge (0,-7.9);
		\path (0,-7.9) edge (2,-7.9);
		\path (0,-7.8) edge (2,-7.8);
		\path (2,-7.9) edge (4,-7.9);
		\path (2,-7.8) edge (4,-7.8);
		\path (2,-7.7) edge (4,-7.7);
		\path (4,-7.9) edge (6,-7.9);
		\path (4,-7.8) edge (6,-7.8);
		\path (4,-7.7) edge (6,-7.7);
		\path (4,-7.6) edge (6,-7.6);
		\path (6,-7.9) edge (8,-7.9);
		\path (6,-7.8) edge (8,-7.8);
		
		\node (b5a) at (5,-7.2) {${ \dbr \dbr  \dbr }$};
		\node (a5t) at (6,-7) {};
		\node (a5b) at (3,-8) {};
		\path[->] (a3b) edge (a5t);

		\path[red,thick] (-5,-9) edge (-5,-10);
		\path[red,thick] (-3,-9) edge (-3,-10);
		\path[red,thick] (-1,-9) edge (-1,-10);
		\path[red,thick] (1,-9) edge (1,-10);
		\path[red,thick] (3,-9) edge (3,-10);
		\path[red,thick] (5,-9) edge (5,-10);
		\path (-5,-9.9) edge (-3,-9.9);
		\path (-3,-9.9) edge (-1,-9.9);
		\path (-3,-9.8) edge (-1,-9.8);
		\path (-1,-9.9) edge (1,-9.9);
		\path (-1,-9.8) edge (1,-9.8);
		\path (-1,-9.7) edge (1,-9.7);
		\path (1,-9.9) edge (3,-9.9);
		\path (1,-9.8) edge (3,-9.8);
		\path (1,-9.7) edge (3,-9.7);
		\path (3,-9.9) edge (5,-9.9);
		\path (3,-9.8) edge (5,-9.8);
		\path (3,-9.7) edge (5,-9.7);
		
		\node (b6a) at (4,-9.2) {${ \dbr  }$};
		\node (b6b) at (2,-9.2) {${ \dbr  }$};
		\node (b6b) at (0,-9.2) {${ \dbr  }$};
		\node (a6t) at (0,-9) {};
		\node (a6b) at (0,-10) {};
		\path[->] (a4b) edge (a6t);
		\path[->] (a5b) edge (a6t);

		\path[red,thick] (-8,-11) edge (-8,-12);
		\path[red,thick] (-6,-11) edge (-6,-12);
		\path[red,thick] (-4,-11) edge (-4,-12);
		\path[red,thick] (-2,-11) edge (-2,-12);
		\path[red,thick] (0,-11) edge (0,-12);
		\path[red,thick] (2,-11) edge (2,-12);
		\path (-8,-11.9) edge (-6,-11.9);
		\path (-6,-11.9) edge (-4,-11.9);
		\path (-6,-11.8) edge (-4,-11.8);
		\path (-4,-11.9) edge (-2,-11.9);
		\path (-4,-11.8) edge (-2,-11.8);
		\path (-2,-11.9) edge (0,-11.9);
		\path (-2,-11.8) edge (0,-11.8);
		\path (0,-11.9) edge (2,-11.9);
		\path (0,-11.8) edge (2,-11.8);
		
		\node (b7a) at (1,-11.2) {${ \dbr \dbr   }$};
		\node (b7b) at (-5,-11.2) {${ \dbr  }$};
		\node (a7t) at (-3,-11) {};
		\node (a7b) at (-6,-12) {};
		\path[->] (a6b) edge (a7t);

		\path[red,thick] (-2,-13) edge (-2,-14);
		\path[red,thick] (0,-13) edge (0,-14);
		\path[red,thick] (2,-13) edge (2,-14);
		\path[red,thick] (4,-13) edge (4,-14);
		\path[red,thick] (6,-13) edge (6,-14);
		\path[red,thick] (8,-13) edge (8,-14);
		\path (-2,-13.9) edge (0,-13.9);
		\path (0,-13.9) edge (2,-13.9);
		\path (0,-13.8) edge (2,-13.8);
		\path (2,-13.9) edge (4,-13.9);
		\path (2,-13.8) edge (4,-13.8);
		\path (2,-13.7) edge (4,-13.7);
		\path (4,-13.9) edge (6,-13.9);
		\path (4,-13.8) edge (6,-13.8);
		\path (6,-13.9) edge (8,-13.9);
		
		\node (b8a) at (3,-13.2) {${ \dbr \dbr }$};
		\node (a8t) at (6,-13) {};
		\node (a8b) at (3,-14) {};
		\path[->] (a6b) edge (a8t);

		\path[red,thick] (-5,-15) edge (-5,-16);
		\path[red,thick] (-3,-15) edge (-3,-16);
		\path[red,thick] (-1,-15) edge (-1,-16);
		\path[red,thick] (1,-15) edge (1,-16);
		\path[red,thick] (3,-15) edge (3,-16);
		\path[red,thick] (5,-15) edge (5,-16);
		\path (-5,-15.9) edge (-3,-15.9);
		\path (-3,-15.9) edge (-1,-15.9);
		\path (-3,-15.8) edge (-1,-15.8);
		\path (-1,-15.9) edge (1,-15.9);
		\path (-1,-15.8) edge (1,-15.8);
		\path (1,-15.9) edge (3,-15.9);
		\path (1,-15.8) edge (3,-15.8);
		\path (3,-15.9) edge (5,-15.9);
		
		\node (b9a) at (2,-15.2) {${ \dbr  }$};
		\node (b9b) at (-2,-15.2) {${ \dbr  }$};
		\node (a9t) at (0,-15) {};
		\node (a9b) at (0,-16) {};
		\path[->] (a7b) edge (a9t);
		\path[->] (a8b) edge (a9t);
		
		\path[red,thick] (-5,-17) edge (-5,-18);
		\path[red,thick] (-3,-17) edge (-3,-18);
		\path[red,thick] (-1,-17) edge (-1,-18);
		\path[red,thick] (1,-17) edge (1,-18);
		\path[red,thick] (3,-17) edge (3,-18);
		\path[red,thick] (5,-17) edge (5,-18);
		\path (-5,-17.9) edge (-3,-17.9);
		\path (-3,-17.9) edge (-1,-17.9);
		\path (-1,-17.9) edge (1,-17.9);
		\path (1,-17.9) edge (3,-17.9);
		\path (3,-17.9) edge (5,-17.9);
		
		\node (b10a) at (4,-17.2) {${ \dbr  }$};
		\node (b10b) at (-4,-17.2) {${ \dbr  }$};
		\node (a10t) at (0,-17) {};
		\path[->] (a9b) edge (a10t);

		\path[red,thick] (-5,-19) edge (-5,-20);
		\path[red,thick] (-3,-19) edge (-3,-20);
		\path[red,thick] (-1,-19) edge (-1,-20);
		\path[red,thick] (1,-19) edge (1,-20);
		\path[red,thick] (3,-19) edge (3,-20);
		\path[red,thick] (5,-19) edge (5,-20);

		\node (11t) at (0,-19) {};
		\node (11b) at (0,-18) {};
		\path[->] (11b) edge (11t);
		
	\end{tikzpicture}
\end{equation*}

\section{Mass deformation and resolved crystals}
\label{sec:massdef}

At this point we turn on real masses for the hypermultiplets and develop the theory of Kashiwara crystals for the resulting partial resolution of $\mC$.\par
Recall from Section \ref{sec:param} our description of the parameter space $\mM ^{(n)} = \bigsqcup_{\lambda} \mM_{\lambda} ^{(n)}$, where $\lvert \lambda \rvert =n$, together with the definitions of $\mX_{\lambda}$ and $\mX$ from \eqref{eq:defresol}. Fix $\lambda =(\lambda_1, \lambda_2, \dots, \lambda_{\ell})$ and let $\mathsf{w}_j (\lambda_k)$ denote the number of hypermultiplets at the $j^{\text{th}}$ node that belong to the group of $\lambda_k$ equal masses. The highest weight tableau $T$, of shape $(n)$ and weight $\wt (T) = (\mathsf{w}_1, \dots , \mathsf{w}_r)$, splits into the direct sum $T (\lambda_1) \oplus T (\lambda_2) \oplus \cdots \oplus  T (\lambda_{\ell})$, with each $T (\lambda_k)$ of shape $(\lambda_k)$ and weight $\wt (T (\lambda_k)) = (\mathsf{w}_1 (\lambda_k), \dots ,\mathsf{w}_r (\lambda_k))$. We stress that the weight is not additive under $\oplus$.\par 
In this way we obtain a crystal $\fX_{\lambda}$ which is the disjoint union of $\ell$ sub-crystals. We borrow the nomenclature from resolution of singularities and refer to $\fX_{\lambda} $ as \emph{(partially) resolved crystal}.\footnote{Disjoint unions of crystals are usually depicted as separated components. However, we want to insist on the correspondence between Kashiwara crystals and symplectic singularities. For this reason, we will depict $\fX_{\lambda}$ as a unique crystal whose entries are direct sums of tableaux, to keep track of the stratification of $\fX_{\lambda}$.} As for the resolution $\mX \to \mC$, we will denote by $\fX$ the piecewise constant function on $\mM^{(n)}$ that gives $\fX_{\lambda}$ when $\um \in \mM^{(n)}_{\lambda}$.\par
\begin{mainthm}\label{thm:massdef}
	Turning on real masses, the symplectic foliation of the partial resolution $\mX \to \mC$ is described by a partially resolved crystal $\fX$, which is piecewise constant on the mass parameter space and jumps at positive-codimensional loci. Explicitly, $\mathfrak{X}_{\lambda} \cong \mathscr{P} (\mX_{\lambda})$ $\forall \lambda \in \mathbb{Y}_n$.
\end{mainthm}
The partition $\lambda$ that labels the stratum of the parameter space, and the corresponding phase diagram, can equivalently be read off from the Hanany--Witten setup. For instance, in $\U{2}$ with $5$ flavours we take three sample cases $\lambda =(5), \lambda =(3,2), \lambda=(3,1^2)$ and find: 
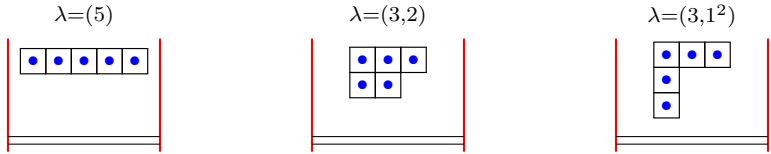
\begin{equation*}
\ytableausetup{smalltableaux}
	\begin{tikzpicture}
		\node (d5l) at (-4,0.7) {$\begin{ytableau} \dbr & \dbr &  \dbr & \dbr & \dbr \end{ytableau}$} ;
		\path[red,thick] (-5,1) edge (-5,-0.5);
		\path[red,thick] (-3,1) edge (-3,-0.5);
		\path[draw] (-5,-0.3) edge (-3,-0.3);
		\path[draw] (-5,-0.4) edge (-3,-0.4);

		\node (d5c) at (0,0.55) {$\begin{ytableau} \dbr & \dbr &  \dbr \\ \dbr & \dbr & \none \end{ytableau}$} ;
		\path[red,thick] (-1,1) edge (-1,-0.5);
		\path[red,thick] (1,1) edge (1,-0.5);
		\path[draw] (-1,-0.3) edge (1,-0.3);
		\path[draw] (-1,-0.4) edge (1,-0.4);

		\node (d5r) at (4,0.45) {$\begin{ytableau} \dbr & \dbr &  \dbr \\ \dbr & \none & \none \\ \dbr & \none & \none \end{ytableau}$} ;
		\path[red,thick] (3,1) edge (3,-0.5);
		\path[red,thick] (5,1) edge (5,-0.5);
		\path[draw] (5,-0.3) edge (3,-0.3);
		\path[draw] (5,-0.4) edge (3,-0.4);

		\node (l1) at (-4,1.3) {${\scriptstyle \lambda = (5)}$};
		\node (l2) at (0,1.3) {${\scriptstyle\lambda = (3,2)}$};
		\node (l3) at (4,1.3) {${\scriptstyle\lambda = (3,1^2)}$};
	\end{tikzpicture}
\end{equation*}
The boxes are immaterial, but we have drawn them for the analogy with Young tableaux.\par

\subsection{Mass-deformed type A examples}
We lay out some examples of the crystal resolution to support the validity of Result \ref{thm:massdef}. Further detailed examples are worked out in Appendix \ref{app:ExamplesC}. As in Section \ref{sec:C=C}, we will adopt the more convenient, equivalent presentation \eqref{eq:bijcrys}.

\subsubsection{$\U{2}$ with four flavours}
Let us start with a simple example to clarify the procedure: $\U{2}$ with $n=4$ flavours. The parameter space decomposes as 
\begin{equation*}
	\mM^{(4)} = \mM_{(1^4)}^{(4)} \sqcup \mM _{(2,1^2)}^{(4)} \sqcup \mM_{(2^2)}^{(4)} \sqcup \mM_{(3,1)}^{(4)} \sqcup \mM_{(4)} ^{(4)}.
\end{equation*}
We now schematically represent the structure of $\mX_{\lambda} \left[ \overset{ 2}{\circ} - \overset{ 4}{\Box}   \right]$ along each stratum $\mM_{\lambda} ^{(4)}$. We draw the brane setup on the left, in the middle the resolved crystal $\fX_{\lambda}$ and on the right the Hasse diagram of $\mX$ with the quaternionic dimension of the leaves explicitly written.
\begin{itemize}
	\item For $\lambda =(1^4)$ $\mathcal{H}$ is lifted and $\mC$ is fully resolved. Correspondingly, the crystal is resolved into $\fX_{(1^4)} =\ytableausetup{smalltableaux} \begin{ytableau} 1 & \none[ \scriptstyle \oplus ] & 1 & \none[ \scriptstyle \oplus ] & 1 & \none[ \scriptstyle \oplus ] & 1 \end{ytableau} $. There are no singularities and hence no transitions.\par
	\item For $\lambda= (2,1^2)$ we have 
	\begin{equation*}
	\ytableausetup{smalltableaux}
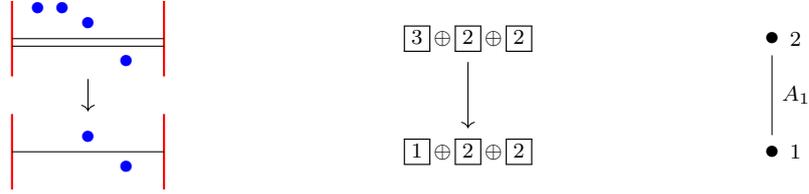
	\begin{tikzpicture}
		\node (t1) at (0,0) {\begin{ytableau} 3 & \none[\scriptstyle \oplus] &  2  & \none[\scriptstyle \oplus] &  2 \end{ytableau}}  ; 
		\node (t2) at (0,-1.5) {\begin{ytableau} 1 & \none[\scriptstyle \oplus] &  2  & \none[\scriptstyle \oplus] &  2 \end{ytableau}}  ; 
		\node (h1) at (4,0) {$\bullet$};
		\node (h2) at (4,-1.5) {$\bullet$};
		\node[anchor=west] (l1) at (4.1,0) {${\scriptstyle 2 }$};
		\node[anchor=west] (l2) at (4.1,-1.5) {${\scriptstyle 1 }$};
		\draw[->] (t1) edge (t2);
		\path[] (h1) edge node[right] {$\scriptstyle A_{1}$} (h2);
		
		\node (d5a) at (-5.5,0.4) {$\dbr \ \dbr $};
		\node (d5c) at (-5,0.2) {$\dbr  $};
		\node (d5d) at (-4.5,-0.3) {$\dbr  $};
		\path[red,thick] (-6,0.5) edge (-6,-0.5);
		\path[red,thick] (-4,0.5) edge (-4,-0.5);
		\path (-6,0) edge (-4,0);
		\path (-6,-0.1) edge (-4,-0.1);

		\node (d5B) at (-5,-1.3) {$\dbr  $};
		\node (d5C) at (-4.5,-1.7) {$\dbr  $};
		\path[red,thick] (-6,-1) edge (-6,-2);
		\path[red,thick] (-4,-1) edge (-4,-2);
		\path (-6,-1.5) edge (-4,-1.5);
		
		\node (aux1) at (-5,-0.4) {};
		\node (aux2) at (-5,-1.1) {};
		\draw[->] (aux1) edge (aux2);
	\end{tikzpicture}
	\end{equation*}
	\item For $\lambda= (2^2)$ we have 
	\begin{equation*}
	\ytableausetup{smalltableaux}
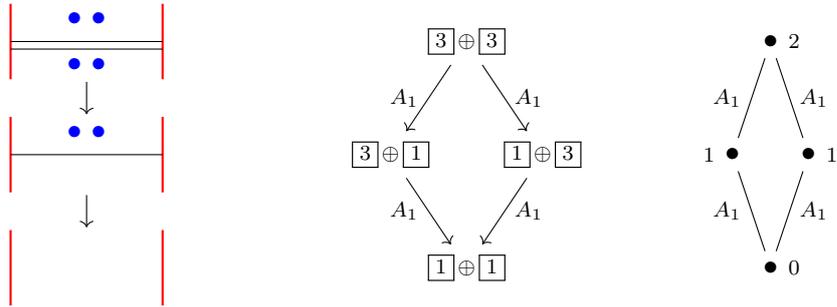
	\begin{tikzpicture}
		\node (t1) at (0,0) {$\begin{ytableau} 3 & \none[ \scriptstyle \oplus ] & 3 \end{ytableau}$};
		\node (t2a) at (-1,-1.5) {$\begin{ytableau} 3 & \none[ \scriptstyle \oplus ] & 1 \end{ytableau}$};
		\node (t2b) at (1,-1.5) {$\begin{ytableau} 1 & \none[ \scriptstyle \oplus ] & 3 \end{ytableau}$};
		\node (t3) at (0,-3) {$\begin{ytableau} 1 & \none[ \scriptstyle \oplus ] & 1 \end{ytableau}$};
		
		\path[->] (t1) edge node[left] {${\scriptstyle A_1}$} (t2a);
		\path[->] (t1) edge node[right] {${\scriptstyle A_1}$} (t2b);
		\path[->] (t2a) edge node[left] {${\scriptstyle A_1}$} (t3);
		\path[->] (t2b) edge node[right] {${\scriptstyle A_1}$} (t3);
		
		\node (h1) at (4,0) {$\bullet$};
		\node (h2a) at (4.5,-1.5) {$\bullet$};
		\node (h2b) at (3.5,-1.5) {$\bullet$};
		\node (h3) at (4,-3) {$\bullet$};
		
		\node[anchor=west] (l1) at (4.1,0) {${\scriptstyle 2}$};
		\node[anchor=west] (l2a) at (4.6,-1.5) {${\scriptstyle 1}$};
		\node[anchor=east] (l2b) at (3.4,-1.5) {${\scriptstyle 1}$};
		\node[anchor=west] (l3) at (4.1,-3) {${\scriptstyle 0}$};
		
		\path[] (h1) edge node[right] {$\scriptstyle A_{1}$} (h2a);
		\path[] (h1) edge node[left] {$\scriptstyle A_{1}$} (h2b);
		\path[] (h2a) edge node[right] {$\scriptstyle A_{1}$} (h3);
		\path[] (h2b) edge node[left] {$\scriptstyle A_{1}$} (h3);

		\node (d5a) at (-5,0.3) {$\dbr \ \dbr  $};
		\node (d5b) at (-5,-0.3) {$\dbr \ \dbr  $};
		\path[red,thick] (-6,0.5) edge (-6,-0.5);
		\path[red,thick] (-4,0.5) edge (-4,-0.5);
		\path (-6,0) edge (-4,0);
		\path (-6,-0.1) edge (-4,-0.1);

		\node (d5B) at (-5,-1.2) {$\dbr \ \dbr $};
		\path[red,thick] (-6,-1) edge (-6,-2);
		\path[red,thick] (-4,-1) edge (-4,-2);
		\path (-6,-1.5) edge (-4,-1.5);
		
		\node (aux1) at (-5,-0.4) {};
		\node (aux2) at (-5,-1.1) {};
		\draw[->] (aux1) edge (aux2);
		
		\path[red,thick] (-6,-2.5) edge (-6,-3.5);
		\path[red,thick] (-4,-2.5) edge (-4,-3.5);
		
		\node (auxC) at (-5,-1.9) {};
		\node (auxD) at (-5,-2.6) {};
		\draw[->] (auxC) edge (auxD);
	\end{tikzpicture}
	\end{equation*}
	Be aware of the difference between the $\lambda=(2,2)$ mass deformation of $\U{2}$ with $n=4$ and the $\U{2} \times \U{2}$ quiver with framing $\un =(2,2)$.
	\item For $\lambda= (3,1)$ we have 
	\begin{equation*}
	\ytableausetup{smalltableaux}
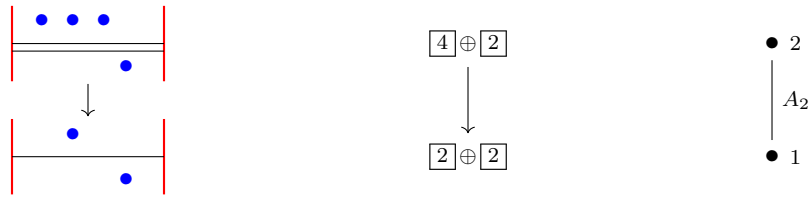
	\begin{tikzpicture}
		\node (t1) at (0,0) {$\begin{ytableau} 4 & \none[ \scriptstyle \oplus ] & 2 \end{ytableau}$}  ; 
		\node (t2) at (0,-1.5) {$\begin{ytableau} 2 & \none[ \scriptstyle \oplus ] & 2 \end{ytableau}$}  ; 
		\node (h1) at (4,0) {$\bullet$};
		\node (h2) at (4,-1.5) {$\bullet$};
		\node[anchor=west] (l1) at (4.1,0) {${\scriptstyle 2 }$};
		\node[anchor=west] (l2) at (4.1,-1.5) {${\scriptstyle 1 }$};
		\draw[->] (t1) edge (t2);
		\path[] (h1) edge node[right] {$\scriptstyle A_{2}$} (h2);
		
		\node (d5a) at (-5.2,0.3) {$\dbr \ \dbr \ \dbr $};
		\node (d5c) at (-4.5,-0.3) {$\dbr  $};
		\path[red,thick] (-6,0.5) edge (-6,-0.5);
		\path[red,thick] (-4,0.5) edge (-4,-0.5);
		\path (-6,0) edge (-4,0);
		\path (-6,-0.1) edge (-4,-0.1);

		\node (d5B) at (-5.2,-1.2) {$\dbr  $};
		\node (d5C) at (-4.5,-1.8) {$\dbr  $};
		\path[red,thick] (-6,-1) edge (-6,-2);
		\path[red,thick] (-4,-1) edge (-4,-2);
		\path (-6,-1.5) edge (-4,-1.5);
		
		\node (aux1) at (-5,-0.4) {};
		\node (aux2) at (-5,-1.1) {};
		\draw[->] (aux1) edge (aux2);
	\end{tikzpicture}
	\end{equation*}
	\item For $\lambda=(4)$ we have 
	\begin{equation*}
	\ytableausetup{smalltableaux}
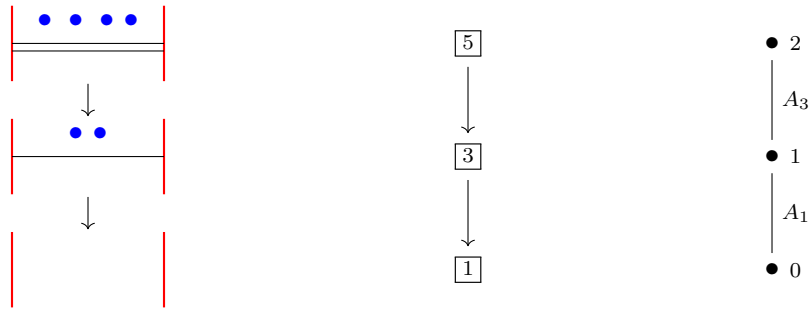
	\begin{tikzpicture}
		\node (t1) at (0,0) {\begin{ytableau} 5 \end{ytableau}}  ; 
		\node (t1a) at (0,-1.5) {\begin{ytableau} 3 \end{ytableau}} ; 
		\node (t1b) at (0,-3) {\begin{ytableau} 1  \end{ytableau}} ; 
		
		\node (h1) at (4,0) {$\bullet$};
		\node (h2) at (4,-1.5) {$\bullet$};
		\node (h3) at (4,-3) {$\bullet$};
		\node[anchor=west] (l1) at (4.1,0) {${\scriptstyle 2}$};
		\node[anchor=west] (l2) at (4.1,-1.5) {${\scriptstyle 1}$};
		\node[anchor=west] (l3) at (4.1,-3) {${\scriptstyle 0}$};
		\draw[->] (t1) edge (t1a);
		\draw[->] (t1a) edge (t1b);
		\path[] (h1) edge node[right] {$\scriptstyle A_{3}$} (h2);
		\path[] (h2) edge node[right] {$\scriptstyle A_{1}$} (h3);

		\node (d5a) at (-5,0.3) {$\dbr \ \dbr \ \dbr \ \dbr $};
		\path[red,thick] (-6,0.5) edge (-6,-0.5);
		\path[red,thick] (-4,0.5) edge (-4,-0.5);
		\path (-6,0) edge (-4,0);
		\path (-6,-0.1) edge (-4,-0.1);

		\node (d5B) at (-5,-1.2) {$\dbr \ \dbr $};
		\path[red,thick] (-6,-1) edge (-6,-2);
		\path[red,thick] (-4,-1) edge (-4,-2);
		\path (-6,-1.5) edge (-4,-1.5);
		
		\node (aux1) at (-5,-0.4) {};
		\node (aux2) at (-5,-1.1) {};
		\draw[->] (aux1) edge (aux2);
		
		\path[red,thick] (-6,-2.5) edge (-6,-3.5);
		\path[red,thick] (-4,-2.5) edge (-4,-3.5);
		
		\node (auxC) at (-5,-1.9) {};
		\node (auxD) at (-5,-2.6) {};
		\draw[->] (auxC) edge (auxD);
	\end{tikzpicture}
	\end{equation*}
\end{itemize}
Moving gradually to higher-codimensional loci in $\mM ^{(n)}$ we observe:
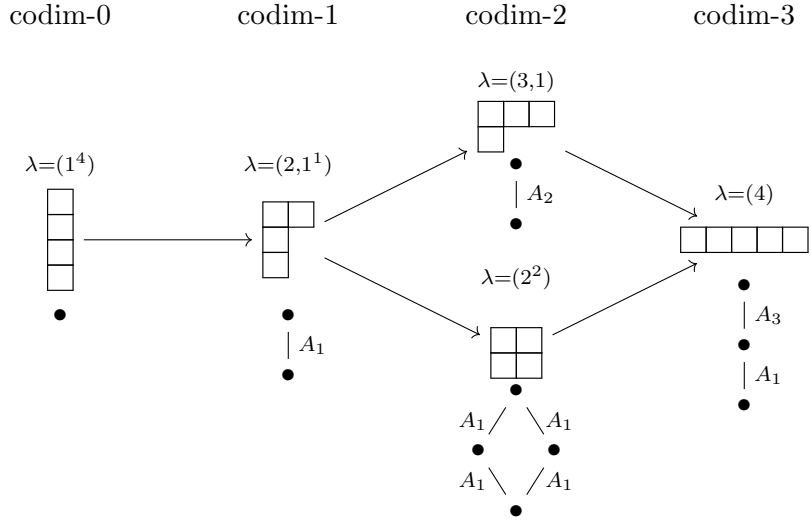
\begin{equation*}
	\begin{tikzpicture}
		\node (t1) at (-6,0) {\begin{ytableau} \ \\ \ \\ \ \\ \ \end{ytableau}}  ; 
		\node (t2) at (-3,0) {$\begin{ytableau} \ & \ \\ \ \\ \ \end{ytableau}$} ; 
		\node (t31) at (0,1.5) {$\begin{ytableau} \ & \ &  \ \\ \ \end{ytableau}$} ; 
		\node (t22) at (0,-1.5) {$\begin{ytableau} \ & \ \\ \ & \ \end{ytableau}$} ; 		
		\node (t4) at (3,0) {$\begin{ytableau} \ &  & \ & \ & \ \end{ytableau}$} ; 
		
		\node (l1) at (3,0.6) {${\scriptstyle \lambda = (4)}$};
		\node (l2a) at (0,2.1) {${\scriptstyle \lambda = (3,1)}$};
		\node (l2b) at (0,-0.5) {${\scriptstyle \lambda = (2^2)}$};
		\node (l3) at (-3,1) {${\scriptstyle \lambda = (2,1^1)}$};
		\node (l4) at (-6,1) {${\scriptstyle \lambda = (1^4)}$};
		
		\node (h11) at (3,-.6) {$\bullet$};
		\node (h12) at (3,-1.4) {$\bullet$};
		\node (h13) at (3,-2.2) {$\bullet$};
		\node (h2a1) at (0,1) {$\bullet$};
		\node (h2a2) at (0,0.2) {$\bullet$};
		\node (h2b1) at (0,-2) {$\bullet$};
		\node (h2b2l) at (-0.5,-2.8) {$\bullet$};
		\node (h2b2r) at (0.5,-2.8) {$\bullet$};
		\node (h2b3) at (0,-3.6) {$\bullet$};
		\node (h31) at (-3,-1) {$\bullet$};
		\node (h32) at (-3,-1.8) {$\bullet$};
		\node (h4) at (-6,-1) {$\bullet$};
		\node (c0) at (-6,3) {codim-0};
		\node (c1) at (-3,3) {codim-1};
		\node (c2) at (0,3) {codim-2};
		\node (c3) at (3,3) {codim-3};

		\draw[->] (t1) edge (t2);
		\draw[->] (t2) edge (t22);
		\draw[->] (t2) edge (t31);
		\draw[->] (t22) edge (t4);
		\draw[->] (t31) edge (t4);
		\path[] (h11) edge node[right] {$\scriptstyle A_{3}$} (h12);
		\path[] (h12) edge node[right] {$\scriptstyle A_{1}$} (h13);
		\path[] (h2a1) edge node[right] {$\scriptstyle A_{2}$} (h2a2);
		\path[] (h2b1) edge node[left] {$\scriptstyle A_{1}$} (h2b2l);
		\path[] (h2b1) edge node[right] {$\scriptstyle A_{1}$} (h2b2r);
		\path[] (h2b2l) edge node[left] {$\scriptstyle A_{1}$} (h2b3);
		\path[] (h2b2r) edge node[right] {$\scriptstyle A_{1}$} (h2b3);
		\path[] (h31) edge node[right] {$\scriptstyle A_{1}$} (h32);
	\end{tikzpicture}
\end{equation*}

\subsubsection{SQCD}
Examples of resolved crystals $\fX \left[ \overset{ N}{\circ} - \overset{ n}{\Box} \right]$ for $\U{N}$ theories with $n$ fundamental flavours are collected in Appendix \ref{app:ExamplesC}. The Hasse diagram of $\mX_{\lambda} \left[ \overset{ 2}{\circ} - \overset{ n}{\Box} \right]$ is: 
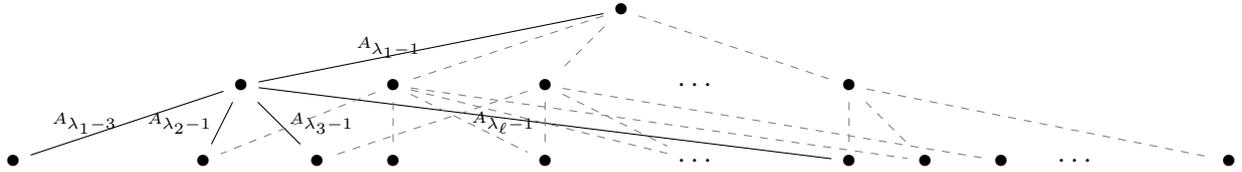
\begin{equation*}
	\begin{tikzpicture}
		\node (top) at (0,0) {$\bullet$};
		\node (l1) at (-5,-1) {$\bullet$};
		\node (l2) at (-3,-1) {$\bullet$};
		\node (l3) at (-1,-1) {$\bullet$};
		\node (1empty) at (1,-1) {$ \cdots $};
		\node (lm) at (3,-1) {$\bullet$};
		\node (l11) at (-8,-2) {$\bullet$};
		\node (l12) at (-5.5,-2) {$\bullet$};
		\node (l13) at (-4,-2) {$\bullet$};
		\node (l22) at (-3,-2) {$\bullet$};
		\node (l23) at (-1,-2) {$\bullet$};
		\node (2empty) at (1,-2) {$ \cdots $};
		\node (lm1) at (3,-2) {$\bullet$};
		\node (lm2) at (4,-2) {$\bullet$};
		\node (lm3) at (5,-2) {$\bullet$};
		\node (lmempty) at (6,-2)  {$\cdots $};
		\node (lmm) at (8,-2) {$\bullet$};

		\path[] (top) edge node[left] {${\scriptscriptstyle A_{\lambda_1 -1} }$} (l1);
		\draw[gray,dashed] (top)--(l2);
		\draw[gray,dashed] (top)--(l3);
		\draw[gray,dashed] (top)--(lm);
		\path[] (l1) edge node[left] {${\scriptscriptstyle A_{\lambda_1 -3} }$} (l11);
		\path[] (l1) edge node[left] {${\scriptscriptstyle A_{\lambda_2 -1} }$} (l12);
		\path[] (l1) edge node[right] {${\scriptscriptstyle A_{\lambda_3 -1} }$} (l13);
		\path[] (l1) edge node[left] {${\scriptscriptstyle A_{\lambda_{\ell} -1} }$} (lm1);
		\draw[gray,dashed] (l2)--(l12);
		\draw[gray,dashed] (l2)--(l22);
		\draw[gray,dashed] (l2)--(l23);
		\draw[gray,dashed] (l3)--(l13);
		\draw[gray,dashed] (l3)--(l23);
		\draw[gray,dashed] (l2)--(lm2);
		\draw[gray,dashed] (l3)--(lm3);
		\draw[gray,dashed] (l2)--(2empty);
		\draw[gray,dashed] (l3)--(2empty);
		\draw[gray,dashed] (lm)--(lm1);
		\draw[gray,dashed] (lm)--(lm2);
		\draw[gray,dashed] (lm)--(lmm);
	\end{tikzpicture}
\end{equation*}
while higher rank theories have Hasse diagrams that continue below in the same fashion. We have written the type of a few transitions explicitly as an example, with other lines gray and dashed to help visualization.\par

\subsubsection{$A_2, \un=(5,2)$}

We now discuss the resolution $\fX \left[ \overset{ 5}{\Box} -  \overset{ 4}{\circ}-  \overset{ 3}{\circ} - \overset{ 2}{\Box} \right]$ of the crystal \eqref{eq:A2n52} when real masses are turned on. We focus on the concrete case $\lambda = (5,2)$. There are three possible highest weight tableaux:
		\begin{equation*}
			\ytableausetup{smalltableaux}
				 \left(  \begin{ytableau} 6 & \none[ \scriptstyle \otimes ] & 1 \end{ytableau} \right) \oplus \left( \begin{ytableau} 1 & \none[ \scriptstyle \otimes ] & 3 \end{ytableau} \right) \qquad \left(  \begin{ytableau} 5 & \none[ \scriptstyle \otimes ] & 2 \end{ytableau} \right) \oplus \left( \begin{ytableau} 2 & \none[ \scriptstyle \otimes ] & 2 \end{ytableau} \right) \qquad   \left(  \begin{ytableau} 4 & \none[ \scriptstyle \otimes ] & 3 \end{ytableau} \right) \oplus \left( \begin{ytableau} 3 & \none[ \scriptstyle \otimes ] & 1 \end{ytableau} \right) 
 		\end{equation*}
 		corresponding to, respectively, 
 		\begin{equation*}
 			\left( \mathsf{w}_1 (\lambda_1), \mathsf{w}_2 (\lambda_1) \right) = \begin{cases} (5,0) , \\ (4,1) , \\ (3,2) ,  \end{cases}  \quad \left( \mathsf{w}_1 (\lambda_2), \mathsf{w}_2 (\lambda_2) \right) = \begin{cases} (0,2) , \\ (1,1) , \\ (2,0) .  \end{cases} 
 		\end{equation*}
 		The three resulting crystals are:	
 		\begin{equation}
	\ytableausetup{smalltableaux}

\label{eq:massdefA2n52C3}
\end{equation}
An arrow is gray and dashed to improve the visualization and might be thought of as winding around the crystal.\par
In all three cases, the outcome matches with the analysis of brane configurations:
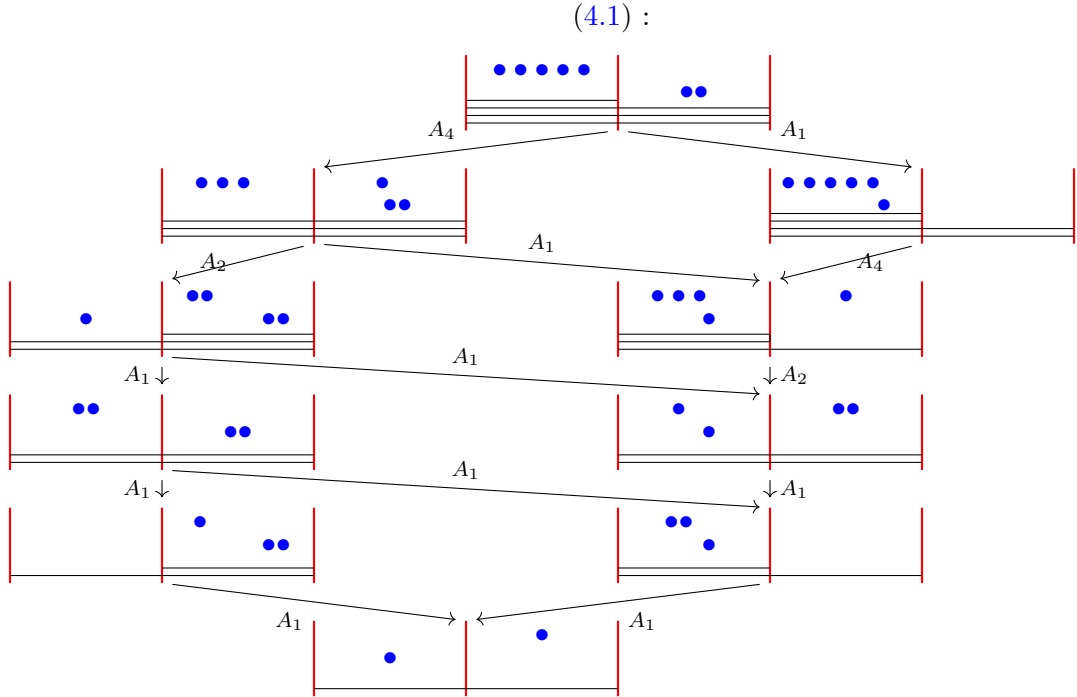
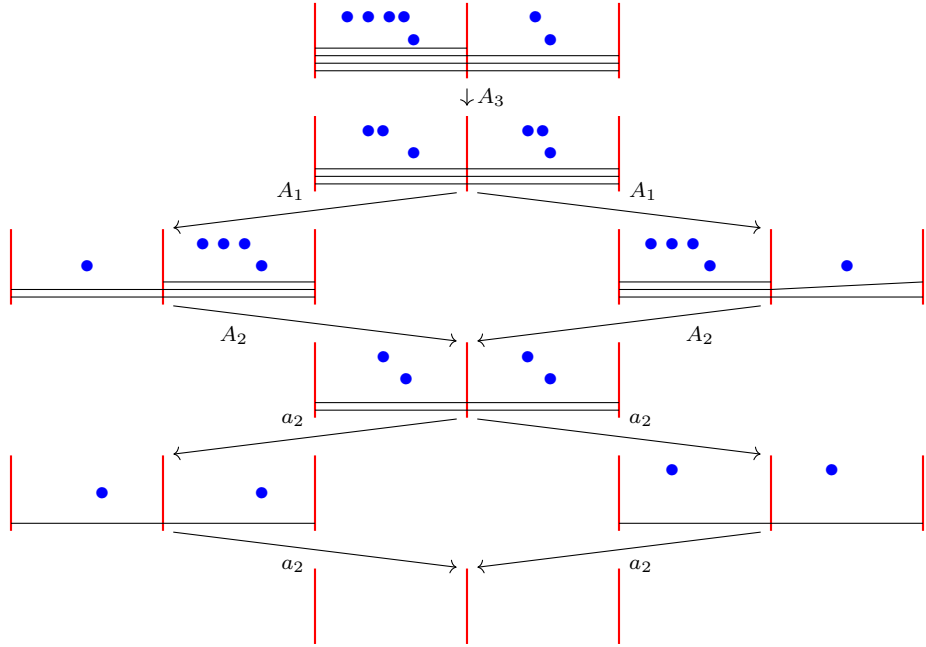
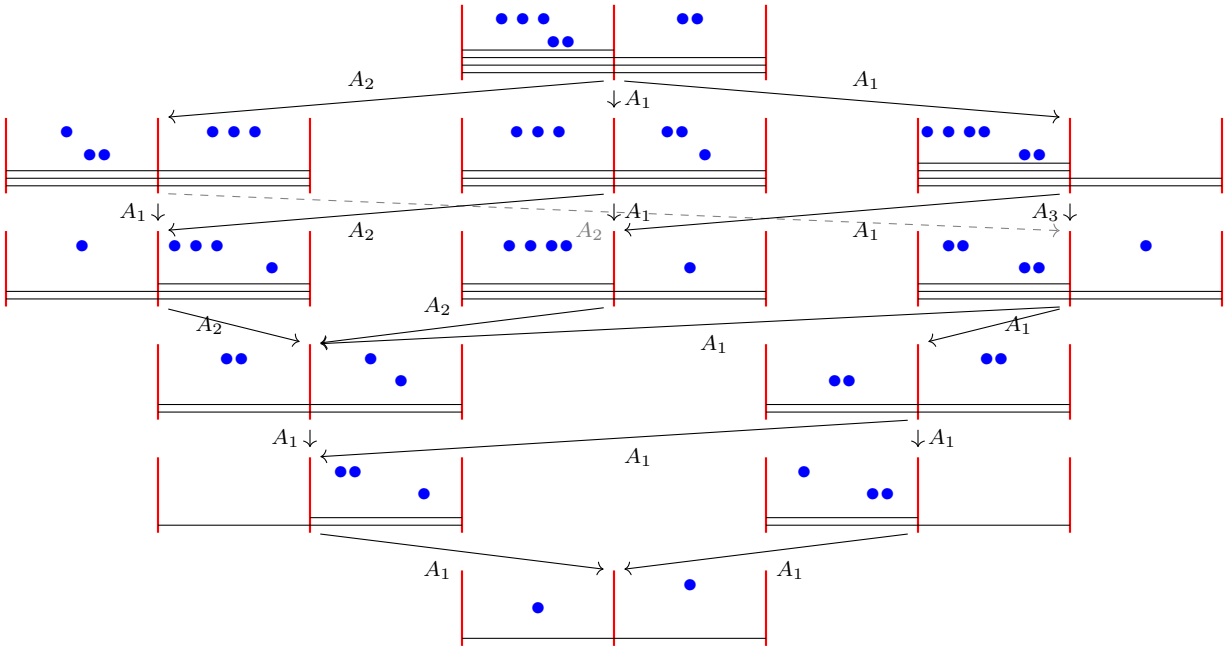
\begin{equation*}
	%
\end{equation*}

\section{Crystals for other classical root systems}
\label{sec:BCD}
Despite having given Result \ref{thm:precise} in full generality, so far the spotlight has been on type A quivers. Nevertheless, Kashiwara crystals exist for type BCD quivers as well \cite{Kashiwara96} (see \cite{Hong,BumpSchilling} for textbook treatments). Crystals for Coulomb branches of type BCD are in bijection with sequences of phases of brane configurations involving ON planes. We refrain from a detailed description, which follows straightforwardly from combining the ideas in Section \ref{sec:crystalBrane} with the analysis of \cite{Bourget:2021siw}.\par
We continue to exploit the map \eqref{eq:bijcrys}, with one additional detail: we will draw the boxes in the same shape as $\mathsf{Q}^{\circ}$. This will make clear our conventions about, for instance, which boxes of a tableaux correspond to which spinor node in type D quivers. Let us remark that, despite the unusual appearance, our crystals agree with the mathematical literature, simply because they satisfy the same set of axioms.\par
As above, we provide evidence in support of Result \ref{thm:precise} through the construction of explicit examples.

\subsection{Type D examples}
\label{sec:Dquivercrystal}
The Dynkin diagram of type D is classical and simply-laced. We proceed as in Section \ref{sec:CrystalAxioms}, where the rules to construct Kashiwara crystals have been stated for all finite or affine $\mathfrak{g}$.\par
A type D quiver $\mathsf{Q}$ may accommodate both A-type and D-type sub-quivers, thus we will see the appearance of a novel kind of transition, whose transverse slice is the closure of a minimal nilpotent orbit $d_r$. It corresponds to the $D_r$ quiver 
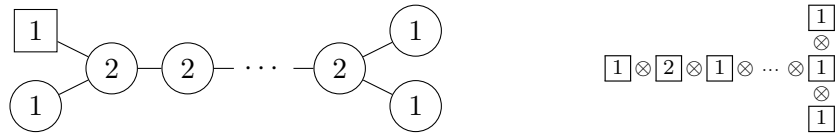
\begin{equation*}
\ytableausetup{smalltableaux} 
\begin{tikzpicture}[{square/.style={regular polygon,regular polygon sides=4}}]

	\node[circle,draw] (c1) at (0,0) {2};
	\node[circle,draw] (c2) at (1,0) {2};
	\node (c3) at (2,0) {$\cdots$};
	\node[circle,draw] (c4) at (3,0) {2};
	\node[circle,draw] (ld) at (-1,-0.5) {1};
	\node[circle,draw] (ru) at (4,0.5) {1};
	\node[circle,draw] (rd) at (4,-0.5) {1};
	\node[square,draw] (lu) at (-1,0.5) { \hspace{4pt} };
	\node[draw=none] (aux) at (-1,0.5) {1};
	
	\draw (c1)--(ld);
	\draw (c1)--(lu);
	\draw (c4)--(rd);
	\draw (c4)--(ru);
	\draw (c1)--(c2);
	\draw (c3)--(c2);
	\draw (c3)--(c4);

	\node (T) at (8,0) {$\begin{ytableau} \none & \none & \none & \none & \none & \none & \none & \none & 1 \\ \none & \none & \none & \none & \none & \none & \none & \none & \none[\scriptstyle \otimes]  \\ 1 & \none[ \scriptstyle \otimes] & 2 & \none[ \scriptstyle \otimes] & 1 & \none[ \scriptstyle \otimes] & \none[ \scriptscriptstyle \cdots] & \none[ \scriptstyle \otimes] &1 \\  \none & \none & \none & \none & \none & \none & \none & \none & \none[\scriptstyle \otimes] \\ \none & \none & \none & \none & \none & \none & \none & \none & 1 \end{ytableau}$};

\end{tikzpicture}
\end{equation*}
whose associated tableau is drawn on the right (note the different conventions for the spinor nodes compared to the linear part).

\subsubsection{Balanced $D_4$ quiver with two flavours}
To exemplify the difference between quivers of type A and D, let us consider the balanced $D_4$ quiver with gauge group $\U{4} \times \U{2}^3$ with two flavours attached at the $\U{4}$ node.\par
We show the quivers describing the various strata of the Coulomb branch on the left and the corresponding crystal on the right:
\begin{equation}
\ytableausetup{smalltableaux} 
\begin{tikzpicture}[{square/.style={regular polygon,regular polygon sides=4}}]

	\node (q51) at (-3.5,0) {$\circ$};
	\node (q52) at (-3,0) {$\circ$};
	\node (q54) at (-3,.5) {$\circ$};
	\node (q55) at (-3,-.5) {$\circ$};
	\node (q57) at (-2.5,0) {$\Box $};
	\node (a51) at (-3.25,0) {$-$};
	\node (a52) at (-2.75,0) {$-$};
	\node (v53) at (-3,.25) {$\scriptstyle \vert $};
	\node (v54) at (-3,-.25) {$\scriptstyle \vert $};
	
	\node (lab51) at (-3.5,-0.2) {$\scriptstyle 2$};
	\node (lab52) at (-2.8,-0.2) {$\scriptstyle 4$};
	\node (lab54) at (-2.5,0.3) {$\scriptstyle 2$};
	\node (lab55) at (-3,0.7) {$\scriptstyle 2$};
	\node (lab58) at (-3,-0.7) {$\scriptstyle 2 $};

	\node (q21) at (-3.5,-2.5) {$\circ$};
	\node (q22) at (-3,-2.5) {$\circ$};
	\node (q24) at (-3,-2) {$\circ$};
	\node (q25) at (-3,-3) {$\circ$};
	\node (q26) at (-2.5,-2) {$\Box $};
	\node (q27) at (-2.5,-3) {$\Box $};
	\node (q23) at (-3.5,-2) {$\Box $};
	\node (a21) at (-3.25,-2.5) {$-$};
	\node (a24) at (-2.75,-2) {$-$};
	\node (a25) at (-2.75,-3) {$-$};
	\node (v23) at (-3,-2.25) {$\scriptstyle \vert $};
	\node (v24) at (-3,-2.75) {$\scriptstyle \vert $};
	\node (v21) at (-3.5,-2.25) {$\scriptstyle \vert $};
	
	\node (lab21) at (-3.5,-2.7) {$\scriptstyle 2$};
	\node (lab22) at (-2.8,-2.7) {$\scriptstyle 3$};
	\node (lab24) at (-3.5,-1.7) {$\scriptstyle 1$};
	\node (lab25) at (-3,-1.8) {$\scriptstyle 2$};
	\node (lab28) at (-3,-3.2) {$\scriptstyle 2 $};
	\node (lab25) at (-2.5,-1.7) {$\scriptstyle 1$};
	\node (lab28) at (-2.5,-3.3) {$\scriptstyle 1 $};

	\node (q31) at (-3.5,-5) {$\circ$};
	\node (q32) at (-3,-5) {$\circ$};
	\node (q34) at (-3,-4.5) {$\circ$};
	\node (q35) at (-3,-5.5) {$\circ$};
	\node (q33) at (-3.5,-4.5) {$\Box $};
	\node (a31) at (-3.25,-5) {$-$};
	\node (v33) at (-3,-4.75) {$\scriptstyle \vert $};
	\node (v34) at (-3,-5.25) {$\scriptstyle \vert $};
	\node (v31) at (-3.5,-4.75) {$\scriptstyle \vert $};
	
	\node (lab31) at (-3.5,-5.2) {$\scriptstyle 2$};
	\node (lab32) at (-2.8,-5.2) {$\scriptstyle 2$};
	\node (lab34) at (-3.5,-4.2) {$\scriptstyle 2$};
	\node (lab35) at (-3,-4.3) {$\scriptstyle 1$};
	\node (lab38) at (-3,-5.7) {$\scriptstyle 1 $};

	\node (q41) at (-3.5,-7.5) {$\circ$};
	\node (q42) at (-3,-7.5) {$\circ$};
	\node (q44) at (-3,-7) {$\circ$};
	\node (q45) at (-3,-8) {$\circ$};
	\node (q43) at (-2.5,-7.5) {$\Box $};
	\node (a41) at (-3.25,-7.5) {$-$};
	\node (a42) at (-2.75,-7.5) {$-$};
	\node (v43) at (-3,-7.75) {$\scriptstyle \vert $};
	\node (v44) at (-3,-7.25) {$\scriptstyle \vert $};
	
	\node (lab41) at (-3.5,-7.7) {$\scriptstyle 1$};
	\node (lab42) at (-2.8,-7.7) {$\scriptstyle 2$};
	\node (lab44) at (-2.5,-7.2) {$\scriptstyle 1$};
	\node (lab45) at (-3,-6.8) {$\scriptstyle 1$};
	\node (lab48) at (-3,-8.2) {$\scriptstyle 1 $};

	\node (qf) at (-3,-10) {$\varnothing $};

	\node (T1) at (3,0) {$\begin{ytableau} \none & \none & 1  \\ \none & \none & \none[\scriptstyle \otimes]  \\ 1 & \none[ \scriptstyle \otimes] & 3 \\ \none & \none & \none[\scriptstyle \otimes]  \\ \none & \none & 1 \end{ytableau}$};
	\node (T2) at (3,-2.5) {$\begin{ytableau} \none & \none & 2  \\ \none & \none & \none[\scriptstyle \otimes]  \\ 2 & \none[ \scriptstyle \otimes] & 1 \\ \none & \none & \none[\scriptstyle \otimes]  \\ \none & \none & 2 \end{ytableau}$};
	\node (T3) at (3,-5) {$\begin{ytableau} \none & \none & 1  \\ \none & \none & \none[\scriptstyle \otimes]  \\ 3 & \none[ \scriptstyle \otimes] & 1 \\ \none & \none & \none[\scriptstyle \otimes]  \\ \none & \none & 1 \end{ytableau}$};
	\node (T4) at (3,-7.5) {$\begin{ytableau} \none & \none & 1  \\ \none & \none & \none[\scriptstyle \otimes]  \\ 1 & \none[ \scriptstyle \otimes] & 2 \\ \none & \none & \none[\scriptstyle \otimes]  \\ \none & \none & 1  \end{ytableau}$};
	\node (T5) at (3,-10) {$\begin{ytableau} \none & \none & 1  \\ \none & \none & \none[\scriptstyle \otimes]  \\ 1 & \none[ \scriptstyle \otimes] & 1 \\ \none & \none & \none[\scriptstyle \otimes]  \\ \none & \none & 1  \end{ytableau}$};

	\path[->] (T1) edge node[right] {$\scriptstyle A_1$} (T2);
	\path[->] (T2) edge node[right] {$\scriptstyle d_3$} (T3);
	\path[->] (T3) edge node[right] {$\scriptstyle A_1$} (T4);
	\path[->] (T4) edge node[right] {$\scriptstyle d_4$} (T5);

\end{tikzpicture}
\label{eq:exD4}
\end{equation}
The second transition, denoted $d_3$, is nothing but an $a_3$ transition involving the two spinor nodes. The transverse slices are isomorphic, $a_3 \cong d_3$, nevertheless the notation $d_3$ identifies which roots are involved and the consequent action on the neighbour boxes. The last step is a $d_4$ transition.\par

\subsubsection{Balanced $D_5$ quiver with four flavours}
The next example is a $D_5$ quiver with gauge group $\U{2} \times \U{4}^3 \times \U{6}$ and four flavours, two at each spinor node. We represent the quiver subtraction pattern on the left and the crystal on the right, showing perfect agreement:
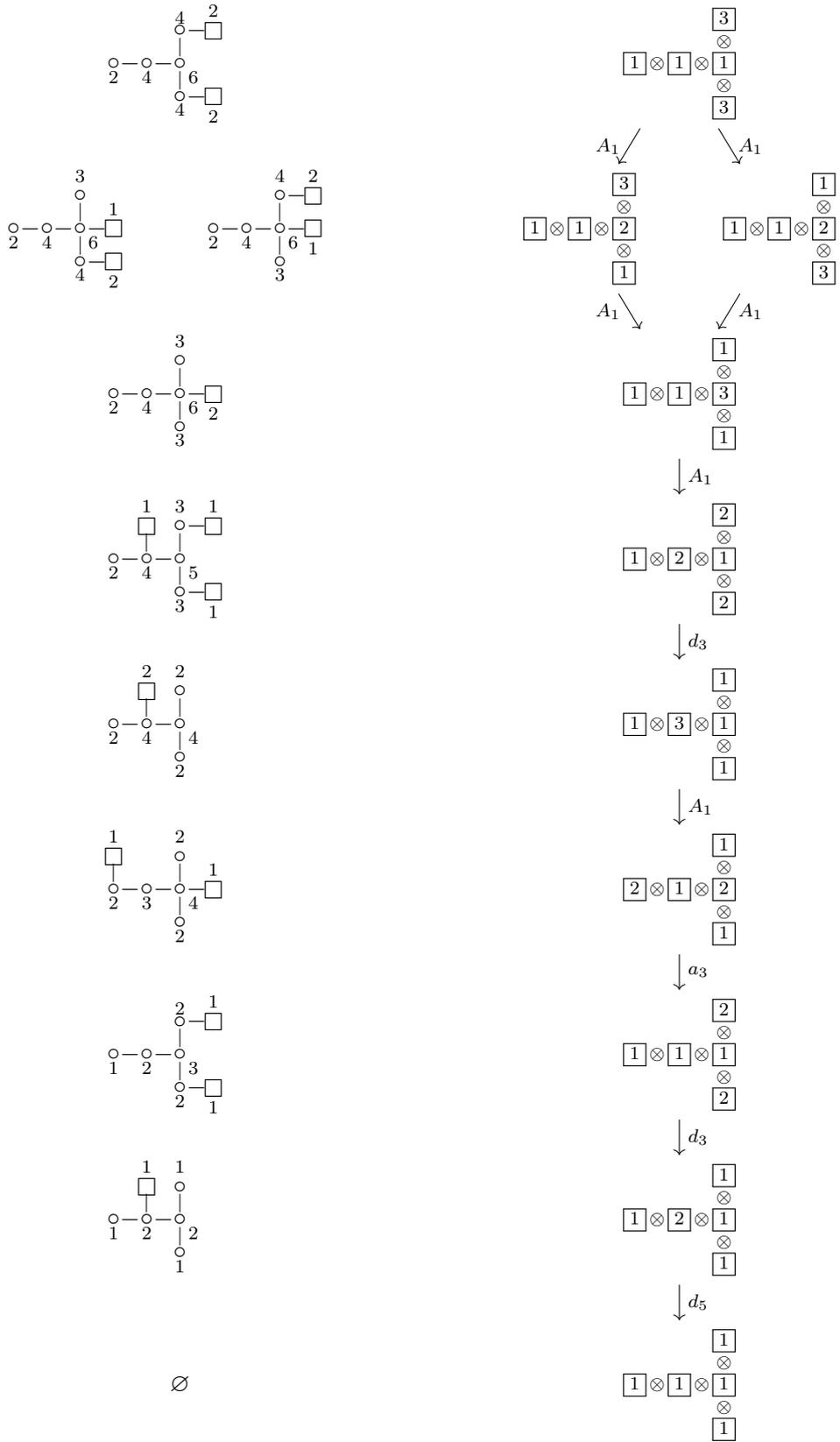
\begin{equation}
\ytableausetup{smalltableaux} 
\begin{tikzpicture}[{square/.style={regular polygon,regular polygon sides=4}}]
	\node (q11) at (-5.5,0) {$\circ$};
	\node (q12) at (-5,0) {$\circ$};
	\node (q13) at (-4.5,0) {$\circ$};
	\node (q14) at (-4.5,0.5) {$\circ$};
	\node (q15) at (-4.5,-0.5) {$\circ$};
	\node (q16) at (-4,-0.5) {$\Box $};
	\node (q17) at (-4,0.5) {$\Box $};
	\node (a11) at (-5.25,0) {$-$};
	\node (a12) at (-4.75,0) {$-$};
	\node (a13) at (-4.25,0.5) {$-$};
	\node (a14) at (-4.25,-0.5) {$-$};
	\node (v13) at (-4.5,0.25) {$\scriptstyle \vert $};
	\node (v14) at (-4.5,-0.25) {$\scriptstyle \vert $};
	
	\node (lab11) at (-5.5,-0.2) {$\scriptstyle 2$};
	\node (lab12) at (-5,-0.2) {$\scriptstyle 4$};
	\node (lab13) at (-4.3,-0.2) {$\scriptstyle 6$};
	\node (lab14) at (-4.5,0.7) {$\scriptstyle 4$};
	\node (lab15) at (-4.5,-0.7) {$\scriptstyle 4$};
	\node (lab16) at (-4,-0.8) {$\scriptstyle 2 $};
	\node (lab17) at (-4,0.8) {$\scriptstyle 2 $};

	\node (q2L1) at (-7,-2.5) {$\circ$};
	\node (q2L2) at (-6.5,-2.5) {$\circ$};
	\node (q2L3) at (-6,-2.5) {$\circ$};
	\node (q2L4) at (-6,-2) {$\circ$};
	\node (q2L5) at (-6,-3) {$\circ$};
	\node (q2L6) at (-5.5,-2.5) {$\Box $};
	\node (q2L7) at (-5.5,-3) {$\Box $};
	\node (a2L1) at (-6.75,-2.5) {$-$};
	\node (a2L2) at (-6.25,-2.5) {$-$};
	\node (a2L3) at (-5.75,-2.5) {$-$};
	\node (a2L4) at (-5.75,-3) {$-$};
	\node (v2L3) at (-6,-2.25) {$\scriptstyle \vert $};
	\node (v2L4) at (-6,-2.75) {$\scriptstyle \vert $};
	
	\node (lab2L1) at (-7,-2.7) {$\scriptstyle 2$};
	\node (lab2L2) at (-6.5,-2.7) {$\scriptstyle 4$};
	\node (lab2L3) at (-5.8,-2.7) {$\scriptstyle 6$};
	\node (lab2L4) at (-6,-1.7) {$\scriptstyle 3$};
	\node (lab2L5) at (-6,-3.2) {$\scriptstyle 4$};
	\node (lab2L6) at (-5.5,-3.3) {$\scriptstyle 2 $};
	\node (lab2L7) at (-5.5,-2.2) {$\scriptstyle 1 $};

	\node (q2R1) at (-4,-2.5) {$\circ$};
	\node (q2R2) at (-3.5,-2.5) {$\circ$};
	\node (q2R3) at (-3,-2.5) {$\circ$};
	\node (q2R4) at (-3,-2) {$\circ$};
	\node (q2R5) at (-3,-3) {$\circ$};
	\node (q2R6) at (-2.5,-2.5) {$\Box $};
	\node (q2R7) at (-2.5,-2) {$\Box $};
	\node (a2R1) at (-3.75,-2.5) {$-$};
	\node (a2R2) at (-3.25,-2.5) {$-$};
	\node (a2R3) at (-2.75,-2.5) {$-$};
	\node (a2R4) at (-2.75,-2) {$-$};
	\node (v2R3) at (-3,-2.25) {$\scriptstyle \vert $};
	\node (v2R4) at (-3,-2.75) {$\scriptstyle \vert $};
	
	\node (lab2R1) at (-4,-2.7) {$\scriptstyle 2$};
	\node (lab2R2) at (-3.5,-2.7) {$\scriptstyle 4$};
	\node (lab2R3) at (-2.8,-2.7) {$\scriptstyle 6$};
	\node (lab2R4) at (-3,-1.7) {$\scriptstyle 4$};
	\node (lab2R5) at (-3,-3.2) {$\scriptstyle 3$};
	\node (lab2R6) at (-2.5,-1.7) {$\scriptstyle 2 $};
	\node (lab2R7) at (-2.5,-2.8) {$\scriptstyle 1 $};

	\node (q31) at (-5.5,-5) {$\circ$};
	\node (q32) at (-5,-5) {$\circ$};
	\node (q33) at (-4.5,-5) {$\circ$};
	\node (q34) at (-4.5,-4.5) {$\circ$};
	\node (q35) at (-4.5,-5.5) {$\circ$};
	\node (q36) at (-4,-5) {$\Box $};
	\node (a31) at (-5.25,-5) {$-$};
	\node (a32) at (-4.75,-5) {$-$};
	\node (a33) at (-4.25,-5) {$-$};
	\node (v33) at (-4.5,-4.75) {$\scriptstyle \vert $};
	\node (v34) at (-4.5,-5.25) {$\scriptstyle \vert $};
	
	\node (lab31) at (-5.5,-5.2) {$\scriptstyle 2$};
	\node (lab32) at (-5,-5.2) {$\scriptstyle 4$};
	\node (lab33) at (-4.3,-5.2) {$\scriptstyle 6$};
	\node (lab34) at (-4.5,-4.2) {$\scriptstyle 3$};
	\node (lab35) at (-4.5,-5.7) {$\scriptstyle 3$};
	\node (lab37) at (-4,-5.3) {$\scriptstyle 2 $};

	\node (q41) at (-5.5,-7.5) {$\circ$};
	\node (q42) at (-5,-7.5) {$\circ$};
	\node (q43) at (-4.5,-7.5) {$\circ$};
	\node (q44) at (-4.5,-7) {$\circ$};
	\node (q45) at (-4.5,-8) {$\circ$};
	\node (q46) at (-4,-8) {$\Box $};
	\node (q47) at (-4,-7) {$\Box $};
	\node (q47) at (-5,-7) {$\Box $};
	\node (a41) at (-5.25,-7.5) {$-$};
	\node (a42) at (-4.75,-7.5) {$-$};
	\node (a43) at (-4.25,-7) {$-$};
	\node (a44) at (-4.25,-8) {$-$};
	\node (v43) at (-4.5,-7.25) {$\scriptstyle \vert $};
	\node (v44) at (-4.5,-7.75) {$\scriptstyle \vert $};
	\node (v45) at (-5,-7.25) {$\scriptstyle \vert $};
	
	\node (lab41) at (-5.5,-7.7) {$\scriptstyle 2$};
	\node (lab42) at (-5,-7.7) {$\scriptstyle 4$};
	\node (lab43) at (-4.3,-7.7) {$\scriptstyle 5$};
	\node (lab44) at (-4.5,-6.7) {$\scriptstyle 3$};
	\node (lab45) at (-4.5,-8.2) {$\scriptstyle 3$};
	\node (lab46) at (-4,-6.7) {$\scriptstyle 1 $};
	\node (lab47) at (-4,-8.3) {$\scriptstyle 1 $};
	\node (lab48) at (-5,-6.7) {$\scriptstyle 1 $};

	\node (q51) at (-5.5,-10) {$\circ$};
	\node (q52) at (-5,-10) {$\circ$};
	\node (q53) at (-4.5,-10) {$\circ$};
	\node (q54) at (-4.5,-9.5) {$\circ$};
	\node (q55) at (-4.5,-10.5) {$\circ$};
	\node (q57) at (-5,-9.5) {$\Box $};
	\node (a51) at (-5.25,-10) {$-$};
	\node (a52) at (-4.75,-10) {$-$};
	\node (v53) at (-4.5,-9.75) {$\scriptstyle \vert $};
	\node (v54) at (-4.5,-10.25) {$\scriptstyle \vert $};
	\node (v55) at (-5,-9.75) {$\scriptstyle \vert $};
	
	\node (lab51) at (-5.5,-10.2) {$\scriptstyle 2$};
	\node (lab52) at (-5,-10.2) {$\scriptstyle 4$};
	\node (lab53) at (-4.3,-10.2) {$\scriptstyle 4$};
	\node (lab54) at (-4.5,-9.2) {$\scriptstyle 2$};
	\node (lab55) at (-4.5,-10.7) {$\scriptstyle 2$};
	\node (lab58) at (-5,-9.2) {$\scriptstyle 2 $};

	\node (q61) at (-5.5,-12.5) {$\circ$};
	\node (q62) at (-5,-12.5) {$\circ$};
	\node (q63) at (-4.5,-12.5) {$\circ$};
	\node (q64) at (-4.5,-12) {$\circ$};
	\node (q65) at (-4.5,-13) {$\circ$};
	\node (q66) at (-5.5,-12) {$\Box $};
	\node (q67) at (-4,-12.5) {$\Box $};
	\node (a61) at (-5.25,-12.5) {$-$};
	\node (a62) at (-4.75,-12.5) {$-$};
	\node (a63) at (-4.25,-12.5) {$-$};
	\node (v63) at (-4.5,-12.75) {$\scriptstyle \vert $};
	\node (v64) at (-4.5,-12.25) {$\scriptstyle \vert $};
	\node (v65) at (-5.5,-12.25) {$\scriptstyle \vert $};
	
	\node (lab61) at (-5.5,-12.7) {$\scriptstyle 2$};
	\node (lab62) at (-5,-12.7) {$\scriptstyle 3$};
	\node (lab63) at (-4.3,-12.7) {$\scriptstyle 4$};
	\node (lab64) at (-4.5,-11.7) {$\scriptstyle 2$};
	\node (lab65) at (-4.5,-13.2) {$\scriptstyle 2$};
	\node (lab66) at (-4,-12.2) {$\scriptstyle 1 $};
	\node (lab68) at (-5.5,-11.7) {$\scriptstyle 1 $};

	\node (q71) at (-5.5,-15) {$\circ$};
	\node (q72) at (-5,-15) {$\circ$};
	\node (q73) at (-4.5,-15) {$\circ$};
	\node (q74) at (-4.5,-14.5) {$\circ$};
	\node (q75) at (-4.5,-15.5) {$\circ$};
	\node (q76) at (-4,-15.5) {$\Box $};
	\node (q77) at (-4,-14.5) {$\Box $};
	\node (a71) at (-5.25,-15) {$-$};
	\node (a72) at (-4.75,-15) {$-$};
	\node (a73) at (-4.25,-14.5) {$-$};
	\node (a74) at (-4.25,-15.5) {$-$};
	\node (v73) at (-4.5,-14.75) {$\scriptstyle \vert $};
	\node (v74) at (-4.5,-15.25) {$\scriptstyle \vert $};
	
	\node (lab71) at (-5.5,-15.2) {$\scriptstyle 1$};
	\node (lab72) at (-5,-15.2) {$\scriptstyle 2$};
	\node (lab73) at (-4.3,-15.2) {$\scriptstyle 3$};
	\node (lab74) at (-4.5,-14.3) {$\scriptstyle 2$};
	\node (lab75) at (-4.5,-15.7) {$\scriptstyle 2$};
	\node (lab76) at (-4,-15.8) {$\scriptstyle 1 $};
	\node (lab77) at (-4,-14.2) {$\scriptstyle 1 $};

	\node (q81) at (-5.5,-17.5) {$\circ$};
	\node (q82) at (-5,-17.5) {$\circ$};
	\node (q83) at (-4.5,-17.5) {$\circ$};
	\node (q84) at (-4.5,-17) {$\circ$};
	\node (q85) at (-4.5,-18) {$\circ$};
	\node (q86) at (-5,-17) {$\Box $};
	\node (a81) at (-5.25,-17.5) {$-$};
	\node (a82) at (-4.75,-17.5) {$-$};
	\node (v83) at (-4.5,-17.75) {$\scriptstyle \vert $};
	\node (v84) at (-4.5,-17.25) {$\scriptstyle \vert $};
	\node (v85) at (-5,-17.25) {$\scriptstyle \vert $};
	
	\node (lab81) at (-5.5,-17.7) {$\scriptstyle 1$};
	\node (lab82) at (-5,-17.7) {$\scriptstyle 2$};
	\node (lab83) at (-4.3,-17.7) {$\scriptstyle 2$};
	\node (lab84) at (-4.5,-16.7) {$\scriptstyle 1$};
	\node (lab85) at (-4.5,-18.2) {$\scriptstyle 1$};
	\node (lab88) at (-5,-16.7) {$\scriptstyle 1 $};
	
	\node (q9) at (-4.5,-20) {$\varnothing $};

	\node (T1) at (3,0) {$\begin{ytableau} \none & \none & \none & \none & 3  \\ \none & \none & \none & \none & \none[\scriptstyle \otimes]  \\ 1 & \none[ \scriptstyle \otimes] & 1  & \none[ \scriptstyle \otimes] & 1 \\ \none & \none & \none & \none & \none[\scriptstyle \otimes]  \\ \none & \none & \none & \none & 3 \end{ytableau}$};
	\node (T2a) at (1.5,-2.5) {$\begin{ytableau} \none & \none & \none & \none & 3  \\ \none & \none & \none & \none & \none[\scriptstyle \otimes]  \\ 1 & \none[ \scriptstyle \otimes] & 1  & \none[ \scriptstyle \otimes] & 2 \\ \none & \none & \none & \none & \none[\scriptstyle \otimes]  \\ \none & \none & \none & \none & 1 \end{ytableau}$};
	\node (T2b) at (4.5,-2.5) {$\begin{ytableau} \none & \none & \none & \none & 1  \\ \none & \none & \none & \none & \none[\scriptstyle \otimes]  \\ 1 & \none[ \scriptstyle \otimes] & 1  & \none[ \scriptstyle \otimes] & 2 \\ \none & \none & \none & \none & \none[\scriptstyle \otimes]  \\ \none & \none & \none & \none & 3 \end{ytableau}$};
	\node (T3) at (3,-5) {$\begin{ytableau} \none & \none & \none & \none & 1  \\ \none & \none & \none & \none & \none[\scriptstyle \otimes]  \\ 1 & \none[ \scriptstyle \otimes] & 1  & \none[ \scriptstyle \otimes] & 3 \\ \none & \none & \none & \none & \none[\scriptstyle \otimes]  \\ \none & \none & \none & \none & 1  \end{ytableau}$};
	\node (T4) at (3,-7.5) {$\begin{ytableau} \none & \none & \none & \none & 2  \\ \none & \none & \none & \none & \none[\scriptstyle \otimes]  \\ 1 & \none[ \scriptstyle \otimes] & 2  & \none[ \scriptstyle \otimes] & 1 \\ \none & \none & \none & \none & \none[\scriptstyle \otimes]  \\ \none & \none & \none & \none & 2 \end{ytableau}$};
	\node (T5) at (3,-10) {$\begin{ytableau} \none & \none & \none & \none & 1  \\ \none & \none & \none & \none & \none[\scriptstyle \otimes]  \\ 1 & \none[ \scriptstyle \otimes] & 3  & \none[ \scriptstyle \otimes] & 1 \\ \none & \none & \none & \none & \none[\scriptstyle \otimes]  \\ \none & \none & \none & \none & 1 \end{ytableau}$};
	\node (T6) at (3,-12.5) {$\begin{ytableau} \none & \none & \none & \none & 1  \\ \none & \none & \none & \none & \none[\scriptstyle \otimes]  \\ 2 & \none[ \scriptstyle \otimes] & 1  & \none[ \scriptstyle \otimes] & 2 \\ \none & \none & \none & \none & \none[\scriptstyle \otimes]  \\ \none & \none & \none & \none & 1 \end{ytableau}$};
	\node (T7) at (3,-15) {$\begin{ytableau} \none & \none & \none & \none & 2 \\ \none & \none & \none & \none & \none[\scriptstyle \otimes]  \\ 1 & \none[ \scriptstyle \otimes] & 1  & \none[ \scriptstyle \otimes] & 1 \\ \none & \none & \none & \none & \none[\scriptstyle \otimes]  \\ \none & \none & \none & \none & 2 \end{ytableau}$};
	\node (T8) at (3,-17.5) {$\begin{ytableau} \none & \none & \none & \none & 1  \\ \none & \none & \none & \none & \none[\scriptstyle \otimes]  \\ 1 & \none[ \scriptstyle \otimes] & 2  & \none[ \scriptstyle \otimes] & 1 \\ \none & \none & \none & \none & \none[\scriptstyle \otimes]  \\ \none & \none & \none & \none & 1 \end{ytableau}$};
	\node (T9) at (3,-20) {$\begin{ytableau} \none & \none & \none & \none & 1  \\ \none & \none & \none & \none & \none[\scriptstyle \otimes]  \\ 1 & \none[ \scriptstyle \otimes] & 1  & \none[ \scriptstyle \otimes] & 1 \\ \none & \none & \none & \none & \none[\scriptstyle \otimes]  \\ \none & \none & \none & \none & 1 \end{ytableau}$};

	\path[->] (T1) edge node[left] {$\scriptstyle A_1$} (T2a);
	\path[->] (T1) edge node[right] {$\scriptstyle A_1$} (T2b);
	\path[->] (T2a) edge node[left] {$\scriptstyle A_1$} (T3);
	\path[->] (T2b) edge node[right] {$\scriptstyle A_1$} (T3);
	\path[->] (T3) edge node[right] {$\scriptstyle A_1$} (T4);
	\path[->] (T4) edge node[right] {$\scriptstyle d_3 $} (T5);
	\path[->] (T5) edge node[right] {$\scriptstyle A_1$} (T6);
	\path[->] (T6) edge node[right] {$\scriptstyle a_3 $} (T7);
	\path[->] (T7) edge node[right] {$\scriptstyle d_3 $} (T8);
	\path[->] (T8) edge node[right] {$\scriptstyle d_5 $} (T9);

\end{tikzpicture}
\label{eq:exD5}
\end{equation}
Note again the distinction between the $d_3$ transition, involving the spinor nodes, and the $a_3$ transition, involving the linear part.

\subsection{Non-simply laced quivers}
\label{sec:BCquivercrystal}
On the field theory side, non-simply laced quivers are elusive from the Lagrangian point of view. However, one can always associate a Coulomb branch $\mC [\mathsf{Q} ]$ to a quiver $\mathsf{Q}$ shaped as a $B_r$ or $C_r$ Dynkin diagram. These Coulomb branches are constructed leveraging the natural inclusions $B_r \hookrightarrow D_{r+1}$ and $C_r \hookrightarrow A_{2r-1}$ \cite{Bumpbook}. Each such map induces the inclusion $\mathsf{Q} \hookrightarrow \widetilde{\mathsf{Q}}$ into a larger, simply-laced quiver. Then, folding $\widetilde{\mathsf{Q}}$ to produce $\mathsf{Q}$ induces a folding operation $\mC [ \widetilde{\mathsf{Q}} ] \to \mC \left[ \mathsf{Q} \right] $ \cite{Dey:2016qqp,Bourget:2020bxh}. Two concrete examples are:
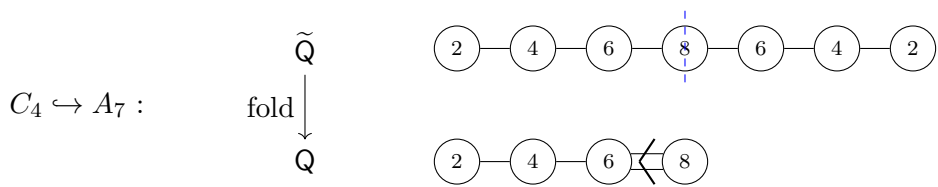
\begin{equation*}
	\begin{tikzpicture}[{square/.style={regular polygon,regular polygon sides=4}}]
		\node[] (CA) at (-11,-.75) {$C_4 \hookrightarrow A_7$ :};
		\node[circle,draw] (2l) at (-6,0) {$\scriptstyle 2 $};
		\node[circle,draw] (4l) at (-5,0) {$\scriptstyle 4 $};
		\node[circle,draw] (6l) at (-4,0) {$\scriptstyle 6 $};
		\node[circle,draw] (8) at (-3,0) {$\scriptstyle 8 $};
		\node[circle,draw] (2r) at (0,0) {$\scriptstyle 2 $};
		\node[circle,draw] (4r) at (-1,0) {$\scriptstyle 4 $};
		\node[circle,draw] (6r) at (-2,0) {$\scriptstyle 6 $};
		\draw (2l)--(4l);
		\draw (4l)--(6l);
		\draw (6l)--(8);
		\draw (8)--(6r);
		\draw (6r)--(4r);
		\draw (4r)--(2r);
		
		\draw[dashed,blue]  (-3,0.5)--(-3,-0.5);
		
		\node[circle,draw] (2f) at (-6,-1.5) {$\scriptstyle 2 $};
		\node[circle,draw] (4f) at (-5,-1.5) {$\scriptstyle 4 $};
		\node[circle,draw] (6f) at (-4,-1.5) {$\scriptstyle 6 $};
		\node[circle,draw] (8f) at (-3,-1.5) {$\scriptstyle 8 $};
		\draw (2f)--(4f);
		\draw (4f)--(6f);
		\path[-] (6f.20) edge (8f.160);
		\path[-] (6f.340) edge (8f.200);
		\draw[thick] (-3.6,-1.5)--(-3.4,-1.2);
		\draw[thick] (-3.6,-1.5)--(-3.4,-1.8);
		
		\node (wQ) at (-8,0) {$\widetilde{\mathsf{Q}}$};
		\node (fQ) at (-8,-1.5) {$\mathsf{Q}$};
		\path[->] (wQ) edge node[left] {fold} (fQ);

	\end{tikzpicture}
\end{equation*}
and 		
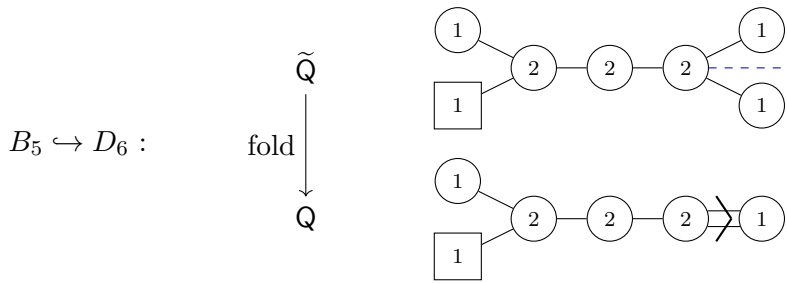
\begin{equation*}
	\begin{tikzpicture}[{square/.style={regular polygon,regular polygon sides=4}}]
			
		\node[] (BD) at (-1,-1) {$B_5 \hookrightarrow D_6$ :};
		\node[circle,draw] (2b) at (4,0.5) {$\scriptstyle 1 $};
		\node[square,draw] (2ex) at (4,-0.5) {$\scriptstyle 1 $};
		\node[circle,draw] (4b) at (5,0) {$\scriptstyle 2 $};
		\node[circle,draw] (6b) at (6,0) {$\scriptstyle 2 $};
		\node[circle,draw] (8b) at (7,0) {$\scriptstyle 2 $};
		\node[circle,draw] (4s) at (8,0.5) {$\scriptstyle 1 $};
		\node[circle,draw] (4c) at (8,-0.5) {$\scriptstyle 1 $};
		\draw (2b)--(4b);
		\draw (2ex)--(4b);
		\draw (4b)--(6b);
		\draw (6b)--(8b);
		\draw (8b)--(4s);
		\draw (8b)--(4c);
		
		\draw[dashed,blue]  (7.3,0)--(8.3,0);
		
		\node[circle,draw] (2g) at (4,-1.5) {$\scriptstyle 1 $};
		\node[square,draw] (2gx) at (4,-2.5) {$\scriptstyle 1 $};
		\node[circle,draw] (4g) at (5,-2) {$\scriptstyle 2 $};
		\node[circle,draw] (6g) at (6,-2) {$\scriptstyle 2 $};
		\node[circle,draw] (8g) at (7,-2) {$\scriptstyle 2 $};
		\node[circle,draw] (4p) at (8,-2) {$\scriptstyle 1 $};
		\draw (2g)--(4g);
		\draw (2gx)--(4g);
		\draw (4g)--(6g);
		\draw (6g)--(8g);
		\path[-] (8g.20) edge (4p.160);
		\path[-] (8g.340) edge (4p.200);
		\draw[thick] (7.4,-1.7)--(7.6,-2);
		\draw[thick] (7.4,-2.3)--(7.6,-2);
		
		\node (wQ) at (2,0) {$\widetilde{\mathsf{Q}}$};
		\node (fQ) at (2,-2) {$\mathsf{Q}$};
		\path[->] (wQ) edge node[left] {fold} (fQ);
		
	\end{tikzpicture}
\end{equation*}\par
Notably, this construction is precisely the one to produce crystals for non-simply laced Lie algebras ${}^L \mathfrak{g}$: starting from an \emph{ambient} crystal $\widetilde{\fC}$, associated to the larger, simply-connected Lie algebra, one obtains a sub-crystal $\fC$ for ${}^L \mathfrak{g}$.\par
\begin{mainthm}\label{thm:nonsimply}
	The phase diagram $\mathscr{P} \left( \mC [\mathsf{Q}] \right)$ of the Coulomb branch $\mC [\mathsf{Q}]$, constructed via quiver folding $\widetilde{\mathsf{Q}} \to \mathsf{Q}$, is congruent to the crystal $\fC [\mathsf{Q}]$ obtained folding the crystal $\fC [ \widetilde{\mathsf{Q}} ]$.
\end{mainthm}
In particular, we observe that the construction via folding and that in Algorithm \ref{algorithm1} produce the same output. This yields an independent check of the validity of the quiver folding procedure to construct $\mC[\mathsf{Q}]$, since the proof of the agreement on the crystal side is combinatorial and independent of the gauge theory interpretation.\par
Folding crystals of type AD will automatically produce  $b_p,c_p$ transitions of tableaux, that are folded versions of the $d_{p+1}, a_{2p-1}$ transitions. We proceed to show the validity of Result \ref{thm:nonsimply} in selected examples.

\subsubsection{$C_2$ quiver with two flavours}
To present the folding procedure \cite{Bourget:2020bxh} at the level of crystals, we start with the simple $C_2$ quiver $\overset{2}{\Box} - \overset{2}{\circ} \Leftarrow \overset{2}{\circ}$, obtained folding the balanced $A_3$ quiver $\overset{2}{\Box} - \overset{2}{\circ} - \overset{2}{\circ} - \overset{2}{\circ} - \overset{2}{\Box}$. The corresponding crystals are:
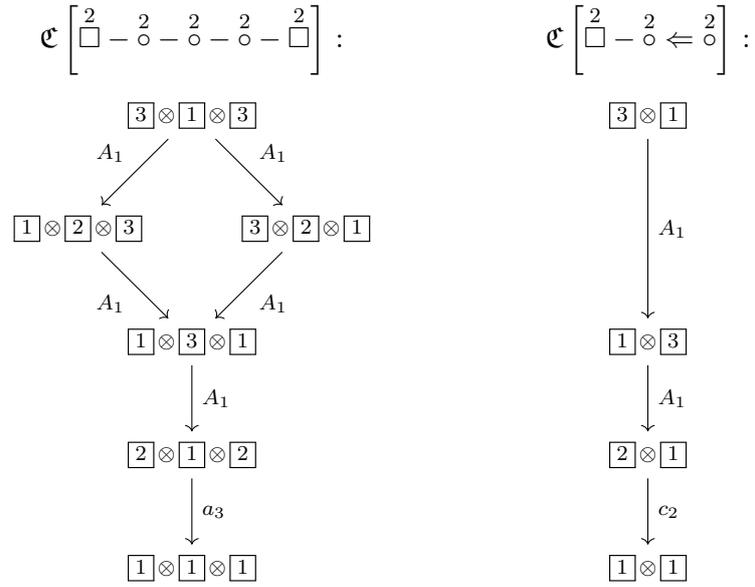
\begin{equation*}
\ytableausetup{smalltableaux} 
	\begin{tikzpicture}
	\node (wQ) at (-3,1) {$\fC \left[ \overset{2}{\Box} - \overset{2}{\circ} - \overset{2}{\circ} - \overset{2}{\circ} - \overset{2}{\Box} \right] $ :};
	\node (fQ) at (3,1) {$\fC \left[  \overset{2}{\Box} - \overset{2}{\circ} \Leftarrow \overset{2}{\circ} \right] $ :};
	
	\node (t1) at (-3,0) {$\begin{ytableau} 3 & \none[\scriptstyle \otimes] & 1 & \none[\scriptstyle \otimes] & 3 \end{ytableau}$};
	\node (t2a) at (-4.5,-1.5) {$\begin{ytableau} 1 & \none[\scriptstyle \otimes] & 2 & \none[\scriptstyle \otimes] & 3 \end{ytableau}$};
	\node (t2b) at (-1.5,-1.5) {$\begin{ytableau} 3 & \none[\scriptstyle \otimes] & 2 & \none[\scriptstyle \otimes] & 1 \end{ytableau}$};
	\node (t3) at (-3,-3) {$\begin{ytableau} 1 & \none[\scriptstyle \otimes] & 3 & \none[\scriptstyle \otimes] & 1 \end{ytableau}$};
	\node (t4) at (-3,-4.5) {$\begin{ytableau} 2 & \none[\scriptstyle \otimes] & 1 & \none[\scriptstyle \otimes] & 2 \end{ytableau}$};
	\node (t5) at (-3,-6) {$\begin{ytableau} 1 & \none[\scriptstyle \otimes] & 1 & \none[\scriptstyle \otimes] & 1 \end{ytableau}$};
	
	\path[->] (t1) edge node[left,anchor=south east] {$\scriptstyle A_1$} (t2a);
	\path[->] (t1) edge node[right,anchor=south west] {$\scriptstyle A_1$} (t2b);
	\path[->] (t2a) edge node[left,anchor=north east] {$\scriptstyle A_1$} (t3);
	\path[->] (t2b) edge node[left,anchor=north west] {$\scriptstyle A_1$} (t3);
	\path[->] (t3) edge node[right] {$\scriptstyle A_1$} (t4);
	\path[->] (t4) edge node[right] {$\scriptstyle a_3$} (t5);
	
	\node (f1) at (3,0) {$\begin{ytableau} 3 & \none[\scriptstyle \otimes] & 1  \end{ytableau}$};
	\node (f3) at (3,-3) {$\begin{ytableau} 1 & \none[\scriptstyle \otimes] & 3 \end{ytableau}$};
	\node (f4) at (3,-4.5) {$\begin{ytableau} 2 & \none[\scriptstyle \otimes] & 1  \end{ytableau}$};
	\node (f5) at (3,-6) {$\begin{ytableau} 1 & \none[\scriptstyle \otimes] & 1  \end{ytableau}$};
	
	\path[->] (f1) edge node[right] {$\scriptstyle A_1$} (f3);
	\path[->] (f3) edge node[right] {$\scriptstyle A_1$} (f4);
	\path[->] (f4) edge node[right] {$\scriptstyle c_2$} (f5);
	
	\end{tikzpicture}
\end{equation*}
Despite its apparent simplicity, this $C_2$ quiver already shows a subtle aspect. As explained in \cite{Bourget:2021siw}, transitions among strata of $\mC$ arise from fine-tuning certain K\"ahler moduli. In this case, there is a modulus, or degree of freedom, left from the next-to-last tableau, which allows for one last $c_2$ transition, even though there is no Higgs branch direction opening in this case. While not immediately obvious from a gauge theoretical perspective, this effect is automatically accounted for in our picture using Kashiwara crystals.

\subsubsection{$C_2$ quiver with six flavours}
The next example is again a $C_2$ quiver, this time unbalanced: 
\begin{equation*}
\begin{tikzpicture}
	\node (wQ) at (-2.25,0) {$\widetilde{\mathsf{Q}}$ :};
	\node (c1) at (-1.5,0) {$\circ$};
	\node (c2) at (-1,0) {$\circ$};
	\node (c3) at (-0.5,0) {$\circ$};
	\node (f1) at (-1,0.5) {$\Box$};
	\node (l1) at (-1.5,-0.2) {$\scriptstyle 1$};
	\node (l2) at (-1,-0.2) {$\scriptstyle 4$};
	\node (l3) at (-0.5,-0.2) {$\scriptstyle 1$};
	\node (lf) at (-0.8,0.5) {$\scriptstyle 6$};
	
	\node (fQ) at (2.75,0) {$\mathsf{Q}$ :};
	\node (a1) at (3.5,0) {$\circ$};
	\node (a2) at (4,0) {$\circ$};
	\node (aF) at (4,0.5) {$\Box$};
	\node (b1) at (3.5,-0.2) {$\scriptstyle 1$};
	\node (b2) at (4,-0.2) {$\scriptstyle 4$};
	\node (bF) at (4.2,0.5) {$\scriptstyle 6$};
	
	\node (varr2) at (4,0.25) {$\scriptstyle\vert $};
	\node (varr1) at (-1,0.25) {$\scriptstyle\vert $};
	\node (harr2) at (-0.75,0) {$-$};
	\node (harr1) at (-1.25,0) {$- $};
	
	\node (arrow) at (3.75,0) {$\Leftarrow$};
	\path[->] (0,0) edge node[above] {fold} (2,0);
	
\end{tikzpicture}
\end{equation*}
The corresponding crystal is:
\begin{equation*}
\ytableausetup{smalltableaux} 
	\begin{tikzpicture}
	\node (wQ) at (-3,1) {$\fC \left[  \widetilde{\mathsf{Q}} \right] $ :};
	\node (fQ) at (3,1) {$\fC \left[  \mathsf{Q} \right] $ :};
	
	\node (t1) at (-3,0) {$\begin{ytableau} 1 & \none[\scriptstyle \otimes] & 7 & \none[\scriptstyle \otimes] & 1 \end{ytableau}$};
	\node (t2) at (-3,-1.5) {$\begin{ytableau} 2 & \none[\scriptstyle \otimes] & 5 & \none[\scriptstyle \otimes] & 2 \end{ytableau}$};
	\node (t3) at (-3,-3) {$\begin{ytableau} 3 & \none[\scriptstyle \otimes] & 3 & \none[\scriptstyle \otimes] & 3 \end{ytableau}$};
	\node (t4a) at (-5.5,-4.5) {$\begin{ytableau} 1 & \none[\scriptstyle \otimes] & 4 & \none[\scriptstyle \otimes] & 3 \end{ytableau}$};
	\node (t4b) at (-3,-4.5) {$\begin{ytableau} 4 & \none[\scriptstyle \otimes] & 1 & \none[\scriptstyle \otimes] & 4 \end{ytableau}$};
	\node (t4c) at (-0.5,-4.5) {$\begin{ytableau} 3 & \none[\scriptstyle \otimes] & 4 & \none[\scriptstyle \otimes] & 1 \end{ytableau}$};
	\node (t5a) at (-5.5,-6) {$\begin{ytableau} 2 & \none[\scriptstyle \otimes] & 2 & \none[\scriptstyle \otimes] & 4 \end{ytableau}$};
	\node (t5b) at (-3,-6) {$\begin{ytableau} 1 & \none[\scriptstyle \otimes] & 5 & \none[\scriptstyle \otimes] & 1 \end{ytableau}$};
	\node (t5c) at (-0.5,-6) {$\begin{ytableau} 4 & \none[\scriptstyle \otimes] & 2 & \none[\scriptstyle \otimes] & 2 \end{ytableau}$};
	\node (t6) at (-3,-7.5) {$\begin{ytableau} 2 & \none[\scriptstyle \otimes] & 3 & \none[\scriptstyle \otimes] & 2 \end{ytableau}$};
	\node (t7) at (-3,-9) {$\begin{ytableau} 3 & \none[\scriptstyle \otimes] & 1 & \none[\scriptstyle \otimes] & 3 \end{ytableau}$};
	
	\path[->] (t1) edge node[right] {$\scriptstyle A_5$} (t2);
	\path[->] (t2) edge node[right] {$\scriptstyle A_3$} (t3);
	\path[->] (t3) edge node[left, anchor=south east] {$\scriptstyle A_1$} (t4a);
	\path[->] (t3) edge node[right, anchor=south west] {$\scriptstyle A_1$} (t4b);
	\path[->] (t3) edge node[right, anchor=south west] {$\scriptstyle A_1$} (t4c);
	\path[->] (t4a) edge node[left] {$\scriptstyle A_2$} (t5a);
	\path[->,gray,dashed] (t4a) edge node[right,pos=0.3,anchor=south west] {$\scriptstyle A_1$} (t5b);
	\path[->] (t4c) edge node[right] {$\scriptstyle A_2$}   (t5c);
	\path[->,gray,dashed] (t4c) edge node[left,pos=0.3,anchor=south east] {$\scriptstyle A_1$} (t5b);
	\path[->] (t4b) edge node[left,pos=0.8,anchor=south east] {$\scriptstyle A_2$} (t5a);
	\path[->] (t4b) edge node[right,pos=0.8,anchor=south west] {$\scriptstyle A_2$} (t5c);
	\path[->] (t5a) edge node[left,anchor=north east] {$\scriptstyle A_2$} (t6);
	\path[->] (t5b) edge node[right] {$\scriptstyle A_2$} (t6);
	\path[->] (t5c) edge node[right,anchor=north west] {$\scriptstyle A_2$} (t6);
	\path[->] (t6) edge node[right] {$\scriptstyle A_1$} (t7);

	\node (f1) at (3,0) {$\begin{ytableau} 1 & \none[\scriptstyle \otimes] & 7  \end{ytableau}$};
	\node (f2) at (3,-1.5) {$\begin{ytableau} 2 & \none[\scriptstyle \otimes] & 5 \end{ytableau}$};
	\node (f3) at (3,-3) {$\begin{ytableau} 3 & \none[\scriptstyle \otimes] & 3 \end{ytableau}$};
	\node (f4a) at (1.5,-4.5) {$\begin{ytableau} 4 & \none[\scriptstyle \otimes] & 2  \end{ytableau}$};
	\node (f4b) at (4.5,-4.5) {$\begin{ytableau} 1 & \none[\scriptstyle \otimes] & 5  \end{ytableau}$};
	\node (f5) at (3,-7.5) {$\begin{ytableau} 2 & \none[\scriptstyle \otimes] & 3  \end{ytableau}$};
	\node (f6) at (3,-9) {$\begin{ytableau} 3 & \none[\scriptstyle \otimes] & 1  \end{ytableau}$};
	
	\path[->] (f1) edge (f2);
	\path[->] (f2) edge (f3);
	\path[->] (f3) edge (f4a);
	\path[->] (f3) edge (f4b);
	\path[->] (f4a) edge (f5);
	\path[->] (f4b) edge (f5);
	\path[->] (f5) edge (f6);
	\end{tikzpicture}
\end{equation*}
The crystal may be continued below in the same vein to get $\fC \left[ \overset{3}{\circ} \Leftarrow \overset{6}{\circ} - \overset{6}{\Box} \right]$.

\subsubsection*{Balanced $B_2$ quiver with four flavours}
The $B_2$ Dynkin diagram is the simplest of the B series and can be obtained folding $D_3 \cong A_3$. Let us present the example of the $B_2$ quiver with gauge group $\U{3} \times \U{2}$, with two flavours attached at the $\U{3}$ node and one flavour attached at the $\U{2}$ node.\par
The quiver folding is 
\begin{equation*}
\begin{tikzpicture}[{square/.style={regular polygon,regular polygon sides=4}}]
	\node (q12) at (0,0) {$\circ$};
	\node (q13) at (0,0.5) {$\circ$};
	\node (q14) at (0,-0.5) {$\circ$};
	\node (q16) at (0.5,-0.5) {$\Box $};
	\node (q17) at (0.5,0.5) {$\Box $};
	\node (q17) at (0.5,0) {$\Box $};
	\node (a11) at (0.25,-0.5) {$-$};
	\node (a12) at (0.25,0.5) {$-$};
	\node (a13) at (0.25,0) {$-$};
	\node (v13) at (0,0.25) {$\scriptstyle \vert $};
	\node (v14) at (0,-0.25) {$\scriptstyle \vert $};
	
	\node (lab11) at (-0.2,0) {$\scriptstyle 3$};
	\node (lab12) at (-0.2,0.5) {$\scriptstyle 2$};
	\node (lab13) at (-0.2,-0.5) {$\scriptstyle 2$};
	\node (lab14) at (0.7,0) {$\scriptstyle 2$};
	\node (lab16) at (0.7,0.5) {$\scriptstyle 1 $};
	\node (lab17) at (0.7,-0.5) {$\scriptstyle 1 $};
	
	\node (arrow) at (2,0) {$\longrightarrow$};
	
	\node (fo) at (4,0) {$\underset{2}{\Box} - \underset{3}{\circ} \Rightarrow \underset{2}{\circ} - \underset{1}{\Box}$};
\end{tikzpicture}	
\end{equation*}
The ambient $D_3$ crystal, its folding into $B_2$ and the quiver subtraction pattern are:
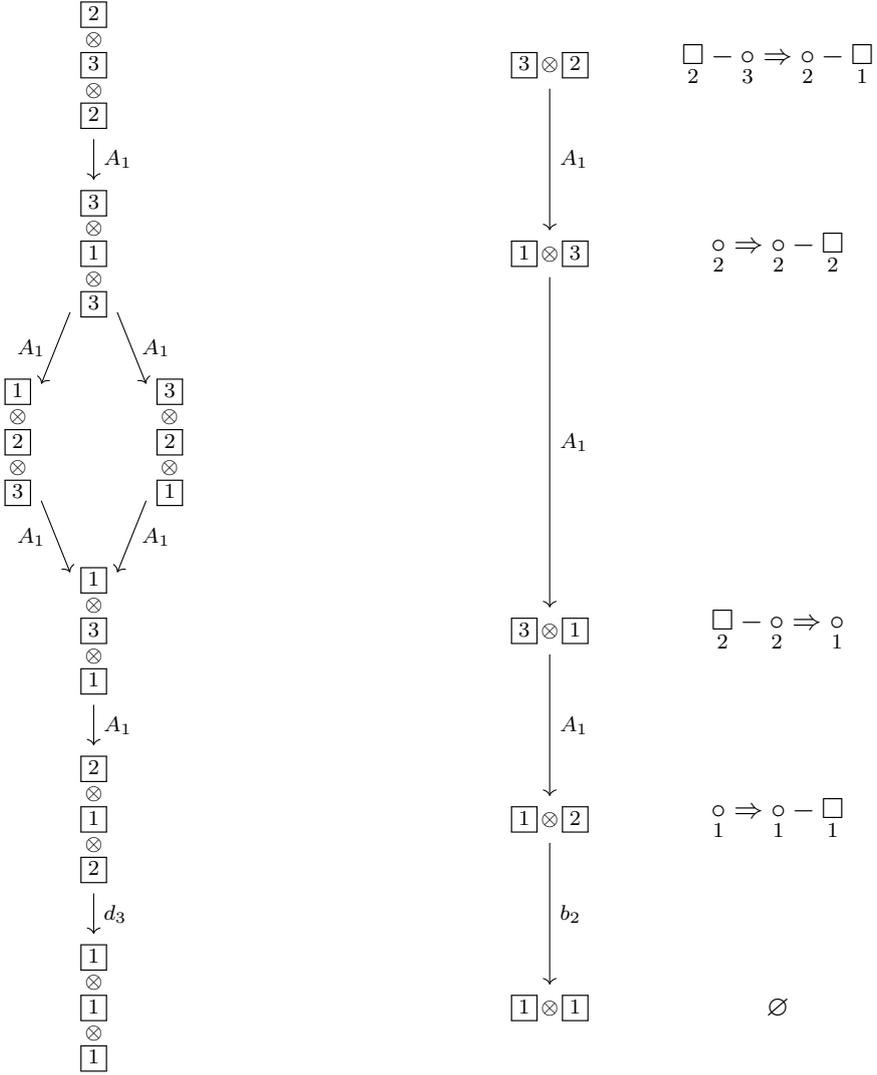
\begin{equation*}
	\ytableausetup{smalltableaux} 
	\begin{tikzpicture}
	
	\node (t1) at (-3,0) {$\begin{ytableau} 2 \\ \none[\scriptstyle \otimes] \\ 3 \\ \none[\scriptstyle \otimes] \\ 2 \end{ytableau}$};
	\node (t2) at (-3,-2.5) {$\begin{ytableau} 3 \\ \none[\scriptstyle \otimes] \\ 1 \\ \none[\scriptstyle \otimes] \\ 3 \end{ytableau}$};
	\node (t3a) at (-4,-5) {$\begin{ytableau} 1 \\ \none[\scriptstyle \otimes] \\ 2 \\ \none[\scriptstyle \otimes] \\ 3 \end{ytableau}$};
	\node (t3b) at (-2,-5) {$\begin{ytableau} 3 \\ \none[\scriptstyle \otimes] \\ 2 \\ \none[\scriptstyle \otimes] \\ 1 \end{ytableau}$};
	\node (t4) at (-3,-7.5) {$\begin{ytableau} 1 \\ \none[\scriptstyle \otimes] \\ 3 \\ \none[\scriptstyle \otimes] \\ 1 \end{ytableau}$};
	\node (t5) at (-3,-10) {$\begin{ytableau} 2 \\ \none[\scriptstyle \otimes] \\ 1 \\ \none[\scriptstyle \otimes] \\ 2 \end{ytableau}$};
	\node (t6) at (-3,-12.5) {$\begin{ytableau} 1 \\ \none[\scriptstyle \otimes] \\ 1 \\ \none[\scriptstyle \otimes] \\ 1 \end{ytableau}$};
	
	\path[->] (t1) edge node[right] {$\scriptstyle A_1$} (t2);
	\path[->] (t2) edge node[left] {$\scriptstyle A_1$} (t3a);
	\path[->] (t2) edge node[right] {$\scriptstyle A_1$} (t3b);
	\path[->] (t3a) edge node[left] {$\scriptstyle A_1$} (t4);
	\path[->] (t3b) edge node[right] {$\scriptstyle A_1$} (t4);
	\path[->] (t4) edge node[right] {$\scriptstyle A_1$} (t5);
	\path[->] (t5) edge node[right] {$\scriptstyle d_3 $} (t6);

	\node (f1) at (3,0) {$\begin{ytableau} 3 & \none[\scriptstyle \otimes] & 2  \end{ytableau}$};
	\node (f2) at (3,-2.5) {$\begin{ytableau} 1 & \none[\scriptstyle \otimes] & 3  \end{ytableau}$};
	\node (f4) at (3,-7.5) {$\begin{ytableau} 3 & \none[\scriptstyle \otimes] & 1 \end{ytableau}$};
	\node (f5) at (3,-10) {$\begin{ytableau} 1 & \none[\scriptstyle \otimes] & 2  \end{ytableau}$};
	\node (f6) at (3,-12.5) {$\begin{ytableau} 1 & \none[\scriptstyle \otimes] & 1  \end{ytableau}$};
	
	\path[->] (f1) edge node[right] {$\scriptstyle A_1$} (f2);
	\path[->] (f2) edge node[right] {$\scriptstyle A_1$} (f4);
	\path[->] (f4) edge node[right] {$\scriptstyle A_1$} (f5);
	\path[->] (f5) edge node[right] {$\scriptstyle b_2 $} (f6);

	\node at (6,0) {$\underset{2}{\Box} - \underset{3}{\circ} \Rightarrow \underset{2}{\circ} - \underset{1}{\Box}$};
	\node at (6,-2.5) {$ \underset{2}{\circ} \Rightarrow \underset{2}{\circ} - \underset{2}{\Box}$};
	\node at (6,-7.5) {$\underset{2}{\Box} - \underset{2}{\circ} \Rightarrow \underset{1}{\circ} $};
	\node at (6,-10) {$ \underset{1}{\circ} \Rightarrow \underset{1}{\circ} - \underset{1}{\Box}$};
	\node at (6,-12.5) {$ \varnothing $};

	\end{tikzpicture}
\end{equation*}

\subsubsection{Balanced $B_4$ quiver with two flavours}
The next example of type B is the balanced $B_4$ quiver $\overset{2}{\circ}-\overset{4}{\circ}-\overset{6}{\circ} \Rightarrow \overset{4}{\circ}- \overset{2}{\Box}$. The corresponding crystal is obtained folding the $D_5$ crystal in \eqref{eq:exD5}:
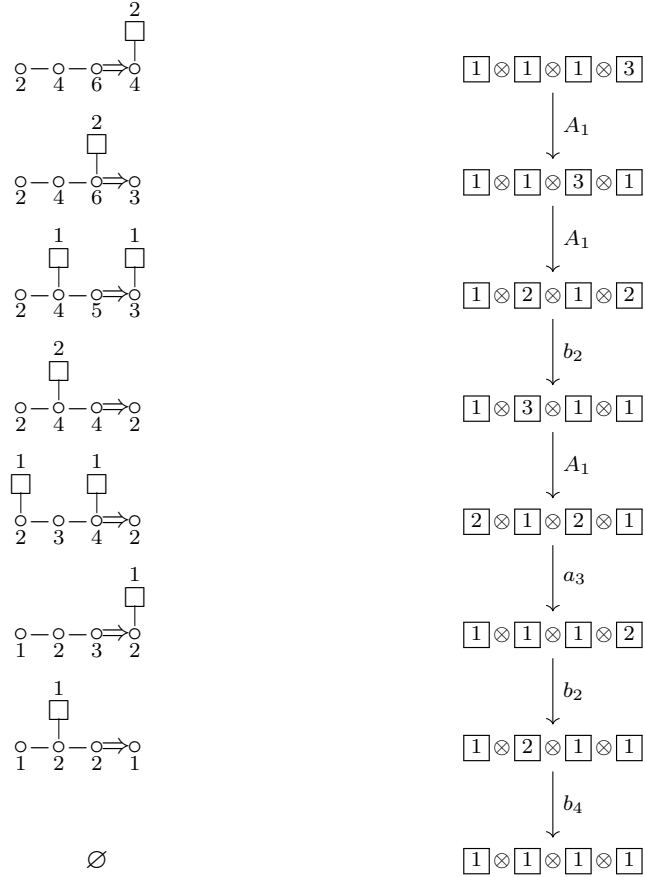
\begin{equation*}
\ytableausetup{smalltableaux} 
\begin{tikzpicture}[{square/.style={regular polygon,regular polygon sides=4}}]

	\node (q11) at (-4,0) {$\circ$};
	\node (q12) at (-3.5,0) {$\circ$};
	\node (q13) at (-3,0) {$\circ$};
	\node (q14) at (-2.5,0) {$\circ$};
	\node (q16) at (-2.5,0.5) {$\Box $};
	\node (a11) at (-3.25,0) {$-$};
	\node (a12) at (-2.75,0) {$\Rightarrow $};
	\node (a13) at (-3.75,0) {$-$};
	\node (v13) at (-2.5,0.25) {$\scriptstyle \vert $};
	
	\node (lab11) at (-4,-0.2) {$\scriptstyle 2$};
	\node (lab12) at (-3.5,-0.2) {$\scriptstyle 4$};
	\node (lab13) at (-3,-0.2) {$\scriptstyle 6$};
	\node (lab14) at (-2.5,-0.2) {$\scriptstyle 4$};
	\node (lab17) at (-2.5,0.8) {$\scriptstyle 2 $};	
	
	\node (q31) at (-4,-1.5) {$\circ$};
	\node (q32) at (-3.5,-1.5) {$\circ$};
	\node (q33) at (-3,-1.5) {$\circ$};
	\node (q34) at (-2.5,-1.5) {$\circ$};
	\node (q36) at (-3,-1) {$\Box $};
	\node (a31) at (-3.25,-1.5) {$-$};
	\node (a32) at (-2.75,-1.5) {$\Rightarrow $};
	\node (a33) at (-3.75,-1.5) {$-$};
	\node (v33) at (-3,-1.25) {$\scriptstyle \vert $};
	
	\node (lab31) at (-4,-1.7) {$\scriptstyle 2$};
	\node (lab32) at (-3.5,-1.7) {$\scriptstyle 4$};
	\node (lab33) at (-3,-1.7) {$\scriptstyle 6$};
	\node (lab34) at (-2.5,-1.7) {$\scriptstyle 3$};
	\node (lab37) at (-3,-0.7) {$\scriptstyle 2 $};

	\node (q41) at (-4,-3) {$\circ$};
	\node (q42) at (-3.5,-3) {$\circ$};
	\node (q43) at (-3,-3) {$\circ$};
	\node (q44) at (-2.5,-3) {$\circ$};
	\node (q45) at (-2.5,-2.5) {$\Box $};
	\node (q46) at (-3.5,-2.5) {$\Box $};
	\node (a41) at (-3.25,-3) {$-$};
	\node (a42) at (-2.75,-3) {$\Rightarrow $};
	\node (a43) at (-3.75,-3) {$-$};
	\node (v42) at (-3.5,-2.75) {$\scriptstyle \vert $};
	\node (v43) at (-2.5,-2.75) {$\scriptstyle \vert $};
	
	\node (lab41) at (-4,-3.2) {$\scriptstyle 2$};
	\node (lab42) at (-3.5,-3.2) {$\scriptstyle 4$};
	\node (lab43) at (-3,-3.2) {$\scriptstyle 5$};
	\node (lab44) at (-2.5,-3.2) {$\scriptstyle 3$};
	\node (lab45) at (-2.5,-2.2) {$\scriptstyle 1 $};
	\node (lab47) at (-3.5,-2.2) {$\scriptstyle 1 $};

	\node (q51) at (-4,-4.5) {$\circ$};
	\node (q52) at (-3.5,-4.5) {$\circ$};
	\node (q53) at (-3,-4.5) {$\circ$};
	\node (q54) at (-2.5,-4.5) {$\circ$};
	\node (q56) at (-3.5,-4) {$\Box $};
	\node (a51) at (-3.25,-4.5) {$-$};
	\node (a52) at (-2.75,-4.5) {$\Rightarrow $};
	\node (a53) at (-3.75,-4.5) {$-$};
	\node (v52) at (-3.5,-4.25) {$\scriptstyle \vert $};
	
	\node (lab51) at (-4,-4.7) {$\scriptstyle 2$};
	\node (lab52) at (-3.5,-4.7) {$\scriptstyle 4$};
	\node (lab53) at (-3,-4.7) {$\scriptstyle 4$};
	\node (lab54) at (-2.5,-4.7) {$\scriptstyle 2$};
	\node (lab57) at (-3.5,-3.7) {$\scriptstyle 2 $};

	\node (q61) at (-4,-6) {$\circ$};
	\node (q62) at (-3.5,-6) {$\circ$};
	\node (q63) at (-3,-6) {$\circ$};
	\node (q64) at (-2.5,-6) {$\circ$};
	\node (q65) at (-4,-5.5) {$\Box $};
	\node (q66) at (-3,-5.5) {$\Box $};
	\node (a61) at (-3.25,-6) {$-$};
	\node (a62) at (-2.75,-6) {$\Rightarrow $};
	\node (a63) at (-3.75,-6) {$-$};
	\node (v61) at (-4,-5.75) {$\scriptstyle \vert $};
	\node (v62) at (-3,-5.75) {$\scriptstyle \vert $};
	
	\node (lab61) at (-4,-6.2) {$\scriptstyle 2$};
	\node (lab62) at (-3.5,-6.2) {$\scriptstyle 3$};
	\node (lab63) at (-3,-6.2) {$\scriptstyle 4$};
	\node (lab64) at (-2.5,-6.2) {$\scriptstyle 2$};
	\node (lab67) at (-4,-5.2) {$\scriptstyle 1 $};
	\node (lab66) at (-3,-5.2) {$\scriptstyle 1 $};

	\node (q71) at (-4,-7.5) {$\circ$};
	\node (q72) at (-3.5,-7.5) {$\circ$};
	\node (q73) at (-3,-7.5) {$\circ$};
	\node (q74) at (-2.5,-7.5) {$\circ$};
	\node (q76) at (-2.5,-7) {$\Box $};
	\node (a71) at (-3.25,-7.5) {$-$};
	\node (a72) at (-2.75,-7.5) {$\Rightarrow $};
	\node (a73) at (-3.75,-7.5) {$-$};
	\node (v72) at (-2.5,-7.25) {$\scriptstyle \vert $};
	
	\node (lab71) at (-4,-7.7) {$\scriptstyle 1$};
	\node (lab72) at (-3.5,-7.7) {$\scriptstyle 2$};
	\node (lab73) at (-3,-7.7) {$\scriptstyle 3$};
	\node (lab74) at (-2.5,-7.7) {$\scriptstyle 2$};
	\node (lab76) at (-2.5,-6.7) {$\scriptstyle 1 $};

	\node (q81) at (-4,-9) {$\circ$};
	\node (q82) at (-3.5,-9) {$\circ$};
	\node (q83) at (-3,-9) {$\circ$};
	\node (q84) at (-2.5,-9) {$\circ$};
	\node (q86) at (-3.5,-8.5) {$\Box $};
	\node (a81) at (-3.25,-9) {$-$};
	\node (a82) at (-2.75,-9) {$\Rightarrow $};
	\node (a83) at (-3.75,-9) {$-$};
	\node (v82) at (-3.5,-8.75) {$\scriptstyle \vert $};
	
	\node (lab81) at (-4,-9.2) {$\scriptstyle 1$};
	\node (lab82) at (-3.5,-9.2) {$\scriptstyle 2$};
	\node (lab83) at (-3,-9.2) {$\scriptstyle 2$};
	\node (lab84) at (-2.5,-9.2) {$\scriptstyle 1$};
	\node (lab86) at (-3.5,-8.2) {$\scriptstyle 1 $};

	\node (q9) at (-3,-10.5) {$\varnothing $};

	\node (T1) at (3,0) {$\begin{ytableau}  1 & \none[ \scriptstyle \otimes] & 1  & \none[ \scriptstyle \otimes] & 1 & \none[ \scriptstyle \otimes] & 3 \end{ytableau}$};
	\node (T3) at (3,-1.5) {$\begin{ytableau}  1 & \none[ \scriptstyle \otimes] & 1  & \none[ \scriptstyle \otimes] & 3 & \none[ \scriptstyle \otimes] & 1 \end{ytableau}$};
	\node (T4) at (3,-3) {$\begin{ytableau} 1 & \none[ \scriptstyle \otimes] & 2  & \none[ \scriptstyle \otimes] & 1 & \none[ \scriptstyle \otimes] & 2 \end{ytableau}$};
	\node (T5) at (3,-4.5) {$\begin{ytableau} 1 & \none[ \scriptstyle \otimes] & 3  & \none[ \scriptstyle \otimes] & 1 & \none[ \scriptstyle \otimes] & 1 \end{ytableau}$};
	\node (T6) at (3,-6) {$\begin{ytableau} 2 & \none[ \scriptstyle \otimes] & 1  & \none[ \scriptstyle \otimes] & 2 & \none[ \scriptstyle \otimes] & 1 \end{ytableau}$};
	\node (T7) at (3,-7.5) {$\begin{ytableau} 1 & \none[ \scriptstyle \otimes] & 1  & \none[ \scriptstyle \otimes] & 1 & \none[ \scriptstyle \otimes] & 2 \end{ytableau}$};
	\node (T8) at (3,-9) {$\begin{ytableau} 1 & \none[ \scriptstyle \otimes] & 2  & \none[ \scriptstyle \otimes] & 1 & \none[ \scriptstyle \otimes] & 1 \end{ytableau}$};
	\node (T9) at (3,-10.5) {$\begin{ytableau} 1 & \none[ \scriptstyle \otimes] & 1  & \none[ \scriptstyle \otimes] & 1 & \none[ \scriptstyle \otimes] & 1 \end{ytableau}$};

	\path[->] (T1) edge node[right] {$\scriptstyle A_1$} (T3);
	\path[->] (T3) edge node[right] {$\scriptstyle A_1$} (T4);
	\path[->] (T4) edge node[right] {$\scriptstyle b_2 $} (T5);
	\path[->] (T5) edge node[right] {$\scriptstyle A_1$} (T6);
	\path[->] (T6) edge node[right] {$\scriptstyle a_3 $} (T7);
	\path[->] (T7) edge node[right] {$\scriptstyle b_2 $} (T8);
	\path[->] (T8) edge node[right] {$\scriptstyle b_4 $} (T9);

\end{tikzpicture}
\end{equation*}

\subsubsection{$G_2$ quivers}
We observe that any $\fC [ \mathsf{Q} ]$ with $\mathsf{Q}$ (i) shaped like a $D_4$ Dynkin diagram, and (ii) with assignment of gauge groups and matter content that preserves the triality symmetry of $D_4$, can be ``tri-folded'' into a crystal of exceptional type $G_2$.

\subsubsection{Two paths to Coulomb branches of non-simply laced quivers}
To sum up, we have provided and tested in various examples two alternative approaches to the construction of Kashiwara crystals $\fC \left[ \mathsf{Q} \right]$ that capture the Coulomb branches $\mC \left[ \mathsf{Q} \right]$ of non-simply laced quivers $\mathsf{Q}$, namely 
\begin{itemize}
\item applying Algorithm \ref{algorithm1} with type B and type C root systems, or
\item folding type D and type A crystals.
\end{itemize}
The existence of a twofold way to get $\fC [\mathsf{Q}]$ is not accidental. Crystals for any $\mathsf{Q}$ have been built in \cite{Krylov}, extending \cite{Braverman99}, starting from generalized slices of the affine Grassmannian. These latter objects, in turn, give $\mC [\mathsf{Q}]$. There exist two proofs of such statement for non-simply laced $\mathsf{Q}$: either via folding \cite{Braverman:2016pwk} or by direct construction \cite{Nakajima:2019olw}. The equivalence of the two proofs as well as the relation with the works \cite{Braverman99,Krylov} appeared in \cite{Nakajima:2019olw}. In the present section, we have closed the circle of ideas, showing the equivalence between folding of crystals and direct construction.\par
Remarkably, the algebro-geometric setup of \cite{Muthiah:2019jif,Nakajima:2019olw} allows to study non-simply laced quivers of more general shape. We leave the investigation of a crystal counterpart of such setup for future work.

\section{Crystals for affine quivers}
\label{sec:affinecrystalquiver}

This section is devoted to the analysis of Coulomb branches of quivers of affine type $\widehat{A}_r$. We label the nodes by $j \in \left\{ 0,1, \dots , r \right\}$, with periodic identification $r+1 \equiv 0 $.\par
Affine crystals have been studied initially for affine type A \cite{Kang92,Shimozono:1998} and later extended to other affine types. Including the coroot $\alpha_0 ^{\vee}$ in the construction of Section \ref{sec:C=C} we obtain a crystal $\fC [\mathsf{Q}]$ which is the amputation of a periodic crystal after $N_j$ Kashiwara operations have been performed on the $j^{\text{th}}$ node, $\forall j=0,1, \dots, r$. In the limit $N_j \to \infty$ we would be left with a periodic crystal.\par
A major difference from the finite $\mathfrak{g}$ case lies in the conservation of the sum of letters in each tableaux, that is, $ \sum_{j=0}^{r} \mathsf{w}_j$ is constant along the crystal.

\subsection{Crystals from affine quiver subtraction and from branes}
Consider an affine framed quiver $\mathsf{Q}$. Repeated application of the quiver subtraction algorithm on $\mathsf{Q}$ \cite{Cabrera:2018ann} produces a sequence of quivers of lower rank. Reasoning as in Section \ref{sec:quiversub}, we identify such quivers with tableaux in the crystal $\fC [\mathsf{Q}]$.\par
\medskip
Moreover, consider the Hanany--Witten setup of Section \ref{sec:crystalBrane}, but now with the direction $x^6$ compactified. That is, the D3 branes now fill the 3d spacetime times a circle. These brane configurations produce the desired affine type A quiver gauge theories. Then, precisely as in Section \ref{sec:crystalBrane}, tableaux in the crystal are in one-to-one correspondence with phases of the Hanany--Witten configuration, once the additional coroot $\alpha_0^{\vee}$ is taken into account to ``affinize'' the crystal.\par
If, in the initial Hanany--Witten configuration, we send the number $(N_1,N_2, \dots, N_r, N_0)$ of D3 branes in each interval to infinity, the brane phase diagram becomes periodic. This would give the full periodic crystal. For a finite number of D3 branes, however, the periodicity is broken after $N_j$ transitions involving the $j^{\text{th}}$ interval have been performed.

\subsection{Affine type A examples}

We now motivate our claim with selected examples. In drawing the Hanany--Witten configurations, D3 branes that wrap around the circle direction are dashed, that is, dashed D3 branes drawn on the left of the leftmost NS5 brane are joined with the dashed D3 branes on the right of the rightmost NS5 brane.\par
We stress that the input for Algorithm \ref{algorithm1} is a \emph{framed} quiver. For balanced, unframed, affine quivers we can only act on tableaux with a combination of simple roots of ${}^L G$ that leaves $\wt (T)$ invariant. For example, for the affine quiver of type A with $\mathbf{G}= \U{N}^{r+1}$ and no framing, we act $N$ times with the trivial combination $\sum_{j=0} ^{r} \alpha_j ^{\vee }$.

\subsubsection{$\widehat{A}_2$ quiver with two flavours}
Consider the following affine crystal of type $\widehat{A}_2$: 
\begin{equation*}
	\ytableausetup{smalltableaux} 

\end{equation*}
This would correspond to a quiver with framing $\un =(0,2,0)$. However, to get a physical Coulomb branch we must specify the ranks $\uN=(N_1, N_2, N_0)$. For instance, the quiver 
\begin{equation}
\label{eq:exA2QF2}
	%
\end{equation}
is associated with the crystal 
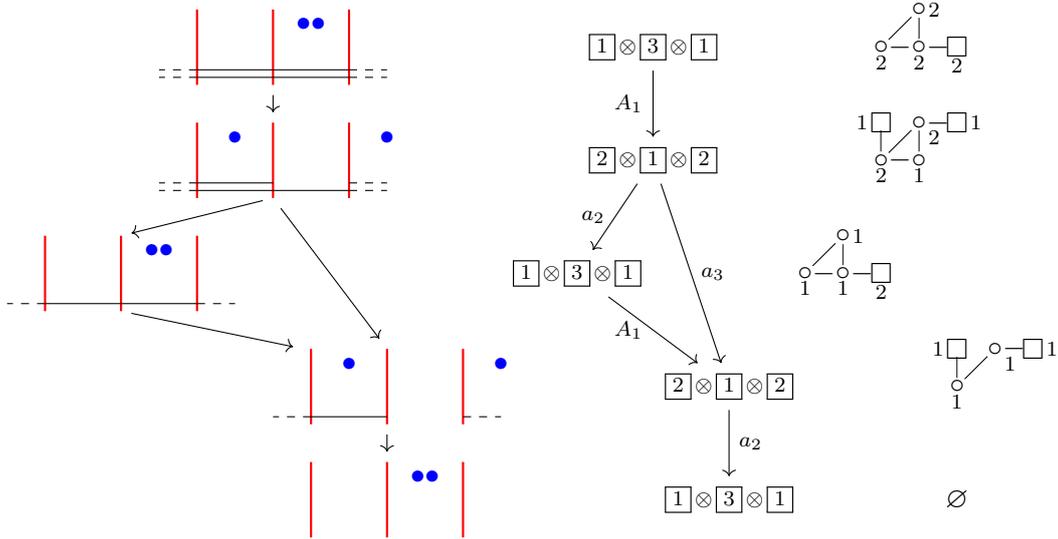
\begin{equation*}
	\ytableausetup{smalltableaux} 
	%
\end{equation*}
where we have represented $\fC [\mathsf{Q}]$ in the middle, the corresponding brane system on the left and the quiver subtraction pattern on the right.

\subsubsection{$\widehat{A}_2$ quiver with six flavours}

For the next example, we add flavours to the $\widehat{A}_2$ quiver \eqref{eq:exA2QF2} and consider 
\begin{equation}
	%
\end{equation}
The associated crystal $\fC[\mathsf{Q}]$ is
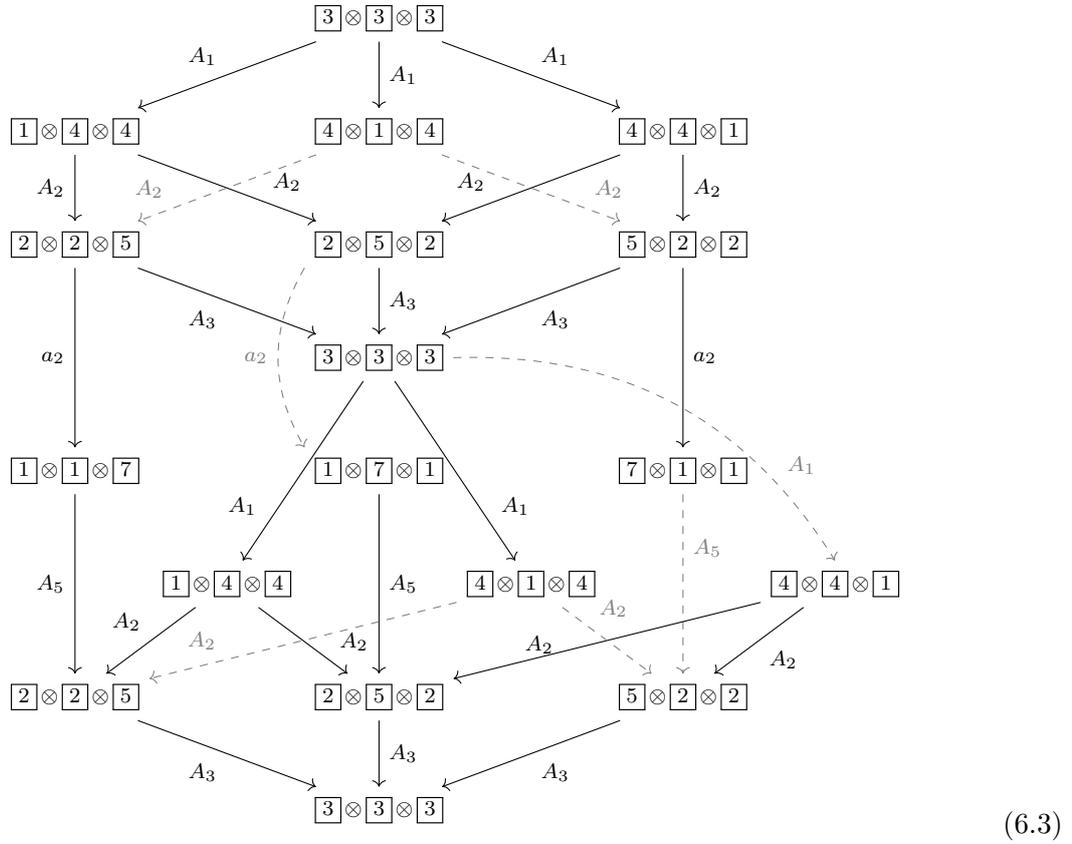
\begin{equation}
\label{eq:CrysA2Qf6}
\ytableausetup{smalltableaux} 
	%
\end{equation}
The triality symmetry of $\mathsf{Q}$ is manifestly inherited by $\fC[\mathsf{Q}]$. Moreover, we observe that, increasing the ranks of the gauge nodes, the crystal would repeat its structure, eventually becoming periodic in the inductive limit $(N_1,N_2,N_0) \to (\infty,\infty,\infty)$, producing an actual affine Kashiwara crystal.\par
The phases predicted by \eqref{eq:CrysA2Qf6} agree with the brane analysis:
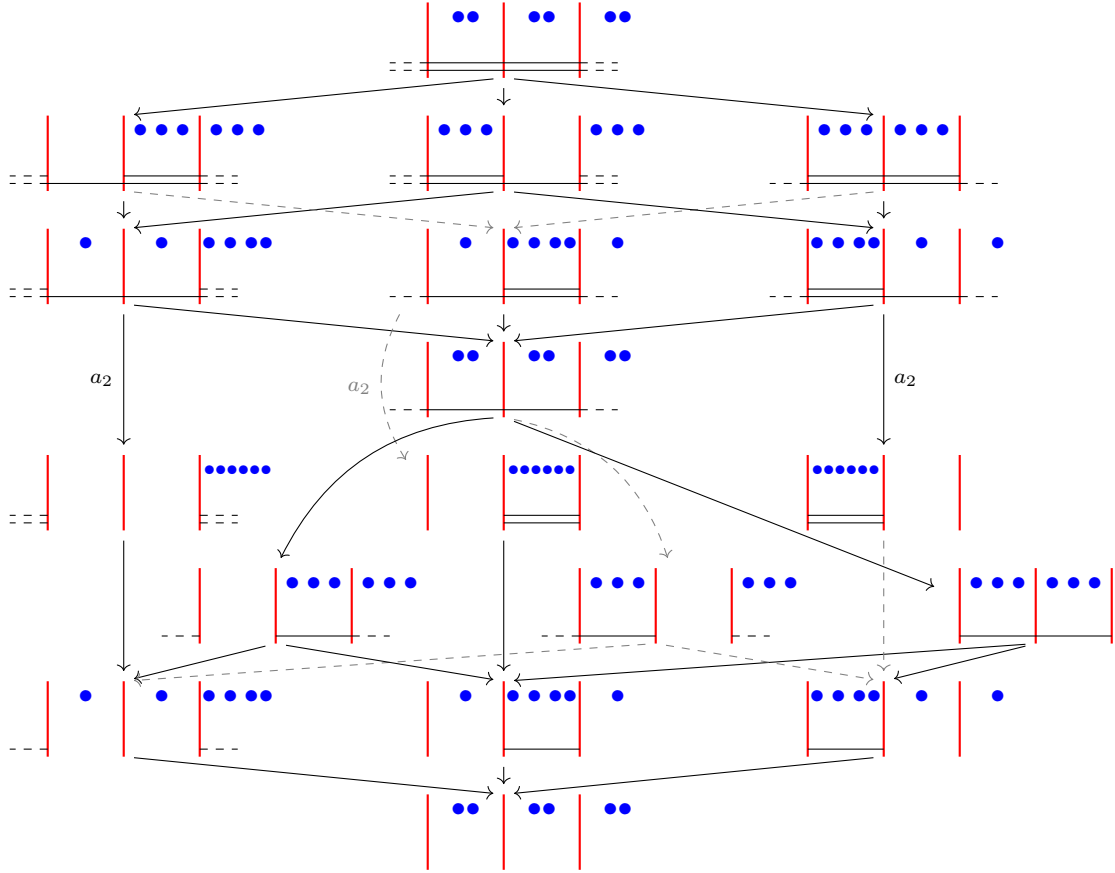
\begin{equation*}	
%
\end{equation*}

\section{Infinite crystals, Verma modules and Hilbert spaces}
\label{sec:Demazure}

In this section we depart from the theory of crystal bases exploited so far, and consider \emph{infinite} crystals $\fC_{\infty} ^{\Phi}$. On one hand, Verma modules admit a realization using infinite crystals \cite{Kashiwara:Demazure}, whilst, on the other hand, the action of monopole operators generates Verma modules of the Coulomb branch coordinate ring $\C [\mC]$ \cite{Bullimore:2016hdc}. The aim of the present section is to explicate this analogy.\par

\subsection{Infinite crystals}
\label{sec:definftycrys}
We now introduce infinite crystals $\fC_{\infty} ^{\Phi}$, following a combinatorial presentation \cite{BumpSchilling}, but we ought to emphasize that the seminal work \cite{Kashiwara:Demazure} introduced them using only the theory of quantum groups. Moreover, a geometric realization modelled after quiver varieties is given in \cite{Kashiwara:97}.\par
Fix a reductive group $G$ of rank $r$, with root system $\Phi$, and let $j \in \mathbb{N}$ run over the index set of $\Phi$. Consider the collection $\left\{ u_j (k) \right\}_{k \in \mathbb{Z}} \equiv \mathfrak{U}_j ^{\Phi} $ and declare that $\wt (u_j (k)) = (0, \dots, 0, k, -k, 0, \dots, 0)$ with $k$ in the $j^{\mathrm{th}}$ entry and $-k$ in the $(j+1)^{\text{th}}$ entry. Defining the action of the Kashiwara operators 
\begin{equation*}
	f_{j} u_{j^{\prime}} (k) = \delta_{j j^{\prime}} u_j (k-1) , \qquad e_{j} u_{j^{\prime}} (k) = \delta_{j j^{\prime}} u_j (k+1) ,
\end{equation*}
so that the weight changes by $\alpha_j \in \Phi$, endows $ \mathfrak{U}_j ^{\Phi}$ with a crystal structure.\par
The ultimate goal is to study moduli spaces of monopole operators in three dimensions. We first follow a textbook treatment of the infinite crystals, in which the gauge group imposes no constraint on the monopoles. Then we expand the discussion to include quivers with more general choices of gauge group.\par

\subsubsection{Infinite crystals: Freely generated Verma modules}
\label{subsec:TSUNVerma}
Let $w_0$ be the long element of the Weyl group of $G$ and let $(j_1, \dots, j_{\ell})$ be the reduced word from the reduced decomposition of $w_0$ \cite{Bumpbook}. For example, if $\Phi =A_{r}$, we take 
\begin{equation*}
	\ell= \frac{r(r+1)}{2} , \qquad (j_1, \dots, j_{\ell}) = (1,2, \dots, r, 1,2, \dots, r-1,  \dots, 3, 1,2 , 1) .
\end{equation*}
To lighten the notation, for any $\ell$-tuple $\underline{k} =(k_1, \dots, k_{\ell})$ let us define \cite{BumpSchilling} 
\begin{equation*}
	u^{\Phi} (\underline{k} ) \equiv u_{j_1} (-k_{1})  \otimes \cdots \otimes u_{j_\ell } (-k_{\ell } ) . 
\end{equation*}
The crystal $\fC_{\infty} ^{\Phi} $ is defined to be the subset 
\begin{equation*}
	\fC_{\infty}  ^{\Phi} \subset \mathfrak{U}_{j_1}  ^{\Phi} \otimes \cdots \otimes \mathfrak{U}_{j_\ell }  ^{\Phi}
\end{equation*}
obtained acting on the element $u ^{\Phi} (\underline{0})$ with Kashiwara operators $f_j$ in any order. Imposing 
\begin{equation*}
	e_j (u^{\Phi} (\underline{k}) ) = 0 \quad \text{ if } k_j =0 
\end{equation*}
yields the crystal structure on $\fC_{\infty} ^{\Phi}$. These crystals are crystal bases for $\mathfrak{g}$-Verma modules. For more details, we refer to the monograph \cite{BumpSchilling} (note that the infinite crystals are usually denoted $ B ( \infty )$ or $\mathcal{B}_{\infty}$ in the literature).\par

\subsubsection{Infinite crystals: Abelian quivers}
\label{subsec:AbelianVerma}
The construction that we have just outlined depends only on $\Phi$ and does not take into account the gauge group, thus works for freely generated Verma modules. Let us now consider Abelian quivers, with focus on $A_r$ for concreteness.\par
To enforce the Abelian group condition, the crystal $\fC_{\infty} ^{\U{1}^r}$ is built as above, but with $\ell =r$ and labels $(1, \dots, r)$. The infinite $A_1$ crystal $\fC_{\infty}^{\U{1}}$ is 
\begin{equation}
\label{eq:CinftySQED}
	{\scriptstyle u (0) } \to {\scriptstyle u (1) } \to {\scriptstyle u (2)  } \to {\scriptstyle u (3)}  \to \cdots  ,
\end{equation}
and the more general $A_r$ crystal $\fC_{\infty} ^{\U{1}^{r}} $ is 
\begin{equation}
\begin{tikzpicture}
	\node (u0c0) at (0,0) {${\scriptstyle u (0,0, \dots, 0) }$};
	\node (u1c0) at (-4,-1) {${\scriptstyle u (1,0, \dots, 0) }$};
	\node (u010) at (-2,-1) {${\scriptstyle u (0,1, \dots, 0) }$};
	\node (u0c1) at (4,-1) {${\scriptstyle u (0,0, \dots, 1) }$};
	\node (cd1) at (0,-1) {${\scriptstyle \cdots  }$};
	
	\path[->] (u0c0) edge node[left,anchor=south east] {${\scriptstyle f_1}$} (u1c0);
	\path[->] (u0c0) edge node[right,anchor=west] {${\scriptstyle f_2}$} (u010);
	\path[->] (u0c0) edge node[right,anchor=south west] {${\scriptstyle f_r}$} (u0c1);
	
	\node (u2c0) at (-6,-2) {${\scriptstyle u (2,0, \dots, 0) }$};
	\node (u110) at (-4,-2) {${\scriptstyle u (1,1, \dots, 0) }$};
	\node (u020) at (-2,-2) {${\scriptstyle u (0,2, \dots, 0) }$};
	\node (u011) at (4,-2) {${\scriptstyle u (0, \dots, 1,1) }$};
	\node (u0c2) at (6,-2) {${\scriptstyle u (0, \dots, 0,2) }$};
	\node (cd2) at (0,-2) {${\scriptstyle \cdots  }$};
	
	\path[->] (u1c0) edge node[left,anchor=south east,pos=0.3] {${\scriptstyle f_1}$} (u2c0);
	\path[->] (u1c0) edge node[right] {${\scriptstyle f_2}$} (u110);
	\path[->] (u010) edge node[left,anchor=south east,pos=0.3] {${\scriptstyle f_1}$} (u110);
	\path[->] (u010) edge node[right] {${\scriptstyle f_2}$} (u020);
	\path[->] (u0c1) edge node[right,anchor=south west] {${\scriptstyle f_r}$} (u0c2);
	\path[->] (u0c1) edge node[left] {${\scriptstyle f_{r-1}}$} (u011);
	
	\node (vda) at (-6,-2.5) {${\scriptstyle \vdots  }$};
	\node (vdb) at (-4,-2.5) {${\scriptstyle \vdots  }$};
	\node (vdc) at (-2,-2.5) {${\scriptstyle \vdots  }$};
	\node (vde) at (4,-2.5) {${\scriptstyle \vdots  }$};
	\node (vdf) at (6,-2.5) {${\scriptstyle \vdots  }$};
\end{tikzpicture}
\label{eq:CInftyArAbelian}
\end{equation}

\subsubsection{Infinite crystals: SQCD}
\label{subsec:infinitenonab}
To study a generic quiver gauge theory we need to decorate each node of the Dynkin diagram of $G$ with arbitrary gauge group $\mathbf{G}$. We now discuss the crystal $\fC_{\infty} ^{\U{N}}$ associated to single-node $\U{N}$ theories. For that, we include the action of Kashiwara operators for $\U{N}$ roots. In other words, we decorate each $u(k)$ in the crystal \eqref{eq:CinftySQED} with an $N$-tuple $\underline{\mathsf{k}} = (\mathsf{k}_1, \dots, \mathsf{k}_N)$ with $\sum_{s=1} ^{N} \mathsf{k}_s = k$. The gauge Kashiwara operators $\mathbf{f}_s$ then act as 
\begin{equation*}
\mathbf{f}_s u (k \vert (\mathsf{k}_1,  \dots, \mathsf{k}_s , \dots, \mathsf{k}_N) ) = u (k \vert (\mathsf{k}_1, \dots, \mathsf{k}_s +1 , \dots , \mathsf{k}_N) ) .
\end{equation*}
In particular each $\mathbf{f}_s$ includes the action $f_1 (u (k)) = u (k+1)$ of the unique $A_1$ Kashiwara operator $f_1$. For example, $ \mathbf{G} = \U{2}$ gives 
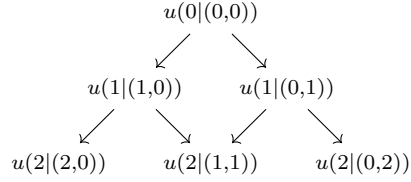
\begin{equation*}
\begin{tikzpicture}
	\node (u00) at (0,0) {${\scriptstyle u(0 \vert (0, 0)) }$};
	\node (u10) at (-1,-1) {${\scriptstyle u (1\vert (1, 0)) }$};
	\node (u01) at (1,-1) {${\scriptstyle u (1\vert (0, 1 )) }$};
	\node (u20) at (-2,-2) {${\scriptstyle u (2\vert (2, 0)) }$};
	\node (u02) at (2,-2) {${\scriptstyle u (2\vert (0, 2 )) }$};
	\node (u11) at (0,-2) {${\scriptstyle u (2\vert (1 ,1)) }$};
	\path[->] (u00) edge (u10);
	\path[->] (u00) edge (u01);
	\path[->] (u10) edge (u20);
	\path[->] (u10) edge (u11);
	\path[->] (u01) edge (u02);
	\path[->] (u01) edge (u11);
\end{tikzpicture}
\end{equation*}
while a gauge group $\mathbf{G} = \U{3}$ gives (we represent it in a 3d view)
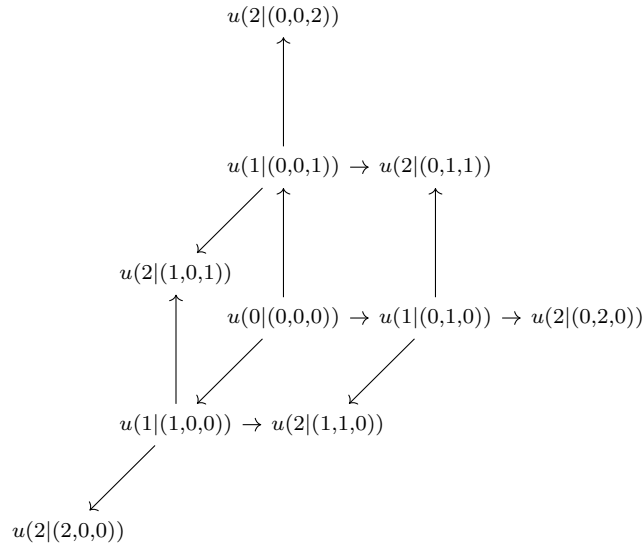
\begin{equation*}
\begin{tikzpicture}
	\node (u000) at (0,0) {${\scriptstyle u (0\vert (0, 0, 0 ) ) }$};
	\node (u100) at (-1.414,-1.414) {${\scriptstyle u (1 \vert (1, 0, 0 ) ) }$};
	\node (u010) at (2,0) {${\scriptstyle u (1 \vert (0, 1, 0 )) }$};
	\node (u001) at (0,2) {${\scriptstyle u (1 \vert (0, 0 , 1)) }$};
	\node (u200) at (-2.828,-2.828) {${\scriptstyle u (2 \vert (2, 0, 0 )) }$};
	\node (u020) at (4,0) {${\scriptstyle u (2 \vert (0, 2, 0 )) }$};
	\node (u002) at (0,4) {${\scriptstyle u (2 \vert (0, 0 , 2)) }$};
	\node (u101) at (-1.414,0.586) {${\scriptstyle u (2 \vert (1, 0, 1 )) }$};
	\node (u011) at (2,2) {${\scriptstyle u (2 \vert (0, 1 , 1)) }$};
	\node (u110) at (0.586,-1.414) {${\scriptstyle u (2 \vert (1, 1,0  )) }$};
	
	\path[->] (u000) edge (u100);
	\path[->] (u000) edge (u010);
	\path[->] (u000) edge (u001);
	\path[->] (u100) edge (u200);
	\path[->] (u100) edge (u110);
	\path[->] (u100) edge (u101);
	\path[->] (u010) edge (u020);
	\path[->] (u010) edge (u110);
	\path[->] (u010) edge (u011);
	\path[->] (u001) edge (u002);
	\path[->] (u001) edge (u101);
	\path[->] (u001) edge (u011);
	
\end{tikzpicture}
\end{equation*}
Both crystals are understood to continue infinitely.\par

\subsubsection{Crystals, characters and Verma modules}
There exists a notion of character of crystals $\fC$, defined as
\begin{equation}
\label{eq:Kostantchar}
	\chi^{\fC} (\underline{t} ) = \sum_{T \in \fC} \underline{t}^{ - \wt (T) + \rho }  =  \sum_{\un \in \Lambda_{\mathrm{w}} ^{\vee} } P^{\fC} (\un ) \underline{t}^{- (\un + \rho) } ,
\end{equation}
where $\rho = \frac{1}{2} \sum_{\alpha_j \in \Phi_{+} } \alpha_j $ is the Weyl vector and the function $P^{\fC} (\un )$, known as Kostant partition function, counts the number of tableaux in $\fC$ of weight $\un$. The multi-variable $\underline{t}$ generically denotes all the fugacities in the problem. For the crystals $\fC_{\infty} ^{\Phi}$ in the bases of freely generated Verma modules, it is well-known and easy to check that the character \eqref{eq:Kostantchar} equals 
\begin{equation}
\label{eq:Prodchar}
	\chi^{\fC_{\infty} ^{\Phi}} (\underline{t}) = \prod_{\alpha \in \Phi_{+}} \left( \underline{t}^{- \alpha /2} - \underline{t}^{\alpha /2} \right)^{-1} .
\end{equation}
More generally, $\chi^{\fC_{\infty} ^{\Phi}} $ equals the character of the corresponding Verma module \cite{Littlemann:1995,Kashiwara:Demazure}.

\subsection{Crystals for Coulomb branch Verma modules}
\label{sec:C&CVerma}
Throughout this subsection, we consider a good 3d $\mN=4$ theory \cite{Gaiotto:2008ak} with gauge group $\mathbf{G}$ as in \eqref{eq:GaugeGroupU}. It is assumed that masses and FI parameters are generic, so that the moduli space of vacua consists of a collection of isolated massive vacua. The theory is placed on $\mathbb{R} \times \C$ with the Omega-background turned on \cite{Bullimore:2015lsa,Bullimore:2016hdc}. In this way, the system is effectively reduced to a quantum mechanics along $\mathbb{R}$, leading to the study of the corresponding Hilbert space. For any choice of vacuum state $\lvert v \rangle$, the action of monopole operators on $\lvert v \rangle$ generates the full Hilbert space $\mathscr{H} (v) $ and endows it with the structure of a Verma module for the Coulomb branch algebra \cite{Bullimore:2016hdc}. In what follows we restrict our attention to balanced quivers.\par
\begin{mainthm}\label{thm:Verma}
	For any $\lvert v \rangle $, $\mathscr{H} (v)$ is described by a crystal $\fC_{\infty} ^{\Phi}$ of type $G$.
\end{mainthm}\par
Within this context, Gaiotto and Okazaki \cite{Gaiotto:2019mmf} (see also \cite{Bullimore:2020jdq}) conjectured that, when generic masses $\um$ and FI parameters $\underline{\zeta}$ are turned on, the three-sphere partition function in a suitable limit takes the form 
\begin{equation}
\label{eq:specialspherePF}
	\mathcal{Z}_{\mathbb{S}^3} (\underline{m} , \underline{\zeta})  = \sum_{v} e^{i \frac{\pi}{2} \kappa_{\mathrm{bg}} (v)   } e^{i 2 \pi \underline{m} \cdot k(v) \cdot \underline{\zeta} } \chi^{\mC} ( \underline{x} ; v)  \chi^{\mH} (\underline{y}  ; v) .
\end{equation}
The sum runs over the isolated vacua of the theory, $ \kappa_{\mathrm{bg}} (v) $ includes background mixed Chern--Simons couplings, $k(v)$ is a matrix of mixed Chern--Simons couplings, and $\chi^{\mC} , \chi^{\mH} $ are twisted traces on the Coulomb and Higgs branch Verma modules in the given vacuum, respectively. We have also introduced the fugacities $\underline{x} \equiv  e^{- 2 \pi \underline{\zeta}}$ and $\underline{y} \equiv e^{- 2 \pi \underline{m}}$ for the Coulomb and Higgs branch global symmetries, respectively.\par
The special limiting procedure of \cite{Gaiotto:2019mmf} introduces a $\mathbb{Z}_2$ twist of the Verma module characters from the centre of the R-symmetry \cite{Bullimore:2020jdq}.\footnote{This twisting factor appears also in the partition function of 3d $\mN=3$ Chern--Simons theories \cite{Santilli:2020snh}.} At the present stage, we do not know how to systematically include such twist in the crystals. Its effect, however, trivializes for balanced quivers, which are the only ones we consider. Under the balancing assumption, we find that 
\begin{equation*}
	\chi^{\fC_{\infty} ^{\Phi} } (\underline{t}) = \chi^{\mC} (\underline{t} ; v) .
\end{equation*}\par
We do not have a direct construction for Higgs branch Verma module characters $\chi^{\mH}$ via crystals. However, we may take mirror symmetry as a working assumption and \emph{define} a Higgs branch crystal as the crystal $\fC_{\infty} ^{\Phi}$ corresponding to the Coulomb branch of the mirror theory. An alternative approach would be to consider the isomorphism between (certain resolutions of) slices in the affine Grassmannian and (resolutions of) closures of nilpotent orbits \cite{MV03} and exploit the equivalence of the crystal bases on the two sides \cite{Dranowski}.\par
For technical reasons, we only consider balanced quivers which moreover belong to one of the three classes in Subsections \ref{subsec:TSUNVerma}, \ref{subsec:AbelianVerma} and \ref{subsec:infinitenonab}. Nonetheless, it is conceivable that one may consider more general non-Abelian quivers, although the balancing conditions seems harder to lift at the present stage.\par
In this way, part of the conjecture of \cite{Gaiotto:2019mmf} holds on general grounds as a consequence of Result \ref{thm:Verma}, mirror symmetry and the equality of characters of crystals and Verma modules. However, to get a full proof of \eqref{eq:specialspherePF} using only the theory of crystal bases, several aspects have to be clarified:
\begin{enumerate}[(i)]
	\item how to treat non-balanced quivers;
	\item how to incorporate the $\mathbb{Z}_2$ R-symmetry twist in the traces;
	\item how to predict the correct mixed Chern--Simons couplings.
\end{enumerate}
We present the explicit match between the characters derived from crystals and those in \cite{Gaiotto:2019mmf}.\footnote{Earlier exact results appeared e.g. in \cite{Kapustin:2010xq,Nishioka:2011dq,Benvenuti:2011ga,Russo:2016ueu,Santilli:2019mtc}.} To lighten the expressions, let us write for short $\fC_{\infty}$ for $\fC_{\infty} ^{A_r}$ and 
\begin{equation}
\label{eq:defDelta}
	\Delta (x,y) \equiv  \left( \left( \frac{x }{y } \right)^{-1/2} - \left( \frac{x }{y} \right)^{1/2} \right)^{-1} .
\end{equation}
As a side observation, note that \eqref{eq:defDelta} satisfies the reproducing property 
\begin{equation*}
	\oint_{\U{1}} \frac{\dd z}{2 \pi i z } \Delta (x,z) \Delta (z,y) = \Delta (x,y) .
\end{equation*}
Physically, to take a Cauchy integral over the fugacity $z$ is to gauge the corresponding $\U{1}$.\par
\medskip
So far we have not discerned among the various crystals $ \fC_{\infty} ^{\Phi}$ associated to different vacua $\lvert v \rangle $, i.e. we have not given a map $v \mapsto \fC_{\infty} ^{\Phi}$. The reason is that they are all isomorphic as crystals. There is nevertheless a simple rule to count them, which corresponds to count the possible ways to express the independent parameters as functions of the fugacities $\underline{x},\underline{y} $.

\subsubsection{SQED}
Consider 3d balanced SQED, i.e. the $A_1$ quiver with $\U{1}$ gauge group and 2 fundamental flavours. The $\mathbf{G} = \U{1}$ crystal $\fC_{\infty} ^{\overset{1}{\circ} - \overset{2}{\Box}}$ was given in \eqref{eq:CinftySQED}. We observe that the structure read off from $\fC_{\infty} ^{\overset{1}{\circ} - \overset{2}{\Box}}$ matches that of $\mathscr{H} (v) $, generated by acting with a single raising monopole operator.\par
To compute the traces $\chi^{\mC}, \chi^{\mH}$, we observe that each weight occurs once in \eqref{eq:CinftySQED}, so that using \eqref{eq:Kostantchar} with $P^{\fC_{\infty} ^{\overset{1}{\circ} - \overset{2}{\Box}}} (\mathsf{w} ) = 1$ $\forall \mathsf{w}$, gives 
\begin{equation}
\label{eq:SQEDcharVerma}
	\chi^{\fC_{\infty} ^{\overset{1}{\circ} - \overset{2}{\Box}}} (\underline{x}) =  \sum_{k=0} ^{\infty}  \left( \frac{x_1}{x_2} \right) ^{k+ \frac{1}{2} } =  \Delta (x_1, x_2) .
\end{equation}
The theory is self-mirror, hence the Higgs branch Verma module is identical to what we have just described.

\subsubsection{$T[\SU{n}]$}
We consider the $T[\SU{n}]$ quiver $\overset{1}{\circ} - \overset{2}{\circ} - \cdots - \overset{(n-1) }{\circ} - \overset{n}{\Box}$. Combining the rules in Subsection \ref{subsec:TSUNVerma} and \ref{subsec:infinitenonab}, elements of the corresponding $A_{n-1}$ crystal $\fC_{\infty} ^{T[\SU{n}]}$ are uniquely specified by a collection of integers $(k_1, \dots , k_{n-1})$ and arrays 
\begin{equation*}
	\left\{ (\mathsf{k}_{j,1}, \dots , \mathsf{k}_{j,j}) \right\}_{j=1, \dots, n-1} \quad \text{such that} \quad \sum_{s=1} ^{j} \mathsf{k}_{j,s} = k_j .
\end{equation*}
This description agrees with the Verma module structure on $\mathscr{H} (v)$ discussed in \cite{Bullimore:2016hdc}.\par
It is known from the onset that Verma modules of the Coulomb branch coordinate ring of $T[\SU{n}]$ are freely generated \cite{Gaiotto:2008ak}. Their character is thus given by \eqref{eq:Prodchar} and reads 
\begin{equation}
\label{eq:TSUNVermachar}
	\chi^{\fC_{\infty} ^{T[\SU{n}]}} (\underline{x}) = \prod_{1 \le j < k \le n} \Delta (x_j, x_k) ,
\end{equation}
with the function $\Delta (x_j,x_k)$ defined in \eqref{eq:defDelta}. It indeed agrees with the sphere partition function computations \cite{Nishioka:2011dq,Benvenuti:2011ga}.\par
A heuristic way to observe that the Verma module is freely generated stems from counting Kashiwara operators: from Subsection \ref{subsec:TSUNVerma} and setting $r=n-1$, we get $\frac{n(n-1)}{2}$ operators $\left\{ f_j  \right\} $, while using the prescription of Subsection \ref{subsec:infinitenonab} at each gauge node gives the same total number of gauge Kashiwara operators $\left\{  \mathbf{f}_{j,s} \right\}$. As acting with a gauge Kashiwara operator increases the weight by one, we see that they do not add elements $u (\underline{k})$ to the crystal.\par
Being $\mathscr{H} (v)$ a freely generated Verma module for $T[\SU{n}]$, the gauge group does not add any further constraint on the construction of the crystal. Therefore, results on Kashiwara crystals apply without modification to the present case. In this context, the map of \cite{Dranowski}, sending $\fC_{\infty} ^{T[\SU{n}]}$ into a basis for Verma modules for $\mH$, can be interpreted as a manifestation of the self-mirror nature of $T[\SU{n}]$.\par
\medskip
It is likely that \eqref{eq:TSUNVermachar} may be derived by induction on $n$ and gluing, in a computation akin to \cite{Hanany:2011db}. Further insight comes from the brane construction. Recall that, in our conventions, the NS5 branes are placed at positions $- \frac{1}{2\pi} \log \underline{x}$ \cite{Hanany:1996ie}. It is straightforward to compute the characters for $n=1$ and $n=2$:
\begin{equation*}
	\begin{tikzpicture}
		\path[red,thick] (-5,1) edge (-5,0);
		\path[black] (-5,0.5) edge (-4,0.5);
		\path[black,dotted] (-4,0.5) edge (-3.6,0.5);
		\node (c1) at (-5,-0.5) {$\chi^{\mC [ T[\SU{1}]]} = 1$};

		\path[red,thick] (0,1) edge (0,0);
		\path[red,thick] (-1,1) edge (-1,0);
		\path[black] (0,0.5) edge (1,0.5);
		\path[black,dotted] (0.99,0.5) edge (1.4,0.5);
		\path[black] (-1,0.4) edge (1,0.4);
		\path[black,dotted] (0.99,0.4) edge (1.4,0.4);
		
		\node (c2) at (0,-0.5) {$\chi^{\mC [ T[\SU{2}]]} = \Delta (x_1,x_2)$};
	\end{tikzpicture}
\end{equation*}
with dotted D3 branes understood to be stretched to infinity. Then, assuming \eqref{eq:TSUNVermachar} holds for $n$, the formula for $n+1$ is obtained by gluing 
\begin{equation*}
\begin{tikzpicture}
	\path[red,thick] (0,1) edge (0,0);
	\path[red,thick] (-1,1) edge (-1,0); 
	\path[red,thick] (-2,1) edge (-2,0); 
	\path[red,thick] (-3,1) edge (-3,0); 
	\path[red,thick] (2,1) edge (2,0); 
	\node (dc) at (1,0.75) {$ \cdots $};
	
	\path[black] (-3,0.1) edge (3,0.1);
	\path[black] (-2,0.2) edge (3,0.2);
	\path[black] (-1,0.3) edge (3,0.3);
	\path[black] (1.6,0.7) edge (3,0.7);
	\path[black] (2,0.8) edge (3,0.8);
	\path[black,dotted] (2.99,0.1) edge (3.4,0.1);
	\path[black,dotted] (2.99,0.2) edge (3.4,0.2);
	\path[black,dotted] (2.99,0.3) edge (3.4,0.3);
	\path[black,dotted] (2.99,0.7) edge (3.4,0.7);
	\path[black,dotted] (2.99,0.8) edge (3.4,0.8);
	
	\node (dol) at (1.75,0.6) {\tiny $\vdots $};
	\node (dor) at (2.25,0.6) {\tiny $\vdots $};

		\path[red,thick] (6,1) edge (6,0);
		\path[black] (5,0.1) edge (7,0.1);
		\path[black,dotted] (7,0.1) edge (7.4,0.1);
		\path[black,dotted] (5,0.1) edge (4.6,0.1);
		\path[black] (5,0.2) edge (7,0.2);
		\path[black,dotted] (7,0.2) edge (7.4,0.2);
		\path[black,dotted] (5,0.2) edge (4.6,0.2);
		\path[black] (5,0.3) edge (7,0.3);
		\path[black,dotted] (7,0.3) edge (7.4,0.3);
		\path[black,dotted] (5,0.3) edge (4.6,0.3);
		\path[black] (5,0.7) edge (7,0.7);
		\path[black,dotted] (7,0.7) edge (7.4,0.7);
		\path[black,dotted] (5,0.7) edge (4.6,0.7);
		\path[black] (5,0.8) edge (7,0.8);
		\path[black,dotted] (7,0.8) edge (7.4,0.8);
		\path[black,dotted] (5,0.8) edge (4.6,0.8);
		\path[black] (6,0.9) edge (7,0.9);
		\path[black,dotted] (7,0.9) edge (7.4,0.9);

		\node (dn0) at (6.25,0.6) {\tiny $\vdots $};
		\node (dn1) at (5.75,0.6) {\tiny $\vdots $};
	
\end{tikzpicture}
\end{equation*}
and assigning a factor $\Delta (x_j,x_{n+1})$ to each D3 brane stretched between the $j^{\text{th}}$ and the newly introduced $(n+1)^{\text{th}}$ NS5 brane. While we do not have a rigorous proof of the gluing rule and of the assignment of factors of $\Delta$, it should be possible to derive it by refining the counting in formula \eqref{eq:Kostantchar} with gauge fugacities, and then average over the gauge group $\mathbf{G}$, which enforces a projection onto gauge invariant operators, in a crystal analogue of \cite{Hanany:2011db} (see also e.g. \cite{Benvenuti:2010pq,Chen:2011wn,Santilli:2020ueh} for related manipulations of unitary matrix integrals).

\subsubsection{Abelian linear quiver}
Our next example is the balanced, Abelian $A_{n-1}$ quiver. Its Coulomb branch is a mirror description of the Higgs branch of SQED with $n$ flavours.\par
The crystal $\fC_{\infty} ^{\U{1}^{n-1}} $ given in \eqref{eq:CInftyArAbelian} captures the Hilbert space of the theory, which is generated by acting with $n-1$ commuting raising monopole operators. The general formula \eqref{eq:Kostantchar} gives the character of \eqref{eq:CInftyArAbelian}: 
\begin{equation}
\label{eq:ArU1charVerma}
	\chi^{\fC_{\infty} ^{\U{1}^{n-1}} } (\underline{y}) = \prod_{j=1}^{n-1}  \sum_{k_j=0} ^{\infty}   \left( \frac{y_j}{y_{j+1}} \right) ^{k_j + \frac{1}{2}} = \prod_{j=1}^{n-1} \Delta (y_j, y_{j+1}) .
\end{equation}
Refining formula \eqref{eq:Kostantchar} by gauge fugacities and then averaging over $\U{1}^{n-1}$ produces the same result.\par
There are $\left(  \begin{matrix} n \\ n-1 \end{matrix} \right) = n $ ways to get $n-1$ independent parameters out of the $n$ fugacities $\underline{y}$, corresponding to the $n$ ways of obtaining a framed $\U{1}^{n-1}$ quiver out of the affine, unframed $\U{1}^{n}$. Putting all the pieces together, we find agreement with the existing literature.\par
We observe that the gluing prescription above works also in this case. We start with SQED with two flavours and add NS5 branes one by one, 
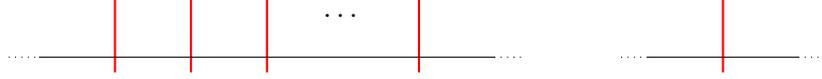
\begin{equation*}
\begin{tikzpicture}
	\path[red,thick] (0,1) edge (0,0);
	\path[red,thick] (-1,1) edge (-1,0); 
	\path[red,thick] (-2,1) edge (-2,0); 
	\path[red,thick] (2,1) edge (2,0); 
	\node (dc) at (1,0.75) {$ \cdots $};
	
	\path[black] (-3,0.2) edge (3,0.2);
	\path[black,dotted] (2.99,0.2) edge (3.4,0.2);
	\path[black,dotted] (-3.4,0.2) edge (-2.99,0.2);
	
	\path[red,thick] (6,1) edge (6,0);
	\path[black] (5,0.2) edge (7,0.2);
	\path[black,dotted] (6.99,0.2) edge (7.4,0.2);
	\path[black,dotted] (5.01,0.2) edge (4.6,0.2);
\end{tikzpicture}
\end{equation*}
Including a factor $\Delta (x_1, x_j)$ when the $j^{\text{th}}$ NS5 brane is added and linked to the first NS5 brane via a D3 brane, we get the character 
\begin{equation*}
	\prod_{j=2}^{n} \Delta (x_1, x_j) .
\end{equation*}
This formula agrees with \eqref{eq:ArU1charVerma} but evaluated at a different vacuum of the theory. The $\underline{x}$-basis and $\underline{y}$-basis are related through $x_k = x_1 \prod_{j=1}^{k} y_{j+1}/y_j$. In other words, formula \eqref{eq:ArU1charVerma} takes as $n-1$ independent variables $y_{j+1} /y_j$ whereas the brane rule expresses the result in the $n-1$ independent variables $x_j/x_1$ (the latter being the normalization chosen in \cite{Gaiotto:2019mmf}).\par

\subsubsection{SQCD}
Consider now $ \U{N}$ with $ 2N $ fundamental flavours. Each element $u (k \vert \underline{\mathsf{k}} ) \in \fC_{\infty} ^{\overset{N}{\circ} - \overset{2N}{\Box}}$ is specified by an integer $k$ and an array $\underline{\mathsf{k}} = (\mathsf{k}_1 , \dots , \mathsf{k}_N) \in \mathbb{N}^{N}$ satisfying $\sum_{s=1} ^{N} \mathsf{k}_s = k$. By the construction in Subsection \ref{subsec:infinitenonab}, 
\begin{equation*}
	k \ : \ \exists (s_1, \dots, s_k) \in \mathbb{N}^{k} \text{ such that } \mathbf{e}_{s_1} \cdots \mathbf{e}_{s_k} u (k \vert \underline{\mathsf{k}}) = u (0) ,
\end{equation*}
where $\mathbf{e}_{s}$ is the $\U{N}$ gauge Kashiwara operator. This is precisely the crystal analogue of the structure of $\mathscr{H} (v)$ \cite{Bullimore:2016hdc}.\par
Let us now derive the Coulomb branch Verma trace. We use that there are $ \left( \begin{matrix} k + N - 1 \\ k \end{matrix} \right) $ elements $u (k \vert \underline{\mathsf{k}}) \in \fC_{\infty}^{\overset{N}{\circ} - \overset{2N}{\Box}} $ at distance $k$ from $u(0)$. Therefore \eqref{eq:Kostantchar} gives
\begin{equation}
\label{eq:SQCDcharVerma}
	\chi^{\fC_{\infty}^{\overset{N}{\circ} - \overset{2N}{\Box}}} (\underline{x})  = \sum_{k=0} ^{\infty} \left( \begin{matrix} k + N - 1 \\ k \end{matrix} \right)  \left( \frac{x_1}{x_2} \right)^{k + \frac{N}{2} } = \Delta (x_1, x_2)^{N} .
\end{equation}\par
This formula is again correctly reproduced by gluing brane configurations:
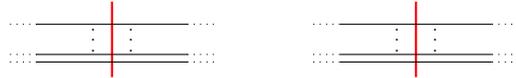
\begin{equation*}
\begin{tikzpicture}
	\path[red,thick] (0,1) edge (0,0);
	
	\path[black] (-1,0.2) edge (1,0.2);
	\path[black] (-1,0.3) edge (1,0.3);
	\path[black] (-1,0.7) edge (1,0.7);
	\path[black,dotted] (-.99,0.2) edge (-1.4,0.2);
	\path[black,dotted] (-.99,0.3) edge (-1.4,0.3);
	\path[black,dotted] (-.99,0.7) edge (-1.4,0.7);
	\path[black,dotted] (.99,0.2) edge (1.4,0.2);
	\path[black,dotted] (.99,0.3) edge (1.4,0.3);
	\path[black,dotted] (.99,0.7) edge (1.4,0.7);
	
	\node (dol) at (.25,0.6) {\tiny $\vdots $};
	\node (dor) at (-.25,0.6) {\tiny $\vdots $};

	\path[red,thick] (4,1) edge (4,0); 
	
	\path[black] (3,0.2) edge (5,0.2);
	\path[black] (3,0.3) edge (5,0.3);
	\path[black] (3,0.7) edge (5,0.7);
	\path[black,dotted] (3,0.2) edge (2.6,0.2);
	\path[black,dotted] (3,0.3) edge (2.6,0.3);
	\path[black,dotted] (3,0.7) edge (2.6,0.7);
	\path[black,dotted] (5,0.2) edge (5.4,0.2);
	\path[black,dotted] (5,0.3) edge (5.4,0.3);
	\path[black,dotted] (5,0.7) edge (5.4,0.7);
	
	\node (dol) at (4.25,0.6) {\tiny $\vdots $};
	\node (dor) at (3.75,0.6) {\tiny $\vdots $};
\end{tikzpicture}
\end{equation*}
and counting a factor $\Delta (x_1,x_2)$ for every D3 brane linking the two NS5 branes.

\section{Outlook}
\label{sec:outlook}

There are several open problems that could be considered, but a treatment of them is out of the scope of the present paper. In this final section we discuss possible directions for which the present study has laid the groundwork for further research.

\paragraph*{Geometric crystals.}
A notion of \emph{geometric} crystal was introduced in \cite{BK}, as opposed to the \emph{combinatorial} crystals that we have considered so far. Geometric crystals are algebraic varieties endowed with additional structures that parallel Kashiwara's construction. The crystals $\fC_{\infty} $ discussed in Section \ref{sec:Demazure} arise as \emph{tropicalizations} of geometric crystals.\par
In the tropical limit description, the elements $u (k_1 , \dots , k_r)$ are represented as convex polytopes, known as MV-polytopes \cite{Anderson03,Kamnitzer:MV1}. A characterizing property of the latter is that they satisfy tropical Pl\"{u}cker relations, see \cite{Kamnitzer:MV1} and subsequent works for in-depth discussion. These MV-polytopes are combinatorial analogues of the MV-cycles (after Mirkovic--Vilonen \cite{MV07}), which are closed subvarieties of $\Gr_{G}$. The crystal structure on MV-cycles and its relevance for 3d $\mN=4$ Coulomb branches were pointed out in \cite{Braverman:2016pwk}.\par
Lifting the correspondence between $\fC_{\infty}$ and Coulomb branch Verma modules to the level of geometric crystals would pave the way to tackle a wealth of problems that revolve around 3d $\mN=4$ Coulomb branches using crystal bases. Here we list a couple of avenues whose systematic exploration is left for future work.

\paragraph*{Homological knot invariants.}
The construction of a knot homology theory starting from the resolution $\mX \to \mC$ of the Coulomb branch of a 3d $\mN=4$ quiver theory has been recently put forward in \cite{Aganagic:2020olg,Aganagic:2021ubp}. A main ingredient in that story is a ``weighted'' generalization of KLR algebras \cite{Webster:2017}. On the other hand, MV-polytopes are in correspondence with representations of KLR algebras \cite{Tingley:2016MV}. In this way, the infinite crystals $\fC_{\infty}$ arise on both sides of the bridge connecting Coulomb branches and knot homologies, although in disguise and playing different roles.\par
Along different lines, Littlemann's path model \cite{Littlemann:path} describes crystal structures as random walks in the (co)weight lattice. This presentation in terms of piecewise linear functions in $\Lambda_{\mathrm{w}} ^{\vee}$ is alternative and equivalent to that using MV-polytopes. A refined version of Littlemann's path model, more fitting in a geometric lift, was discussed in \cite{Biane:LPM}. One result of \cite{Biane:LPM} is, roughly, that acting with a preferred sequence of Kashiwara $e_{j}$ operators on a Littlemann path one gets a Brownian motion in the Weyl chamber of $G$. Aspects of the latter, in turn, are computed in some cases by colored unknot invariants in $\mathbb{S}^3$ \cite{deHaro:2004id}.\par
It would be interesting to explore further the web of relations among geometric crystals and their tropical limit, knot invariants, and 3d $\mN=4$ Coulomb branches. In particular, a question worth asking is whether certain path models for geometric crystals (or a categorification thereof) may produce homological knots invariants, paralleling the way in which certain polynomial knot invariants are computed by combinatorial path models \cite{deHaro:2004id}.

\paragraph*{5d Higgs branches.}
MV-polytopes bear a close resemblance with (non-triangulated) toric polygons. The latter are dual to tropical curves and can be used to describe 5d supersymmetric gauge theories. The tropical geometry approach has been recently adopted in \cite{vanBeest:2020kou} to understand the structure of 5d Higgs branches at strong coupling. Another useful tool to analyze 5d Higgs branches is provided by the magnetic quivers \cite{Ferlito:2017xdq,Cabrera:2018jxt,Closset:2020scj}. These devices are suitable 3d $\mN=4$ quivers whose Coulomb branch is isomorphic to (a component of) the Higgs branch of the 5d theory.\par
Because of their role in describing 3d $\mN=4$ Coulomb branches on one hand, and their roots into tropical geometry on the other, it seems plausible that MV-polytopes and the crystals $\fC_{\infty}$ may be used to shed light on 5d Higgs branches from a new angle. Finding a crystal analogue of the construction of \cite{vanBeest:2020kou} is an intriguing open problem that we leave for the future.

\vspace{0.5cm}
\subsection*{Acknowledgements}
We thank Julius Grimminger and Zhenghao Zhong for useful discussions and comments. The work of LS is supported by the Funda\c{c}\~{a}o para a Ci\^{e}ncia e a Tecnologia (FCT) through the doctoral grant SFRH/BD/129405/2017. The work is also financially supported by FCT Project PTDC/MAT-PUR/30234/2017 and by Comunidad de Madrid (grant QUITEMAD-CM, ref. P2018/TCS-4342).
\vspace{0.5cm}

\begin{appendix}

	\section{Examples of crystals for Coulomb branches}
	\label{app:ExamplesC}
In this appendix we collect further examples of resolved crystals $\fX$ that reproduce the resolution of Coulomb branches $\mX \to \mC$ as one moves along the parameter space. The appendix is meant to complement Section \ref{sec:massdef} and to provide further insight into Result \ref{thm:massdef}.

\subsection{More mass-deformed type A examples}

\subsubsection{Conventions}
Throughout this appendix, for the sake of clarity in the pictures, whenever two lines cross each other, one of them is drawn gray and dashed. Moreover, to reduce clutter, when a subset of the boxes is frozen and does not contribute to any transition, we replace all such boxes by a single black box.

\subsubsection{$\U{2}$ with $6$ flavours}

The parameter space $\mM^{(6)} = \bigsqcup _{\lambda} \mM_{\lambda} ^{(6)}$ is stratified as 
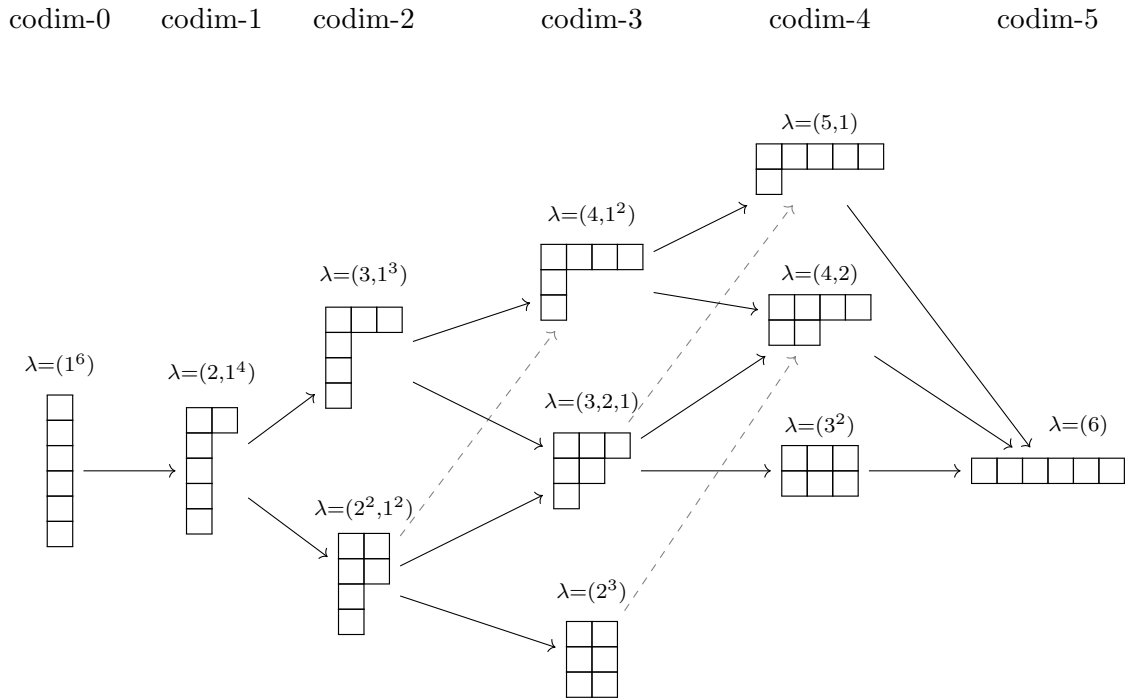
\begin{equation*}
\ytableausetup{smalltableaux}
	\begin{tikzpicture}
		\node (t1) at (-7,0) {\begin{ytableau} \ \\ \ \\ \ \\ \ \\ \ \\ \ \end{ytableau}}  ; 
		\node (t2) at (-5,0) {$\begin{ytableau} \ & \ \\ \ \\ \ \\ \ \\ \ \end{ytableau}$} ; 
		\node (t31) at (-3,1.5) {$\begin{ytableau} \ & \ &  \ \\ \ \\ \ \\ \ \end{ytableau}$} ; 
		\node (t22) at (-3,-1.5) {$\begin{ytableau} \ & \ \\ \ & \ \\ \ \\ \ \end{ytableau}$} ; 		
		\node (t23) at (0,-2.5) {$\begin{ytableau} \ & \ \\ \ & \ \\ \ & \ \end{ytableau}$} ; 
		\node (t32) at (0,0) {$\begin{ytableau} \ & \ &  \ \\ \ & \ \\ \  \end{ytableau}$} ;
		\node (t41) at (0,2.5) {$\begin{ytableau} \ & \ &  \ & \  \\ \ \\ \   \end{ytableau}$} ; 
		\node (t33) at (3,0) {$\begin{ytableau} \ & \ &  \ \\ \ & \ & \  \end{ytableau}$} ;
		\node (t42) at (3,2) {$\begin{ytableau} \ & \ &  \ & \  \\ \ & \   \end{ytableau}$} ; 
		\node (t51) at (3,4) {$\begin{ytableau} \ & \ &  \ & \  &\ \\ \   \end{ytableau}$} ; 
		\node (t6) at (6,0) {$\begin{ytableau} \ & \ &  \ & \  & \ & \   \end{ytableau}$} ; 
		
		\node (l1) at (-7,1.4) {${\scriptstyle \lambda = (1^6)}$};
		\node (l2) at (-5,1.3) {${\scriptstyle \lambda = (2,1^4)}$};
		\node (l31) at (-3,2.6) {${\scriptstyle \lambda = (3,1^3)}$};
		\node (l22) at (-3,-0.5) {${\scriptstyle \lambda = (2^2,1^2)}$};
		\node (l23) at (0,-1.6) {${\scriptstyle \lambda = (2^3)}$};
		\node (l32) at (0,.9) {${\scriptstyle \lambda = (3,2,1)}$};
		\node (l41) at (0,3.4) {${\scriptstyle \lambda = (4,1^2)}$};
		\node (l33) at (3,.6) {${\scriptstyle \lambda = (3^2)}$};
		\node (l42) at (3,2.6) {${\scriptstyle \lambda = (4,2)}$};
		\node (l51) at (3,4.6) {${\scriptstyle \lambda = (5,1)}$};
		\node (l6) at (6.4,0.6) {${\scriptstyle \lambda = (6)}$};

		\node (c0) at (-7,6) {codim-0};
		\node (c1) at (-5,6) {codim-1};
		\node (c2) at (-3,6) {codim-2};
		\node (c3) at (0,6) {codim-3};
		\node (c4) at (3,6) {codim-4};
		\node (c5) at (6,6) {codim-5};
		
		\draw[->] (t1) edge (t2);
		\draw[->] (t2) edge (t22);
		\draw[->] (t2) edge (t31);
		\draw[->] (t22) edge (t23);
		\draw[->,gray,dashed] (t22) edge (t41);
		\draw[->,gray,dashed] (t23) edge (t42);
		\draw[->] (t22) edge (t32);
		\draw[->] (t31) edge (t32);
		\draw[->] (t31) edge (t41);
		\draw[->] (t32) edge (t33);
		\draw[->,gray,dashed] (t32) edge (t51);
		\draw[->] (t32) edge (t42);
		\draw[->] (t41) edge (t42);
		\draw[->] (t41) edge (t51);
		\draw[->] (t42) edge (t6);
		\draw[->] (t51) edge (t6);
		\draw[->] (t33) edge (t6);
	\end{tikzpicture}
\end{equation*}

\begin{itemize}
\item $\lambda =(1^6)$. $\mH$ is lifted and $\mC$ is fully resolved.
\item $\lambda =(2,1^4)$. $\mX$ has an $A_1$ singularity.
	\begin{equation*}
	\ytableausetup{smalltableaux}
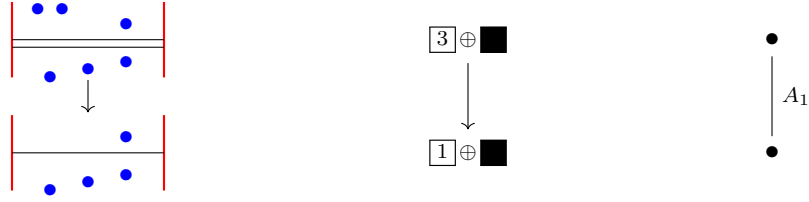
	\begin{tikzpicture}
		\node (t1) at (0,0) {$\begin{ytableau} 3 & \none[ \scriptstyle \oplus ] & *(black)  \end{ytableau}$}  ; 
		\node (t2) at (0,-1.5) {$\begin{ytableau} 1 & \none[ \scriptstyle \oplus ] &  *(black) \end{ytableau}$}  ; 
		\node (h1) at (4,0) {$\bullet$};
		\node (h2) at (4,-1.5) {$\bullet$};
		\draw[->] (t1) edge (t2);
		\path[] (h1) edge node[right] {$\scriptstyle A_{1}$} (h2);
		
		\node (d5a1) at (-5.5,0.4) {$\dbr \ \dbr $};
		\node (d5a3) at (-4.5,0.2) {$\dbr  $};
		\node (d5a4) at (-4.5,-0.3) {$\dbr  $};
		\node (d5a5) at (-5,-0.4) {$\dbr  $};
		\node (d5a6) at (-5.5,-0.5) {$\dbr  $};
		\path[red,thick] (-6,0.5) edge (-6,-0.5);
		\path[red,thick] (-4,0.5) edge (-4,-0.5);
		\path (-6,0) edge (-4,0);
		\path (-6,-0.1) edge (-4,-0.1);

		\node (d5b3) at (-4.5,-1.3) {$\dbr  $};
		\node (d5b4) at (-4.5,-1.8) {$\dbr  $};
		\node (d5b5) at (-5,-1.9) {$\dbr  $};
		\node (d5b6) at (-5.5,-2) {$\dbr  $};
		\path[red,thick] (-6,-1) edge (-6,-2);
		\path[red,thick] (-4,-1) edge (-4,-2);
		\path (-6,-1.5) edge (-4,-1.5);
		
		\node (aux1) at (-5,-0.4) {};
		\node (aux2) at (-5,-1.1) {};
		\draw[->] (aux1) edge (aux2);
	\end{tikzpicture}
	\end{equation*}
\item $\lambda =(3,1^3)$. $\mX$ has an $A_2$ singularity.
	\begin{equation*}
	\ytableausetup{smalltableaux}
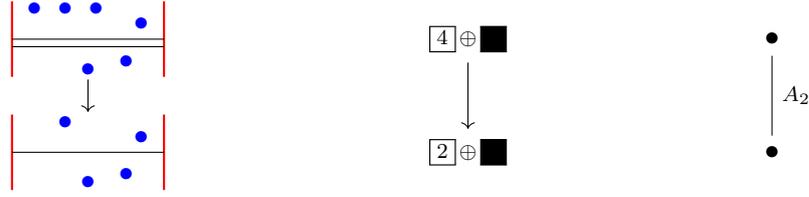
	\begin{tikzpicture}
		\node (t1) at (0,0) {$\begin{ytableau} 4 & \none[ \scriptstyle \oplus ] & *(black)  \end{ytableau}$}  ; 
		\node (t2) at (0,-1.5) {$\begin{ytableau} 2 & \none[ \scriptstyle \oplus ] & *(black)  \end{ytableau}$}  ; 
		\node (h1) at (4,0) {$\bullet$};
		\node (h2) at (4,-1.5) {$\bullet$};
		\draw[->] (t1) edge (t2);
		\path[] (h1) edge node[right] {$\scriptstyle A_{2}$} (h2);
		
		\node (d5a1) at (-5.3,0.4) {$\dbr \ \dbr \ \dbr $};
		\node (d5a3) at (-4.3,0.2) {$\dbr  $};
		\node (d5a4) at (-4.5,-0.3) {$\dbr  $};
		\node (d5a5) at (-5,-0.4) {$\dbr  $};
		\path[red,thick] (-6,0.5) edge (-6,-0.5);
		\path[red,thick] (-4,0.5) edge (-4,-0.5);
		\path (-6,0) edge (-4,0);
		\path (-6,-0.1) edge (-4,-0.1);
		
		\node (d5b1) at (-5.3,-1.1) {$\dbr$};
		\node (d5b3) at (-4.3,-1.3) {$\dbr  $};
		\node (d5b4) at (-4.5,-1.8) {$\dbr  $};
		\node (d5b5) at (-5,-1.9) {$\dbr  $};
		\path[red,thick] (-6,-1) edge (-6,-2);
		\path[red,thick] (-4,-1) edge (-4,-2);
		\path (-6,-1.5) edge (-4,-1.5);
		
		\node (aux1) at (-5,-0.4) {};
		\node (aux2) at (-5,-1.1) {};
		\draw[->] (aux1) edge (aux2);
	\end{tikzpicture}
	\end{equation*}
\item $\lambda= (2^2,1^2)$. $\mX$ has two separated $A_1$ singularities, with two one-quaternionic dimensional flat directions opening. $\fX_{(2^2,1^2)}$ is very similar to the crystal $\fX_{(2^2)}$ for $n=4$.
\item $\lambda =(4,1^2)$. Again, $\fX_{(4,1^2)}$ is essentially analogous to $\fX_{(4)}$.
\item $\lambda =(3,2,1)$.
	\begin{equation*}
	\ytableausetup{smalltableaux}
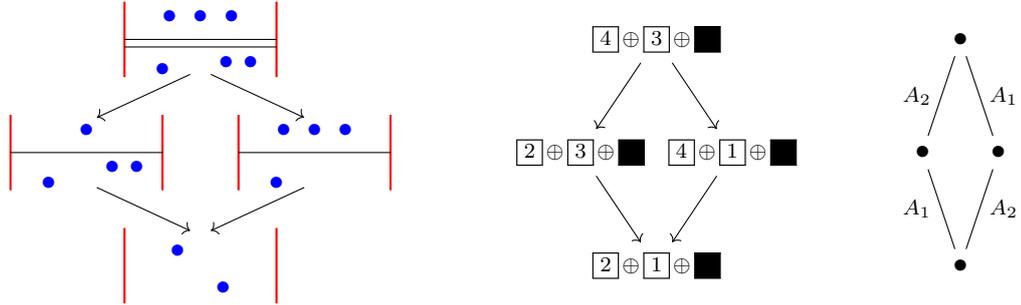
	\begin{tikzpicture}
		\node (t1) at (0,0) {$\begin{ytableau} 4 & \none[ \scriptstyle \oplus ] & 3 & \none[ \scriptstyle \oplus ] & *(black)  \end{ytableau}$}  ; 
		\node (t2a) at (-1,-1.5) {$\begin{ytableau} 2 & \none[ \scriptstyle \oplus ] & 3 & \none[ \scriptstyle \oplus ] & *(black)  \end{ytableau}$}  ; 
		\node (t2b) at (1,-1.5) {$\begin{ytableau} 4 & \none[ \scriptstyle \oplus ] & 1 & \none[ \scriptstyle \oplus ] & *(black)  \end{ytableau}$}  ; 
		\node (t3) at (0,-3) {$\begin{ytableau} 2 & \none[ \scriptstyle \oplus ] & 1 & \none[ \scriptstyle \oplus ] & *(black)  \end{ytableau}$}  ; 
		\node (h1) at (4,0) {$\bullet$};
		\node (h2a) at (3.5,-1.5) {$\bullet$};
		\node (h2b) at (4.5,-1.5) {$\bullet$};
		\node (h3) at (4,-3) {$\bullet$};
		
		\draw[->] (t1) edge (t2a);
		\draw[->] (t1) edge (t2b);
		\draw[->] (t2a) edge (t3);
		\draw[->] (t2b) edge (t3);
		\path[] (h1) edge node[left] {$\scriptstyle A_{2}$} (h2a);
		\path[] (h2a) edge node[left] {$\scriptstyle A_{1}$} (h3);
		\path[] (h1) edge node[right] {$\scriptstyle A_{1}$} (h2b);
		\path[] (h2b) edge node[right] {$\scriptstyle A_{2}$} (h3);
		
		\node (d5a1) at (-6,0.3) {$ \dbr \ \dbr \ \dbr  $};
		\node (d5a4) at (-5.5,-0.3) {$\dbr \ \dbr $};
		\node (d5a5) at (-6.5,-0.4) {$\dbr  $};
		\path[red,thick] (-7,0.5) edge (-7,-0.5);
		\path[red,thick] (-5,0.5) edge (-5,-0.5);
		\path (-7,0) edge (-5,0);
		\path (-7,-0.1) edge (-5,-0.1);

		\node (d5b0) at (-7.5,-1.2) {$\dbr  $};
		\node (d5b1) at (-7,-1.7) {$\dbr \ \dbr $};
		\node (d5b4) at (-8,-1.9) {$\dbr  $};
		\path[red,thick] (-8.5,-1) edge (-8.5,-2);
		\path[red,thick] (-6.5,-1) edge (-6.5,-2);
		\path (-8.5,-1.5) edge (-6.5,-1.5);
		
		\node (d5b2) at (-4.5,-1.2) {$\dbr \ \dbr \ \dbr $};
		\node (d5b3) at (-5,-1.9) {$\dbr  $};
		\path[red,thick] (-5.5,-1) edge (-5.5,-2);
		\path[red,thick] (-3.5,-1) edge (-3.5,-2);
		\path (-5.5,-1.5) edge (-3.5,-1.5);
		
		\node (d5C1) at (-6.3,-2.8) {$\dbr  $};
		\node (d5C2) at (-5.7,-3.3) {$\dbr  $};
		\path[red,thick] (-7,-2.5) edge (-7,-3.5);
		\path[red,thick] (-5,-2.5) edge (-5,-3.5);

		\node (aux1) at (-6,-0.4) {};
		\node (aux2a) at (-7.5,-1.1) {};
		\node (aux3a) at (-7.5,-1.9) {};
		\node (aux2b) at (-4.5,-1.1) {};
		\node (aux3b) at (-4.5,-1.9) {};
		\node (aux4) at (-6,-2.6) {};
		
		\draw[->] (aux1) edge (aux2a);
		\draw[->] (aux1) edge (aux2b);
		\draw[->] (aux3a) edge (aux4);
		\draw[->] (aux3b) edge (aux4);
	\end{tikzpicture}
	\end{equation*}	
\item $\lambda= (2^3)$. $\mX_{(2^3)}$ has three separated, indistinguishable $A_1$ singularities. However, due to the limitation on the rank (equivalently, on the number of D3 branes), only two transitions are available. $\fX$ in this case is recovered amputating the crystal for $\mathrm{U} (3)$ with $6$ flavours and $\lambda =(2^3)$ given in Appendix \ref{app:U3n6} below.
\item $\lambda =(4,2)$. $\mX_{(4,2)}$ has an $A_3$ and an $A_1$ singularity. Due to the limitation on the rank, only two out of the three potential transitions can be realized. $\fX$ is obtained amputating the bottom of the crystal for $\U{3}$ with $6$ flavours and $\lambda =(4,2)$.
\item $\lambda=(3^2)$. The $A_5$ singularity of the massless Coulomb branch splits into two separated $A_2$ singularities.
	\begin{equation*}
	\ytableausetup{smalltableaux}
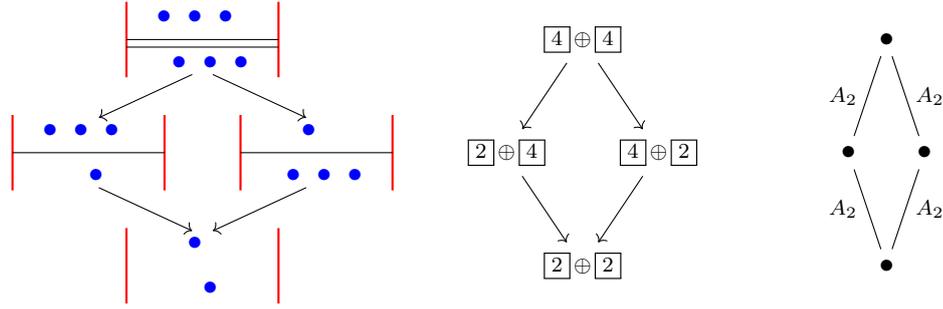
	\begin{tikzpicture}
		\node (t1) at (0,0) {$\begin{ytableau} 4 & \none[ \scriptstyle \oplus ] & 4 \end{ytableau}$}  ; 
		\node (t2a) at (-1,-1.5) {$\begin{ytableau} 2 & \none[ \scriptstyle \oplus ] & 4 \end{ytableau}$}  ; 
		\node (t2b) at (1,-1.5) {$\begin{ytableau} 4 & \none[ \scriptstyle \oplus ] & 2 \end{ytableau}$}  ; 
		\node (t3) at (0,-3) {$\begin{ytableau} 2 & \none[ \scriptstyle \oplus ] & 2 \end{ytableau}$}  ; 
		\node (h1) at (4,0) {$\bullet$};
		\node (h2a) at (3.5,-1.5) {$\bullet$};
		\node (h2b) at (4.5,-1.5) {$\bullet$};
		\node (h3) at (4,-3) {$\bullet$};

		\draw[->] (t1) edge (t2a);
		\draw[->] (t1) edge (t2b);
		\draw[->] (t2a) edge (t3);
		\draw[->] (t2b) edge (t3);
		\path[] (h1) edge node[left] {$\scriptstyle A_{2}$} (h2a);
		\path[] (h1) edge node[right] {$\scriptstyle A_{2}$} (h2b);
		\path[] (h2a) edge node[left] {$\scriptstyle A_{2}$} (h3);
		\path[] (h2b) edge node[right] {$\scriptstyle A_{2}$} (h3);
		
		\node (d5a1) at (-5.1,0.3) {$ \dbr \ \dbr \ \dbr  $};
		\node (d5a4) at (-4.9,-0.3) {$\dbr \ \dbr \ \dbr  $};
		\path[red,thick] (-6,0.5) edge (-6,-0.5);
		\path[red,thick] (-4,0.5) edge (-4,-0.5);
		\path (-6,0) edge (-4,0);
		\path (-6,-0.1) edge (-4,-0.1);

		\node (d5b1A) at (-6.6,-1.2) {$\dbr \ \dbr \ \dbr $};
		\node (d5b2A) at (-6.4,-1.8) {$\dbr   $};
		\path[red,thick] (-7.5,-1) edge (-7.5,-2);
		\path[red,thick] (-5.5,-1) edge (-5.5,-2);
		\path (-7.5,-1.5) edge (-5.5,-1.5);
		
		\node (d5b1B) at (-3.4,-1.8) {$\dbr \ \dbr \ \dbr $};
		\node (d5b2B) at (-3.6,-1.2) {$\dbr   $};
		\path[red,thick] (-4.5,-1) edge (-4.5,-2);
		\path[red,thick] (-2.5,-1) edge (-2.5,-2);
		\path (-4.5,-1.5) edge (-2.5,-1.5);

		\node (d5b1) at (-5.1,-2.7) {$\dbr  $};
		\node (d5b2) at (-4.9,-3.3) {$\dbr  $};
		\path[red,thick] (-6,-2.5) edge (-6,-3.5);
		\path[red,thick] (-4,-2.5) edge (-4,-3.5);
		
		\node (aux1) at (-5,-0.4) {};
		\node (aux2a) at (-6.5,-1.1) {};
		\node (aux2b) at (-3.5,-1.1) {};
		\node (aux3a) at (-6.5,-1.9) {};
		\node (aux3b) at (-3.5,-1.9) {};
		\node (aux4) at (-5,-2.6) {};
		
		\draw[->] (aux1) edge (aux2a);
		\draw[->] (aux1) edge (aux2b);
		\draw[->] (aux3a) edge (aux4);
		\draw[->] (aux3b) edge (aux4);
	\end{tikzpicture}
	\end{equation*}	

\item $\lambda =(5,1)$. $\mX_{(5,1)}$ has a singularity which is minimally resolved by the mass deformation.
	\begin{equation*}
	\ytableausetup{smalltableaux}
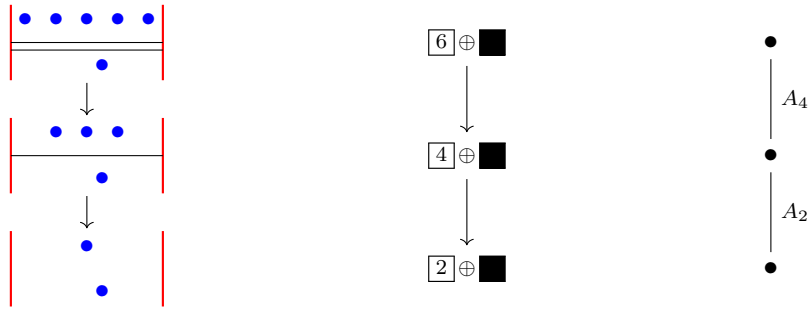
	\begin{tikzpicture}
		\node (t1) at (0,0) {$\begin{ytableau} 6 & \none[ \scriptstyle \oplus ] & *(black)  \end{ytableau}$}  ; 
		\node (t2) at (0,-1.5) {$\begin{ytableau} 4 & \none[ \scriptstyle \oplus ] & *(black)  \end{ytableau}$}  ; 
		\node (t3) at (0,-3) {$\begin{ytableau} 2 & \none[ \scriptstyle \oplus ] & *(black)  \end{ytableau}$}  ; 
		\node (h1) at (4,0) {$\bullet$};
		\node (h2) at (4,-1.5) {$\bullet$};
		\node (h3) at (4,-3) {$\bullet$};
		
		\draw[->] (t1) edge (t2);
		\draw[->] (t2) edge (t3);
		\path[] (h1) edge node[right] {$\scriptstyle A_{4}$} (h2);
		\path[] (h2) edge node[right] {$\scriptstyle A_{2}$} (h3);
		
		\node (d5a1) at (-5,0.3) {$ \dbr \ \dbr \ \dbr \ \dbr \ \dbr $};
		\node (d5a4) at (-4.8,-0.3) {$\dbr  $};
		\path[red,thick] (-6,0.5) edge (-6,-0.5);
		\path[red,thick] (-4,0.5) edge (-4,-0.5);
		\path (-6,0) edge (-4,0);
		\path (-6,-0.1) edge (-4,-0.1);
		
		\node (d5b1) at (-5,-1.2) {$\dbr \ \dbr \ \dbr  $};
		\node (d5b3) at (-4.8,-1.8) {$\dbr  $};
		\path[red,thick] (-6,-1) edge (-6,-2);
		\path[red,thick] (-4,-1) edge (-4,-2);
		\path (-6,-1.5) edge (-4,-1.5);
		
		\node (d5C1) at (-5,-2.7) {$ \dbr  $};
		\node (d5C2) at (-4.8,-3.3) {$ \dbr  $};
		\path[red,thick] (-6,-2.5) edge (-6,-3.5);
		\path[red,thick] (-4,-2.5) edge (-4,-3.5);
		
		\node (aux1) at (-5,-0.4) {};
		\node (aux2) at (-5,-1.1) {};
		\node (aux3) at (-5,-1.9) {};
		\node (aux4) at (-5,-2.6) {};
		\draw[->] (aux1) edge (aux2);
		\draw[->] (aux3) edge (aux4);
	\end{tikzpicture}
	\end{equation*}	
	
\item $\lambda =(6)$. This is the massless case and $\fX$ is found amputating the bottom of the crystal for $\U{3}$ with $\lambda=(6)$ given in Appendix \ref{app:U3n6}.
\end{itemize}

\subsubsection{$\U{3}$ with $6$ flavours}
\label{app:U3n6}
The stratification of the parameter space is independent of the rank, thus most of the singularity structure of $\mX $ agrees with the rank two case. Here we report only on those cases that differ from the lower rank case.\par

\begin{itemize}
	\item $\lambda=(2^3)$. The three singularities are related by triality, manifestly inherited by $\fX_{(2^3)}$:
		\begin{equation*}
		\ytableausetup{smalltableaux}

	\end{equation*}

	\item $\lambda= (6)$. The singularity $\mathcal{C}$ is not resolved and $\fX_{(6)} = \fC$, 
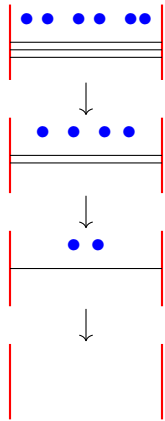
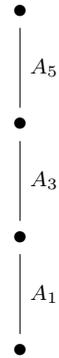
		\begin{equation*}
		\ytableausetup{smalltableaux}
	%
	\end{equation*}	
\end{itemize}

\subsubsection{$\U{3}$ with $7$ flavours}
When the number of hypermultiplets is $n=7$, the parameter space $\bigsqcup_{\lambda} \mM_{\lambda} ^{(7)}$ is stratified as:
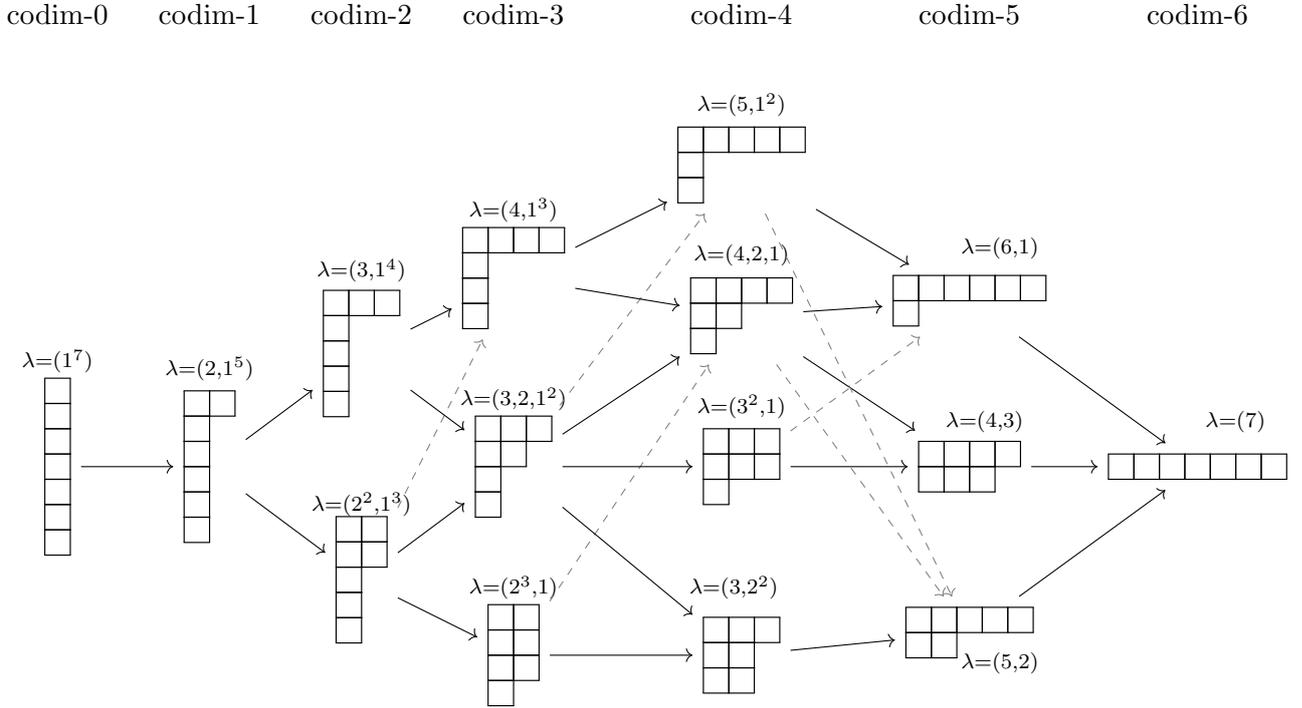
\begin{equation*}
\ytableausetup{smalltableaux}
	%
\end{equation*}
Partitions of $7$ that differ from a partition of $6$ by addition of a row of a single box do not add singular loci, thus do not affect the symplectic foliation and we neglect them. For those $\lambda$, in fact, $\mC$ is partly resolved into the Coulomb branch of the $\lvert \lambda \rvert=6$ theory.\par
	
\begin{itemize}
	\item $\lambda=(3,2^2)$.
		\begin{equation*}
		\ytableausetup{smalltableaux}
	%
	\end{equation*}	
	with corresponding brane configurations 
	\begin{equation*}
		%
	\end{equation*}	
		
\end{itemize}

\subsubsection{Balanced $A_r$ quiver of rank $2r$}
Consider the balanced $A_r$ quiver with gauge group $\U{2}^r$ and framing $\un =(2, 0, \dots, 0,2)$, 
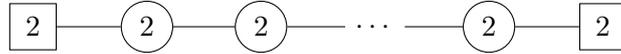
\begin{equation*}
	%
\end{equation*}
and assume $r \ge 4$. The crystal $\fX$ varies along $\mM^{(4)}$ as follows.
\begin{itemize}
\item $\lambda=(1^4)$. $\mC$ is fully resolved and $\fX_{(1^4)}$ is resolved into the direct sum of four single-box tableaux.
\item $\lambda =(2,1^2)$. There exist two partial resolution in this case, depending on whether 
\begin{equation*}
\left( \mathsf{w}_1 (\lambda_1), \mathsf{w}_2 (\lambda_1) \right) = (2,0) \quad \text{ or } \quad \left( \mathsf{w}_1 (\lambda_1), \mathsf{w}_2 (\lambda_1) \right) = (1,1) ,
\end{equation*}
that is, the two equal masses belong to the same node or are at the opposite sides of the quiver: 
	\begin{equation*}
		\ytableausetup{smalltableaux}

	\end{equation*}
\item $\lambda =(2,2)$. Again there are two resolved crystals, depending on the assignment of masses.
	\begin{equation*}
		\ytableausetup{smalltableaux,nobaseline}
	%
	\end{equation*}

\end{itemize}

\subsubsection{Unbalanced $A_3$ quiver of rank five with four flavours}
\label{app:A3G5F4}
Our next example is the unbalanced $A_3$ quiver with gauge group $\U{1}^2 \times \U{3}$ and framing $\un =(0,4,0)$, 
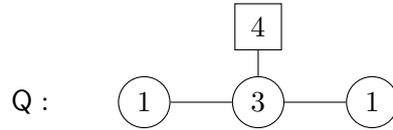
\begin{equation*}
	%
\end{equation*}
As usual, generic masses $\lambda=(1^4)$ yield a non-singular resolution $\mX$ of $\mC$ and the crystal $\fX_{(1^4)}$ consists of the direct sum of single-box tableaux.

\begin{itemize}
	\item $\lambda=(2,1^2)$. 
		\begin{equation*}
		\ytableausetup{smalltableaux}
		%
	\end{equation*}
	The last transitions, indicated by dotted lines, would appear in the crystal $\fX_{(2^2)} \left[ \mathsf{Q}^{\prime} \right]$ for the balanced quiver 
	\begin{equation}
	%
	\label{eq:defQprimeQuiverApp}
	\end{equation}
	The unbalanced quiver $\mathsf{Q}$ under consideration can be realized on $\mX \left[ \mathsf{Q}^{\prime} \right]$ and its crystal is obtained amputating the dotted transitions.
	
\item $\lambda =(3,1)$.
	\begin{equation*}
		\ytableausetup{smalltableaux}
		%
	\end{equation*}
	with corresponding brane configurations 
	\begin{equation*}
		%
	\end{equation*}

\end{itemize}

\subsubsection{Balanced $A_5$ quiver of rank fourteen with four flavours}

Consider the family of balanced $A_r$ quivers with gauge group $\U{2}^2 \times \U{4}^{r-2}$, $r \ge 3$, 
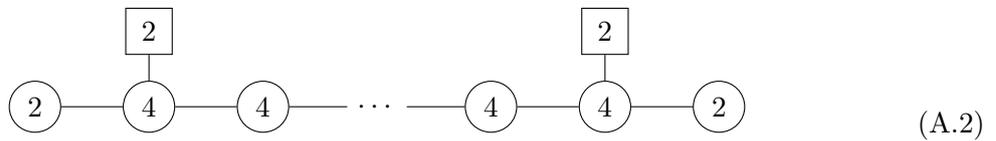
\begin{equation}
	%
	\label{eq:defG4rF22QuiverApp}
	\end{equation}
	The unbalanced quiver studied in Appendix \ref{app:A3G5F4} is realized as a union of strata on the Coulomb branch of the $r=3$ member of the family \eqref{eq:defG4rF22QuiverApp}, given in \eqref{eq:defQprimeQuiverApp}. We now discuss another example, for $r=5$:
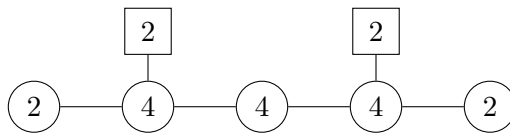
\begin{equation*}
	%
\end{equation*}

\begin{itemize}
\item For $\lambda=(2,1^2)$ the are two resolution patterns, depending on whether or not the two equal masses belong to the same flavour node, 
		\begin{equation*}
		\ytableausetup{smalltableaux}
		%
	\end{equation*}
\item For $\lambda=(2^2)$ the are again two resolved crystals: 
	\begin{equation*}
	\hspace*{-2cm}
		\ytableausetup{smalltableaux,nobaseline}
		%
	\end{equation*}
	where we used the shorthand $\left( \begin{ytableau}  1  \end{ytableau} \right)^{\otimes 5 }  \equiv  \underbrace{ \begin{ytableau}  1 & \none[ \scriptstyle \otimes ] & \none[ \scriptstyle \cdots ] & \none[ \scriptstyle \otimes ] & 1 \end{ytableau} }_{5} $.
\item For $ \lambda=(3,1) $ there exists a single crystal $\fX_{(3,1)}$: 
	\begin{equation*}
		\ytableausetup{smalltableaux,nobaseline}
		%
	\end{equation*}

\item Finally, the non-resolved case $\lambda =(4)$ shows the richest structure: 
	\begin{equation*}
		\ytableausetup{smalltableaux,nobaseline}
		%
	\end{equation*}

\end{itemize}

\section{Induction on the crystal}
\label{app:indlim}
In this appendix we discuss a limit of the setup of Section \ref{sec:C=C} together with its interpretation in type IIB string theory. We work with type A quivers.\par
The highest weight tableau for a quiver with framing $\un = (\mathsf{w}_1, \mathsf{w}_2, \dots, \mathsf{w}_r )$ is 
\begin{equation*}
	\ytableausetup{smalltableaux} 
	%
\end{equation*}
We then take the limit 
\begin{equation*}
	 T_{(\infty)} \equiv \lim_{\mathsf{w}_1 \to \infty} \lim_{\mathsf{w}_2 \to \infty}  \cdots  \lim_{\mathsf{w}_r \to \infty} \begin{ytableau} {\scriptstyle 1 }  & \none[ \scriptstyle \cdots ] & {\scriptstyle 1}  & {\scriptstyle 2} & \none[ \scriptstyle \cdots ] & {\scriptstyle 2}  & \none[ \scriptstyle \cdots ]  & {\scriptstyle r} & \none[ \scriptstyle \cdots ] & {\scriptstyle r} \end{ytableau} 
\end{equation*}
and declare $T_{(\infty)} $ to be the highest weight tableau of an $A_r$ crystal. Acting with Kashiwara operators $f_j$ on $T_{(\infty)} $ generates the whole $\Gr _{\mathrm{PSL} (r+1)}$ from a top-down perspective. From the correspondence between tableaux in a crystal and phases of Hanany--Witten brane configurations, detailed in Section \ref{sec:crystalBrane}, the top-down approach to $\Gr _{\mathrm{PSL} (r+1)}$ corresponds to send the number of D5 and D3 branes in each interval to infinity, keeping the number $r+1$ of NS5 branes fixed. Then, one starts considering the sequence of all allowed phases in the Hanany--Witten setup.\par
\medskip
In addition, the crystal structure of Section \ref{sec:C=C} provides a bottom-up approach to $\Gr _{\mathrm{PSL}(r+1)}$. The starting point is the obvious observation that any highest weight tableau $T$ can be embedded in a larger tableau with arbitrarily many $0$s on the left and arbitrarily many $(r+1)$s on the right, 
\begin{equation*}
\ytableausetup{smalltableaux} 
T = \begin{ytableau} {\scriptstyle 1 }  & \none[ \scriptstyle \cdots ] & {\scriptstyle 1}  & {\scriptstyle 2} & \none[ \scriptstyle \cdots ] & {\scriptstyle 2}  & \none[ \scriptstyle \cdots ]  & {\scriptstyle r} & \none[ \scriptstyle \cdots ] & {\scriptstyle r} & \none & \none[  \cong  ] & \none & \none[ {\scriptstyle \cdots } ] & {\scriptstyle 0} & {\scriptstyle 0 } & {\scriptstyle 1 }  & \none[ \scriptstyle \cdots ] & {\scriptstyle 1}  & {\scriptstyle 2} & \none[ \scriptstyle \cdots ] & {\scriptstyle 2}  & \none[ \scriptstyle \cdots ]  & {\scriptstyle r} & \none[ \scriptstyle \cdots ]  & {\scriptstyle r} & { \scriptstyle  \underset{+1}{r} } & {\scriptstyle \underset{+1}{r} }  &  \none[ {\scriptstyle \cdots} ] \end{ytableau} 
\end{equation*}
Then, highest weight tableaux $T^{\prime}$ with $\wt (T^{\prime}) > \wt (T)$ are generated acting on $T$ with Kashiwara operators $e_j$. The crucial point is that it is possible to construct the full $\Gr _{\mathrm{PSL}(r+1)}$ in this way via an inductive limit (cf. the discussion at the end of \cite{Bourget:2021siw}). For example, the bottom-up construction of the $A_1$ crystal starting from $T= \varnothing $ (left) or $T= \begin{ytableau} {\scriptstyle 1 } \end{ytableau} $ (right) is:
\begin{equation*}
	\ytableausetup{smalltableaux} 
	\begin{tikzpicture}
		\node (T0r) at (3,0) {$\begin{ytableau} \none[ {\scriptstyle \cdots } ] & {\scriptstyle 0} & {\scriptstyle 0 } & {\scriptstyle 1 }  & {\scriptstyle 2 } & {\scriptstyle 2 } & \none[ {\scriptstyle \cdots }]\end{ytableau}$};
		\node (T1r) at (3,-1) {$\begin{ytableau} \none[ {\scriptstyle \cdots } ] & {\scriptstyle 0} & {\scriptstyle 0 } & {\scriptstyle 1 } & {\scriptstyle 1 } & {\scriptstyle 1 } & {\scriptstyle 2 } & {\scriptstyle 2 } & \none[ {\scriptstyle \cdots }]\end{ytableau}$};
		\node (T2r) at (3,-2) {$\begin{ytableau} \none[ {\scriptstyle \cdots } ] & {\scriptstyle 0} & {\scriptstyle 0 } & {\scriptstyle 1 }  & {\scriptstyle 1 } & {\scriptstyle 1 }  & {\scriptstyle 1 } & {\scriptstyle 1 } & {\scriptstyle 2 } & {\scriptstyle 2 } & \none[ {\scriptstyle \cdots }]\end{ytableau}$};
		\node (T0l) at (-3,0) {$\begin{ytableau} \none[ {\scriptstyle \cdots } ] & {\scriptstyle 0} & {\scriptstyle 0 }  & {\scriptstyle 2 } & {\scriptstyle 2 } & \none[ {\scriptstyle \cdots }]\end{ytableau}$};
		\node (T1l) at (-3,-1) {$\begin{ytableau} \none[ {\scriptstyle \cdots } ] & {\scriptstyle 0} & {\scriptstyle 0 } & {\scriptstyle 1 } & {\scriptstyle 1 } & {\scriptstyle 2 } & {\scriptstyle 2 } & \none[ {\scriptstyle \cdots }]\end{ytableau}$};
		\node (T2l) at (-3,-2) {$\begin{ytableau} \none[ {\scriptstyle \cdots } ] & {\scriptstyle 0} & {\scriptstyle 0 } & {\scriptstyle 1 } & {\scriptstyle 1 }  & {\scriptstyle 1 } & {\scriptstyle 1 } & {\scriptstyle 2 } & {\scriptstyle 2 } & \none[ {\scriptstyle \cdots }]\end{ytableau}$};
		\node (ar) at (3,-3) {$\vdots$};
		\node (al) at (-3,-3) {$\vdots$};
		
		\path[->] (T0r) edge (T1r);
		\path[->] (T1r) edge (T2r);
		\path[->] (T0l) edge (T1l);
		\path[->] (T1l) edge (T2l);
		\path[->] (T2r) edge (ar);
		\path[->] (T2l) edge (al);
	\end{tikzpicture}
\end{equation*}
which, using \eqref{eq:bijcrys}, is an upside-down view of \eqref{eq:PSU2ex}. For $T=\begin{ytableau} {\scriptstyle 1 } & {\scriptstyle 2 } \end{ytableau} $ and $A_2$ root system, we get instead 
\begin{equation*}
	\ytableausetup{smalltableaux} 
	\begin{tikzpicture}
		\node (T0) at (0,0) {$\begin{ytableau} \none[ {\scriptstyle \cdots } ] & {\scriptstyle 0} & {\scriptstyle 0 } & {\scriptstyle 1 }  & {\scriptstyle 2 }  & {\scriptstyle 3 } & {\scriptstyle 3 } & \none[ {\scriptstyle \cdots }]\end{ytableau}$};
		\node (T1l) at (-2.5,-1) {$\begin{ytableau} \none[ {\scriptstyle \cdots } ] & {\scriptstyle 0} & {\scriptstyle 0 } & {\scriptstyle 1 } & {\scriptstyle 1 } & {\scriptstyle 1 }  & {\scriptstyle 3 } & {\scriptstyle 3 } & \none[ {\scriptstyle \cdots }]\end{ytableau}$};
		\node (T1r) at (2.5,-1) {$\begin{ytableau} \none[ {\scriptstyle \cdots } ] & {\scriptstyle 0} & {\scriptstyle 0 } & {\scriptstyle 2 }  & {\scriptstyle 2 } & {\scriptstyle 2 }  & {\scriptstyle 3 } & {\scriptstyle 3 } & \none[ {\scriptstyle \cdots }]\end{ytableau}$};
		\node (T2) at (0,-2) {$\begin{ytableau} \none[ {\scriptstyle \cdots } ] & {\scriptstyle 0} & {\scriptstyle 0 } & {\scriptstyle 1 } & {\scriptstyle 1 }  & {\scriptstyle 2 }  & {\scriptstyle 2 }  & {\scriptstyle 3 } & {\scriptstyle 3 } & \none[ {\scriptstyle \cdots }]\end{ytableau}$};
		\node (T3l) at (-2.5,-3) {$\begin{ytableau} \none[ {\scriptstyle \cdots } ] & {\scriptstyle 0} & {\scriptstyle 0 } & {\scriptstyle 1 } & {\scriptstyle 1 } & {\scriptstyle 1 } & {\scriptstyle 1 } & {\scriptstyle 2 }  & {\scriptstyle 3 } & {\scriptstyle 3 } & \none[ {\scriptstyle \cdots }]\end{ytableau}$};
		\node (T3r) at (2.5,-3) {$\begin{ytableau} \none[ {\scriptstyle \cdots } ] & {\scriptstyle 0} & {\scriptstyle 0 } & {\scriptstyle 1 }  & {\scriptstyle 2 } & {\scriptstyle 2 }  & {\scriptstyle 2 } & {\scriptstyle 2 }  & {\scriptstyle 3 } & {\scriptstyle 3 } & \none[ {\scriptstyle \cdots }]\end{ytableau}$};
		\node (ar) at (5,-3.5) {$\vdots$};
		\node (al) at (-5,-3.5) {$\vdots$};
		\node (ac) at (0,-3.5) {$\vdots$};

		\path[->] (T0) edge (T1l);
		\path[->] (T0) edge (T1r);
		\path[->] (T1r) edge (T2);
		\path[->] (T1l) edge (T2);
		\path[->] (T2) edge (T3l);
		\path[->] (T2) edge (T3r);
		\path[->] (T3r) edge (ar);
		\path[->] (T3l) edge (al);
		\path[->] (T3r) edge (ac);
		\path[->] (T3l) edge (ac);

\end{tikzpicture}
\end{equation*}
which agrees with a bottom-up view of the corresponding slice in $\Gr_{\mathrm{PSL}(3)}$.\par
By the result in Section \ref{sec:crystalBrane}, embedding a tableau into a larger one with $0$s on the left and $(r+1)$s on the right is the same as including arbitrarily many D5 branes at infinity on the left of the leftmost NS5 and on the right of the rightmost NS5. This completes the proof of equivalence of our construction of the affine Grassmannian via Kashiwara crystals with the brane construction of \cite{Bourget:2021siw}.

\end{appendix}

	\bibliography{CBandHB,HilbertSeries,AG_crystals}
	
\end{document}